\documentclass[journal]{vgtc}                     

\onlineid{9456}

\vgtccategory{IEEE PacificVis Journal Track}

\title{SurfPatch: Enabling Patch Matching for Exploratory\\Stream Surface Visualization}

\author{%
\authororcid{Delin An}{0000-0002-7945-0201} and
  \authororcid{Chaoli Wang}{0000-0002-0859-3619}
}

\authorfooter{
  \item D.\ An and C.\ Wang are with 
the Department of Computer Science and Engineering, University of Notre Dame, Notre Dame, IN 46556, USA.
E-mail: \{dan3, chaoli.wang\}@nd.edu.
}

\abstract{Unlike their line-based counterparts, surface-based techniques have yet to be thoroughly investigated in flow visualization due to their significant placement, speed, perception, and evaluation challenges. This paper presents SurfPatch, a novel framework supporting exploratory stream surface visualization. To begin with, we translate the issue of surface placement to surface selection and trace a large number of stream surfaces from a given flow field dataset. Then, we introduce a three-stage process: vertex-level classification, patch-level matching, and surface-level clustering that hierarchically builds the connection between vertices and patches and between patches and surfaces. This bottom-up approach enables fine-grained, multiscale patch-level matching, sharply contrasts surface-level matching offered by existing works, and provides previously unavailable flexibility during querying. We design an intuitive visual interface for users to conveniently visualize and analyze the underlying collection of stream surfaces in an exploratory manner. SurfPatch is not limited to stream surfaces traced from steady flow datasets. We demonstrate its effectiveness through experiments on stream surfaces produced from steady and unsteady flows as well as isosurfaces extracted from scalar fields. The code is available at \url{https://github.com/adlsn/SurfPatch}.}


\graphicspath{{figs/}{figures/}{pictures/}{images/}{./}} 

\usepackage{tabu}                      
\usepackage{booktabs}                  
\usepackage{lipsum}                    
\usepackage{mwe}                       

\usepackage{mathptmx}                  

\usepackage{graphicx}
\usepackage{times}
\usepackage{algorithm}
\usepackage{algorithmic}
\usepackage{amsfonts}
\usepackage{booktabs}         
\usepackage{gensymb}
\usepackage{amsmath}
\usepackage{comment}
\usepackage{times} 
\usepackage{caption}
\usepackage{bm}
\usepackage{multirow}
\usepackage{color} 
\usepackage{anyfontsize}
\usepackage{soul}
\usepackage{makecell}

\DeclareMathOperator*{\uagg}{Uagg}
\DeclareMathOperator*{\umap}{UMAP}
\DeclareMathOperator*{\expp}{exp}

\newenvironment{myitemize}{
\begin{itemize}
 \setlength{\itemsep}{1pt}
 \setlength{\parskip}{0pt}
 \setlength{\parsep}{0pt}}{\end{itemize}
 
}

\begin{document}


\firstsection{Introduction}

\maketitle

Flow visualization~\cite{VHB-2005} has long been an established branch of scientific visualization research. 
Among {\em glyph}-, {\em geometry}-, and {\em texture}-based methods, geometry-based methods, which work by integrating flow {\em lines}, {\em surfaces}, and {\em volumes} in 1D, 2D, and 3D, have become the predominant means to visualize the underlying three-dimensional flow fields~\cite{McLoughlin-CGF10}. 
Their popularity can be attributed to their superior capability to display continuous flow features and patterns in 3D (unlike glyph-based methods) while mitigating visual occlusion and clutter (unlike texture-based methods)~\cite{Yu-TVCG12}. 

Within the realm of geometry-based flow visualization, line-based flow visualization is extensively studied, partially due to the relative straightforwardness of the problem (e.g., seeding a streamline is much simpler than seeding a stream surface) and partially because the sparse line representation makes it more visually scalable than surface or volume representation~\cite{Tao-TVCG2018}. 
Despite the flourishing of line-based flow visualization, {\em surface-based flow visualization} (SBFV)~\cite{Edmunds-CG2012} generates stream surfaces by tracing an array of flow lines along a seeding rake (or curve) and connecting them in between. 
The resulting stream surfaces go beyond flow lines by depicting folding, shearing, and twisting behaviors, enhancing visual observation of complicated flow structures, thus helping intuitive understanding of flow geometry~\cite{Garth-VISSYM04}. 
Nevertheless, SBFV still faces significant challenges, including surface placement, computation speed, visual perception, and user evaluation~\cite{Edmunds-CG2012}. 
Seeding a stream surface is more challenging than seeding a streamline due to its dependency on the length and orientation of the seeding curve, which directly impacts the surface quality and subsequent analysis~\cite{Tao-TVCG2018}.  
Moreover, interpreting stream surfaces can be difficult, as some visual characteristics may result from artifacts introduced during seeding rather than faithfully representing the underlying physical behavior.

As an alternative to {\em surface placement}, we can first randomly trace many (e.g., hundreds or thousands) stream surfaces from a given flow dataset and then identify interesting flow features via {\em surface selection}, akin to {\em streamline selection}~\cite{Ma-VDA13, Tao-TVCG13}. 
Different parts of a stream surface typically exhibit dramatically different flow patterns (e.g., flat vs.\ convoluted). 
Existing works on surface clustering and selection, however, only partition individual surfaces~\cite{Han-SurfNet} and then group surfaces as a whole~\cite{Han-FlowNet, Han-SurfNet}. 
This precludes matching and querying surfaces at a fine scale and across multiple surfaces. 
Prior work on flow line exploration~\cite{Tao-FlowString, Tao-TVCG16} suggests that enabling fine-grained partial surface (i.e., {\em patch}) matching and querying is beneficial. 
However, achieving this goal for stream surfaces poses a significant challenge as such line-based partitioning and matching techniques cannot be easily extended to surfaces to yield satisfactory results. Unlike 1D line segments, 2D surface patches cannot be uniformly produced along the streamline and timeline directions due to their potentially distinct flow characteristics. 
Finally, we lack interactive visual interfaces for exploring a large set of stream surfaces toward effective visual analysis of the underlying flow fields. 
In this work, we aim to improve visual perception and user interaction for SBFV to advance the state of the art. 

This paper presents SurfPatch, an interactive framework for exploratory stream surface visualization. 
We advocate a bottom-up strategy to build the connections between {\em vertices} and {\em patches}, and between {\em patches} and {\em surfaces}, from which we design three pillars, {\em vertex-level classification}, {\em patch-level matching}, and {\em surface-level clustering}, to enable effective partial surface matching. These three pillars serve different purposes. 
Vertex-level classification groups vertices based on their local neighborhood's shape and topological features to form patches, the central building block for subsequent visual exploration.
Patch-level matching supports flexible querying of similar flow patterns at varying levels of details, invariant to translation, scaling, and rotation.
Surface-level clustering organizes the collection of stream surfaces hierarchically based on similarity, facilitating the retrieval of surfaces of interest for patch query across multiple surfaces. 
Moreover, we present an intuitive visual interface for users to conveniently perform patch matching and querying for exploratory stream surface visualization and analysis. 
We illustrate the effectiveness of SurfPatch using several flow datasets of various sizes and characteristics. 
We primarily focus on stream surfaces generated from steady flow. In addition, we provide experimental results on stream surfaces traced from unsteady flow and isosurfaces extracted from scalar fields (refer to Section~\ref{subsec:more}), highlighting the usefulness of SurfPatch for analyzing different kinds of surfaces from scientific visualization applications.

\begin{figure*}[htbp]
    \centering
    \includegraphics[width=0.95\linewidth]{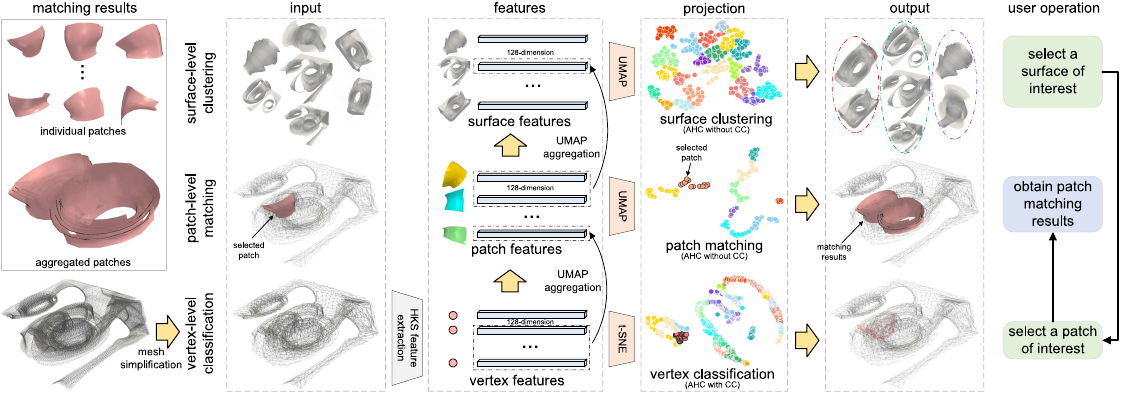} 
    \vspace{-0.1in}
    \caption{The SurfPatch framework consists of three stages: vertex-level classification, patch-level matching, and surface-level clustering. We aggregate HKS features from vertices to patches and from patches to surfaces, enabling patch matching within a single surface and across multiple surfaces.}
    \vspace{-0.125in}
    \label{fig1}
\end{figure*}

The contributions of our SurfPatch are as follows:
\begin{myitemize}
\item We design a three-stage bottom-up approach to connect vertices, patches, and surfaces, enabling flexible patch matching and querying at runtime. 
\item We present an intuitive visual interface and provide a suite of interactions to support exploratory stream surface visualization and analysis.
\item We conduct a comprehensive study to elucidate our design choices and show the advantages of SurfPatch over other methods. 
\item We experiment with SurfPatch using stream surfaces from vector fields (steady and unsteady flow) and isosurfaces from scalar fields, demonstrating its broad applicability.
\end{myitemize}

\vspace{-0.1in}
\section{Related work}

{\bf Stream surface construction and rendering.} Methods for stream surface construction encompass {\em point}-, {\em triangle}-, and {\em quad}-based techniques.  Point-based surfaces are simple and easy to construct. For instance, Schafhitzel et al.\ \cite{Schafhitzel-GI2007} employed a point-based stream with a path surface technique for surface generation.  Triangle-based surfaces use an index array to represent vertices, reducing graphics bandwidth significantly. For example, van Wijk~\cite{Wijk-Vis1993} and Westermann et al.\ \cite{Westermann-Vis2000} constructed triangular stream surfaces through implicit techniques and level sets.  Quad-based surfaces, generated by bounding streamlines and timelines, require fewer computational resources than their triangle-based counterparts. Several works in this category~\cite{McLoughlin-CGI2009, Schneider-CGF2009, McLoughlin-SCCG2010} face issues such as sheering quads and t-junctions. To respond, Peikert and Sadlo~\cite{Peikert-SCCG2009} proposed an enhanced method for generating quad-based surface meshes by incrementally improving the initial curve structure. These methods focus on the efficient construction of stream surfaces.

Stream surface rendering methods employ contour lines, transparency, texturing, etc., to mitigate visual occlusion or clutter and enhance visual examination of flow structures. For illustrative rendering, Born et al.\ \cite{Born-TVCG2010} utilized contour lines to depict surface shapes and demonstrated the effectiveness of illustrative surface streamlines. Hummel et al.\ \cite{Hummel-TVCG2010} conveyed directional and shape information through adjustments in transparency and texturing. Other works solve the occlusion and clutter issue via opacity optimization~\cite{Carnecky-TVCG13, Gunther-CGF14, Gunther-CGF17}. 
Building on these works, we partition stream surfaces into patches and use color, transparency, lighting, and silhouette to differentiate patches and enhance perception, addressing visual occlusion and clutter challenges.

{\bf Feature extraction and representation learning.} Accurate feature extraction and representation learning from individual surfaces are crucial for understanding their structure and shape. Two main categories emerge when analyzing the shape patterns of 3D objects: {\em point-based} and {\em geometry-based}. Point-based approaches utilize 3D point clouds for feature extraction, primarily in computer vision tasks like 3D object classification. In contrast, geometry-based methods, commonly used in computer graphics, process 3D mesh data and rely on surface features such as curvature and normals.

Point cloud data, which is readily accessible, is widely used to represent various 3D objects. Early research focused on hand-designed features, as demonstrated by Wang et al.\ \cite{Wang-TGRS2014}. The advent of deep learning, exemplified by PointNet~\cite{Qi-CVPR2017}, has significantly advanced point-based feature extraction. However, due to their sparse, irregular nature, point clouds lack geometry and topology, making them unsuitable for analyzing complex surfaces like stream surfaces.

In computer graphics, 3D objects are commonly represented as meshes. Traditional methods for feature extraction from meshes involve calculating volume, moments, and Fourier transform coefficients~\cite{Zhang-ICIP2001}. Following the convolutional neural network (CNN) paradigm known for exceptional performance in computer vision, Monti et al.\ \cite{Monti-CVPR2017} introduced a mixture model CNN for processing graphs and manifolds. Wang et al.\ \cite{Wang-TOG2019} devised EdgeConv for extracting intrinsic features, while Haque et al.\ \cite{Haque-arXiv2022} introduced a self-supervised contrastive method for 3D mesh segmentation. Graph neural networks (GNNs) represent another crucial direction. Han and Wang~\cite{Han-SurfNet} demonstrated promising performance in extracting geometric representations of surfaces, including isosurfaces and stream surfaces. Their SurfNet, composed of a graph convolutional network (GCN), outperforms the CNN-based FlowNet~\cite{Han-FlowNet} in training efficiency. Despite its lightweight structure, SurfNet effectively extracts geometric features at the node level, enabling surface-level analysis through aggregating node features.
However, these methods may fall short in analyzing complex surfaces, as point-based methods miss surface topology, and geometry-based methods, relying on self-supervised learning, can be unstable for fine-grained mesh analysis.

To address these limitations, we employ the {\em heat kernel signature} (HKS)~\cite{Sun-HKS} for mesh feature extraction. HKS enables multiscale patch retrieval based on shape and structure, enhancing analysis through dimensionality reduction (DR) and clustering. This reliable method requires minimal training data and extends beyond stream surfaces to isosurfaces, showcasing its versatility for diverse surface data.

{\bf Exploratory interface and interaction.} A visual interface is crucial for a flow visualization system, whether presenting the data through streamlines or stream surfaces. Sketch- and touch-based interfaces~\cite{Wei-PVIS10, Schroeder-SBIM2010, Klein-CGF2012, Tao-TVCG2018} enable users to paint on data, sketch templates, or touch directly for intuitive exploration.
For flow lines, Angelelli and Hauser~\cite{Angelelli-TVCG11} proposed straightening tubular flow for informative visual analysis in arterial blood flow and tubular gas flow applications. Tao et al.\ \cite{Tao-FlowString, Tao-TVCG16} developed an exploratory interface for matching partial streamlines and pathlines.
Concerning stream surfaces, interactive interfaces have been designed for clustering stream surfaces and selecting representatives, as shown in~\cite{Han-FlowNet, Han-SurfNet}. Zhang et al.\ \cite{Zhang-SurfRiver} proposed a visual interface that flattens stream surfaces for comparative visualization. Similar to~\cite{Tao-FlowString, Han-SurfNet}, we develop a visual interface that supports interactive partial stream surface query and representative selection.

\vspace{-0.05in}
\section{SurfPatch}

To explore stream surfaces generated from vector field data, we propose a three-stage solution:
(1) classifying vertices based on shape and topological features in their neighborhoods to generate fine-grained patches that partition the surface,
(2) matching similar patches within a single surface via patch-level features, and
(3) clustering surfaces using surface-level features and enabling convenient patch query across multiple surfaces.
The key challenges lie in
(1) how to extract vertex-level intrinsic features to enable multiscale, flexible patch generation and
(2) how to aggregate fine-level features into coarse-level ones to support effective clustering and querying.
We address these challenges by employing machine learning methods for vertex-level classification, patch-level matching, and surface-level clustering.

Figure~\ref{fig1} shows our SurfPatch framework.
First, we simplify each stream surface to reduce subsequent computational costs while maintaining mesh quality~\cite{Garland-QEM}.
Next, we extract HKS features for each vertex on the surface, followed by vertex-level classification using {\em agglomerative hierarchical clustering} (AHC) with {\em connectivity constraints} (CC)
to generate patches that partition the surface.
We generate patch-level features by aggregating the corresponding vertex-level features. 
These features are used to query similar patches on a single surface or across multiple surfaces using AHC without CC. 
Finally, we aggregate patch-level features into surface-level features for clustering surfaces to facilitate surface selection for patch query across multiple selected surfaces.

\begin{figure}[htb]
    \centering
    \includegraphics[width=0.85\linewidth]{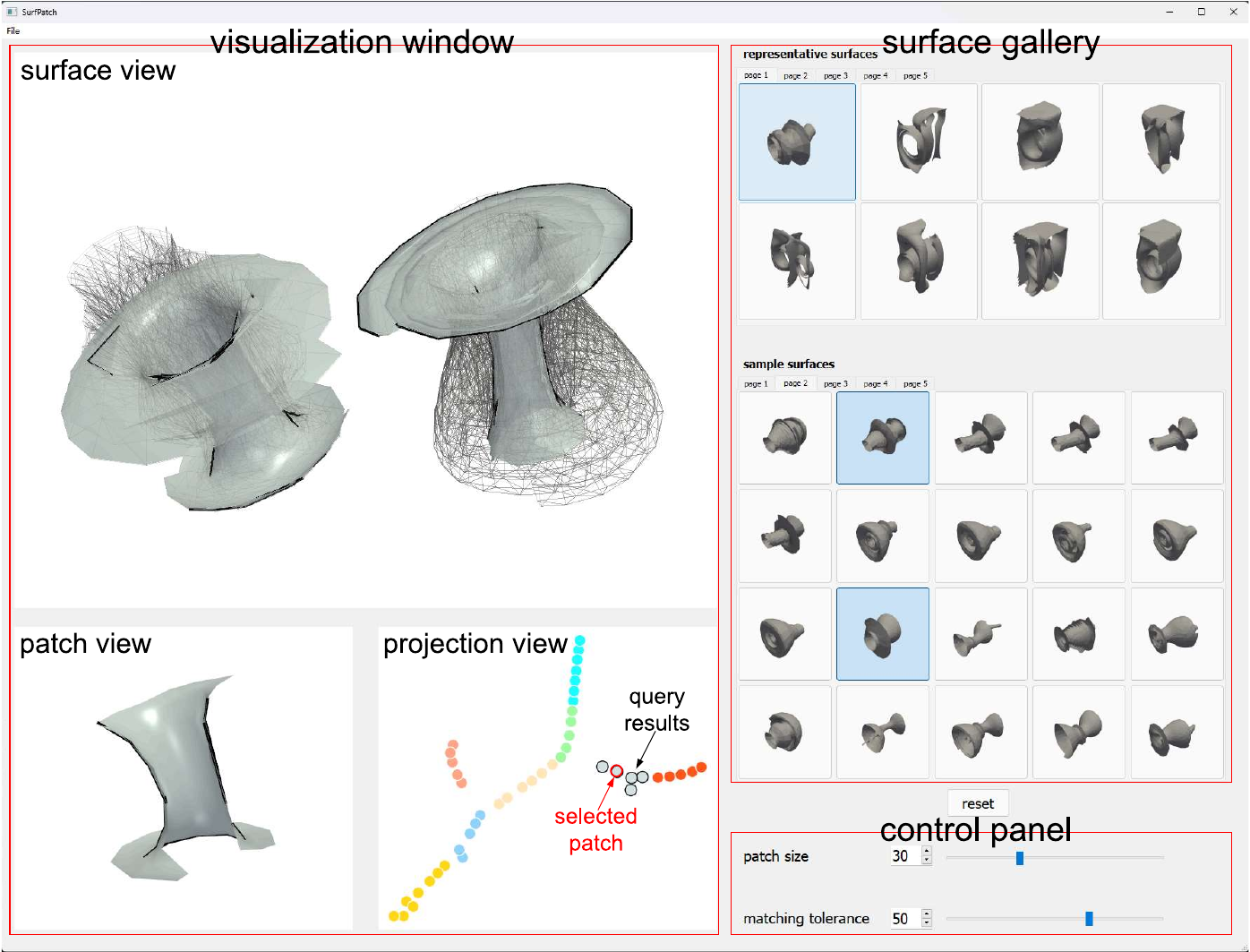} 
    \vspace{-0.1in}
    \caption{The screenshot of SurfPatch interface. Based on the selected patch, we query similar patches across two surfaces of interest retrieved from sample surfaces (highlighted with a light blue background).}
    \vspace{-0.1in}
    \label{fig3}
\end{figure}

\vspace{-0.05in}
\subsection{Interface and Workflow}

We begin by introducing the SurfPatch interface. As shown in Figure~\ref{fig3}, the interface comprises three main components: a visualization window, a surface gallery, and a control panel. The visualization window includes the {\em surface view} on the top (showing 3D stream surfaces and patch-matching results), the {\em patch view} on the bottom left, and the {\em projection view} on the bottom right (showing patch-level or surface-level 2D UMAP projection). The surface gallery showcases representative surfaces (ordered by the number of sample surfaces within) on the top and sample surfaces within each cluster (ordered by the similarity to the representative surface) on the bottom. The control panel provides parameters for users to fine-tune patch generation and matching results.

As shown on the right side of Figure~\ref{fig1}, the typical workflow for using SurfPatch starts with surface selection, patch selection, patch matching, and finally, parameter adjustment. We refer to SurfNet's interface when designing ours, as it shares similar objectives with SurfPatch. However, SurfPatch provides more fine-grained patches and improved patch-matching results, which prompts us to incorporate a dedicated patch view in our interface. Referring to Figure~\ref{fig3}, the steps are detailed below:

\begin{myitemize}
\vspace{-0.05in}
\item {\bf Surface selection.} Users start by clicking on a {\em representative surface} from the top gallery to choose a cluster, then select a {\em sample surface} within that cluster in the bottom gallery. Alternatively, users can select a surface by clicking a point in the projection view, which shows the {\em surface-level} UMAP projection. Once a surface is selected, it is displayed in the surface view, and the projection view updates to show the {\em patch-level} UMAP projection, where each point represents a patch of the selected surface.
\item {\bf Patch selection.} Users select interest patches in the projection view once a surface of interest is identified. We support point-clicking and lasso selection. Clicking on a single point renders the corresponding patch in the surface and patch views. Lasso selection allows users to choose multiple patches simultaneously. All selected patches are displayed in the surface view, while the patch view only highlights the last selected patch.
\item {\bf Patch matching.} After choosing a single patch, similar patches from the same surface are displayed in the surface view. To match patches across multiple surfaces, users can add additional surfaces from the bottom gallery to the surface view, and the system updates patch-matching results on the fly.
\item {\bf Parameter adjustment.} 
We provide two sliders in the control panel for parameter adjustment. The “patch size” slider is associated with the AHC's distance threshold $\delta_1$ for the {\em vertex-level} classification. Increasing $\delta_1$ results in a larger patch size for the query. The “matching tolerance” slider is related to the AHC's $\delta_2$ for the {\em patch-level} classification. Increasing $\delta_2$ leads to matching less similar patches, thus, a less refined query result.
\vspace{-0.05in}
\end{myitemize}

\vspace{-0.05in}
\subsection{Vertex Classification}

Vertex classification generates meaningful patches that partition the stream surface, each characterized by specific shape and topological features.
We approach this task as an unsupervised classification problem, considering the following criteria:
(1) vertex features should include local feature representation and shape description and should be translation, scaling, and rotation invariant and
(2) vertex classification should not only classify vertices based on threshold parameters but also consider their connectivity information.
This work uses HKS features and AHC to meet these criteria.
HKS features, being scale and rotation invariant, adeptly capture local shape information. 
AHC effectively considers vertex connectivity for meaningful vertex classification.

{\bf HKS features.}
HKS~\cite{Sun-HKS} is crucial for our analysis as it robustly captures the local geometric features of 3D meshes, making it ideal for representing surface and patch characteristics. HKS leverages the heat diffusion process, which is sensitive to the underlying geometry, thus providing a powerful descriptor invariant to surface transformations. It begins by assigning each vertex an initial heat distribution $\mathrm{u}_0$ and then measures heat diffusion across the mesh surface. By discretizing the Laplace-Beltrami operator, we represent heat diffusion as a vector-valued function influenced by both time and space. This feature's sensitivity to the mesh's geometric intricacy enables effective differentiation and characterization of surfaces and patches, which is essential for the subsequent steps involving UMAP aggregation and t-SNE projection.
Following~\cite{Sun-HKS}, the HKS is defined as
\begin{equation}
    \frac{\partial \mathrm{u}(x,t)}{\partial t} = -\Delta \mathrm{u}(x,t),
\end{equation}
where $\mathrm{u}(x,t)$ represents the heat distribution at position $x$ at time $t$. $t$ refers to a mathematical concept of ``time'' that controls the extent of the heat diffusion process on the surface. As $t$ increases, the heat diffusion progresses and spreads further from its initial distribution. $\Delta$ is the Laplace-Beltrami operator, which governs the diffusion process on the surface. The solution to this equation can be expressed as
\begin{equation}
    \mathrm{u}(x,t) = \int \mathrm{h}_t(x,y) \mathrm{u}_0(y) \,dy.
\end{equation}
$\mathrm{h}_t(x,y)$ is depicted as
\begin{equation}
    \mathrm{h}_t(x,y) = \sum_{i=0}^{\infty} \expp(-\lambda_i t) \mathrm{\phi}_i(x) \mathrm{\phi}_i(y),
\end{equation}
where $\lambda_i$ and $\phi_i$ are the $i$-th eigenvalues and eigenfunctions of the Laplace-Beltrami operator on the surface. 
Refer to the appendix, which shows the effectiveness of HKS features compared with other features. 

{\bf AHC.}
AHC is a bottom-up clustering method that begins with each vertex as an individual cluster.
It iteratively merges the closest clusters until reaching a set distance threshold $\delta$ or the specified number of clusters.
AHC comes with four merge strategies: Ward, single linkage, average linkage, and complete linkage. 
The comparison of these strategies can be found in the appendix.  
We opt for the Ward strategy due to its variance-minimizing property.
The distance metric used is the Euclidean distance, defined as
\begin{equation}
    \mathrm{D}_t(v_i, v_j) = \sqrt{\sum_{k=1}^{n} (\mathrm{h}_t(v_i,k) - \mathrm{h}_t(v_j,k))^2},
\end{equation}
where $\mathrm{D}_t(v_i, v_j)$ is the distance between vertices $v_i$ and $v_j$ at time $t$, $n$ is the number of vertices, and $\mathrm{h}_t(v_i,k)$ represents the value of the $k$-th feature of vertex $v_i$ at time $t$.

\begin{figure}[htb]
    \centering
    $\begin{array}{c@{\hspace{0.1in}}c}
            \includegraphics[width=0.425\linewidth, height=0.95in]{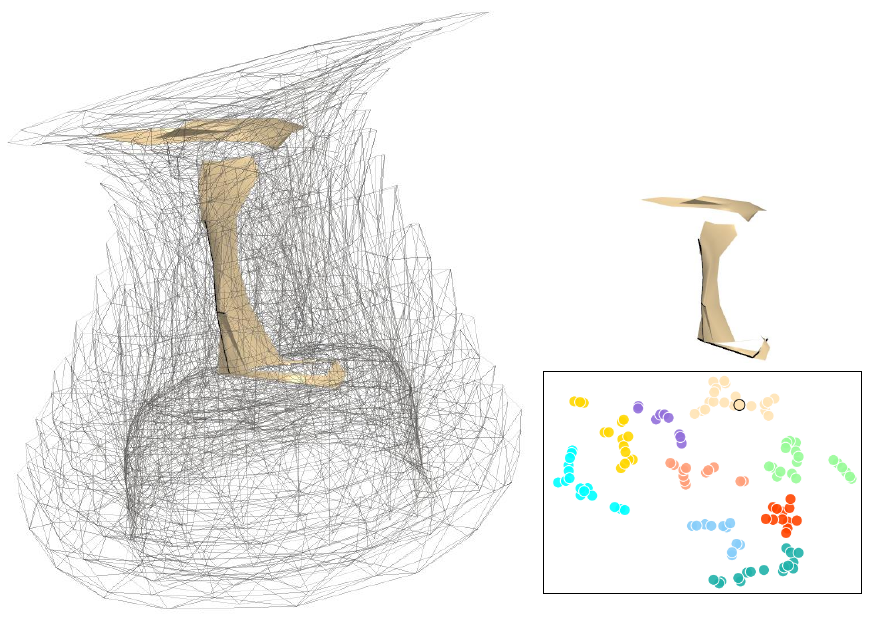} & 
            \includegraphics[width=0.425\linewidth, height=0.95in]{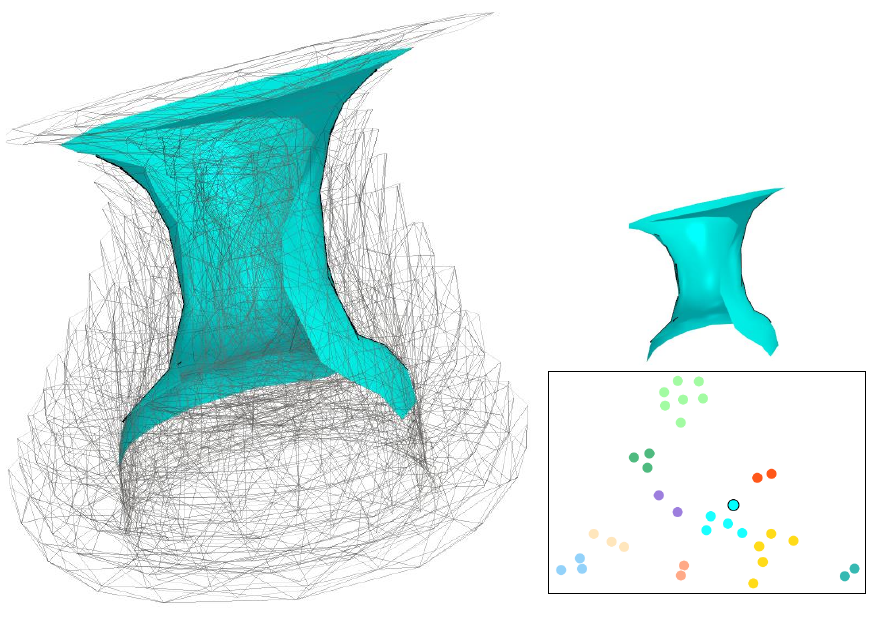} \\ 
            \mbox{\footnotesize (a) without CC}     & \mbox{\footnotesize (b) with CC (ours)}
        \end{array}$
    \vspace{-0.1in}
    \caption{An example patch generated using AHC. Each point in the UMAP projection view represents a patch; the selected patch is highlighted with a black boundary.}
    \vspace{-0.1in}
    \label{connectivity}
\end{figure}

To partition the stream surface into cohesive patches, we enhance AHC with CC.
These constraints restrict vertex merging to only adjacent clusters, utilizing a connectivity sparse matrix extracted from the surface.
Figure~\ref{connectivity} shows an example of a generated patch using AHC without and with CC. 
The patch generated without CC is discontinuous, and that with is continuous.
HKS features are first dimensionally reduced from $d$ to 2 using t-SNE~\cite{vanderMaaten-JMLR08} (we set $d=128$), which provides the most separated distribution, and subsequently clustered through AHC. The resulting clusters are used to delineate patches. 
Refer to Section~\ref{subsec:vlc}, which illustrates the effectiveness of AHC with CC compared with other clustering methods.

\vspace{-0.05in}
\subsection{Patch Matching}
\label{sec:patch_matching}

To support interactive patch query within a single stream surface, we classify, in a preprocessing step, all patches based on patch-level features, which are aggregated vertex-level HKS features. The challenge lies in effective aggregation, given each patch's varying number of vertices. Straightforward methods use average, min, max, and sum. However, they are unsuitable for our scenario because scale and rotation invariant properties are ignored, resulting in information loss. To address this, we propose using UMAP~\cite{McInnes-UMAP} to aggregate vertex-level features into patch-level ones by composing a matrix consisting of vertices as rows and features as columns and applying DR to consolidate vertices instead of features. Using UMAP not only preserves essential properties like scale and rotation invariance but also consolidates feature values to minimize information loss and enhance patch classification accuracy. We refer to this process as {\em UMAP aggregation}.
UMAP is particularly effective among DR methods for producing distinct clusters, which is defined as
\begin{equation}
    \uagg(\mathbf{M})=\umap(\mathbf{M}^T)^T,
\end{equation}
where $\uagg$ denotes UMAP aggregation and $\mathbf{M}$ is a matrix of dimension $n \times d$, where $n$ is 
the number of vertices in each patch (which varies across different patches) 
and $d$ is the fixed length of the HKS feature vector. We set $d=128$ empirically based on matching accuracy and computational efficiency, as increasing $d$ beyond 128 only incurs computational overhead without furthering the accuracy. 
The output of this $\uagg$ function is a $d$-dimensional feature vector.
Refer to the appendix, which demonstrates the effectiveness of UMAP aggregation compared with other aggregations. 


Next, we employ AHC without CC for patch classification. This is because we want intrinsically similar patches to be classified into the same cluster for meaningful subsequent patch querying, regardless of whether they are adjacent. 
By tuning $\delta$, we adjust the similarity degree for matched patches.

\begin{figure}[htb]
    \centering
    $\begin{array}{c@{\hspace{0.025in}}c}
            \includegraphics[width=0.475\linewidth]{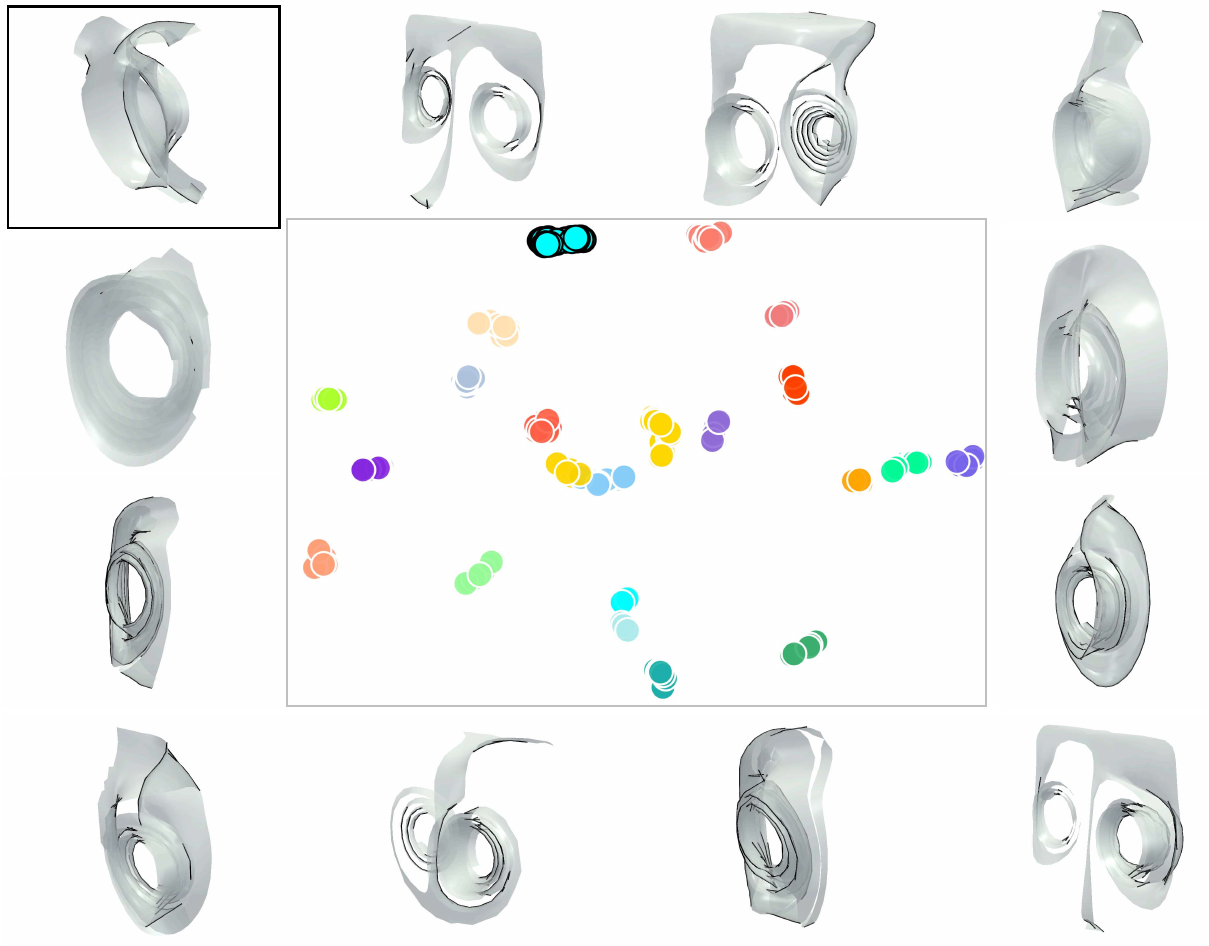} & 
            \includegraphics[width=0.475\linewidth]{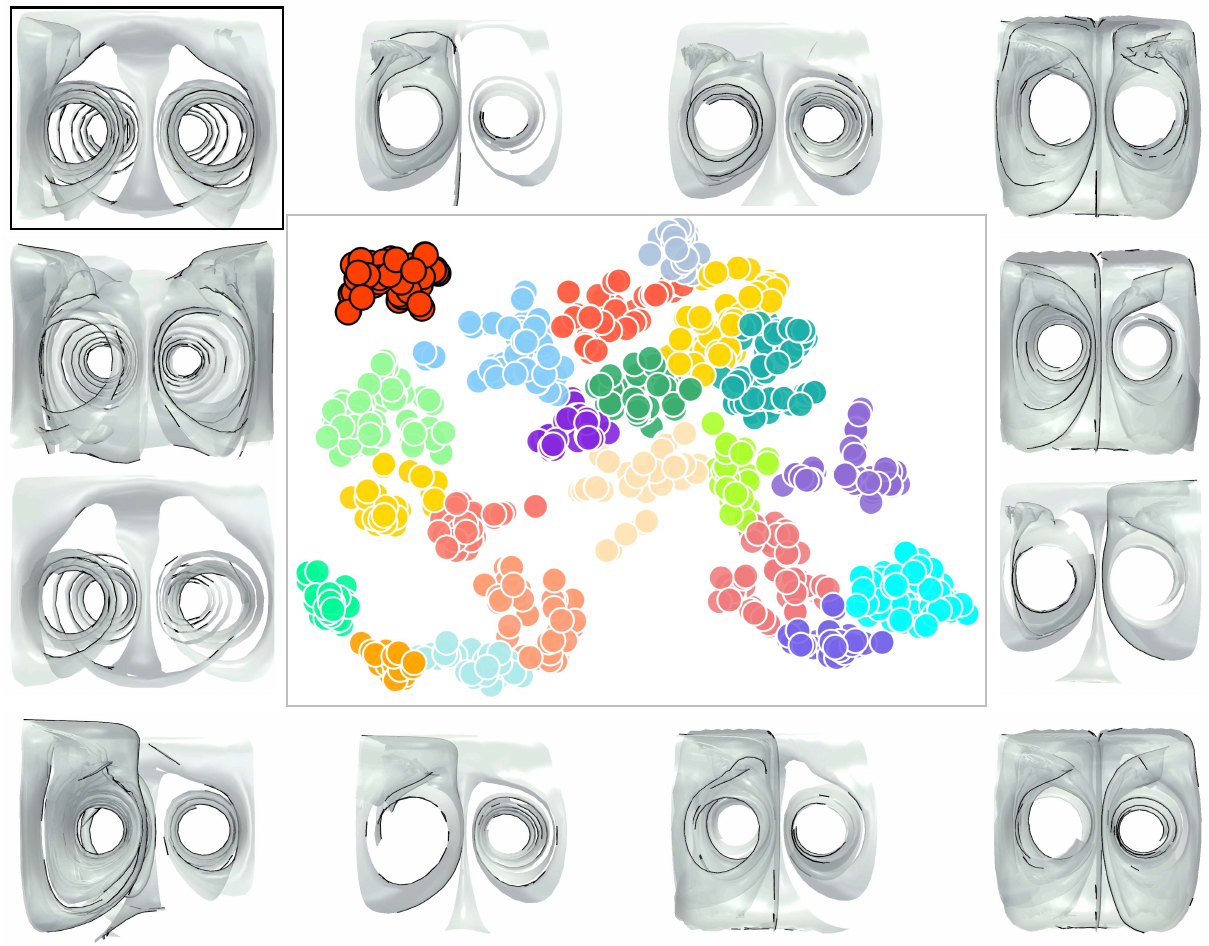} \\ 
            \mbox{\footnotesize (a) vertex-level aggregation} & \mbox{\footnotesize (b) patch-level aggregation}
        \end{array}$
    \vspace{-0.1in}
    \caption{Comparing different aggregations of HKS features in surface clustering. 
Each point in the UMAP projection view represents a surface. 
The selected representative surface is highlighted with the black bounding box, and a subset of the sample surfaces in that cluster are displayed.}
    \vspace{-0.1in}
    \label{patchVSvertex}
\end{figure}

\vspace{-0.05in}
\subsection{Surface Clustering}
\label{sec:surface_clustering}

The large number of stream surfaces we explore (in the order of thousands) poses a challenge for user selection and query.
To simplify the exploration, in a preprocessing step, we cluster these surfaces to identify representatives.
Similar to patch-level classification, obtaining appropriate surface-level features is crucial for surface clustering.
We can aggregate vertex-level HKS features into surface-level ones, but our experiments reveal that doing so is less effective.
Instead, we aggregate patch-level HKS features into surface-level ones.

Figure~\ref{patchVSvertex} shows a comparison between vertex- and patch-level aggregations.
    The results demonstrate that even though aggregating vertex-level features into surface-level ones leads to well-separated clusters in the projection view, examples of sample surfaces in the same cluster encompass dissimilar shapes.
    In contrast, aggregating patch-level features into surface-level ones results in more distributed points within each cluster, facilitating the distinction between similar and dissimilar surfaces in the cluster.

We still use UMAP aggregation to consolidate patch-level features into surface-level ones, followed by surface classification using AHC without CC.
A representative surface is defined as the centroid of each resulting cluster.
Choosing a representative surface allows users to identify others with similar shapes, facilitating convenient comparisons between samples and querying across multiple selected surfaces.

\vspace{-0.05in}
\section{Results and Evaluation}
\label{sec:rad}

We conducted experiments mainly on flow field datasets and presented the results from the vertex-, patch-, and surface-level perspectives.

At the {\em vertex} level, we evaluated vertex feature extraction and clustering methods, confirming the efficacy of multiscale patch generation.

At the {\em patch} level, we validated vertex aggregation and compared different DR and patch classification methods. Then, we presented patch-matching results under various matching tolerances and across multiple surfaces, reporting interesting shape-matching outcomes.

At the {\em surface} level, we compared different DR and surface classification methods, validating patch matching across sample surfaces.

Unless stated otherwise, all figures in this section use UMAP to generate the projection views where each point represents a patch.

Please refer to the accompanying video, which includes the recording of SurfNet's visual interface and interactions.

\begin{table}[htb]
    \caption{The datasets, their stream/iso surfaces, and average numbers of vertices before simplification (BS) and after simplification (AS).}
    \vspace{-0.125in}
    \centering
    \huge
    \resizebox{\columnwidth}{!}{
        \begin{tabular}{cccrrr}
            data                &                      & volume dimension                     & \# surface & average \#    & average \#    \\
            type                & dataset              & ($x\times y \times z \times t$)      & instances  & vertices (BS) & vertices (AS) \\\hline
            steady flow         & B{\'e}nard flow      & 128$\times$32$\times$64$\times$1     & 1,000      & 11,387        & 2,312         \\
            steady flow         & five critical points & 51$\times$51$\times$51$\times$1      & 1,062      & 2,072         & 229           \\
            steady flow         & solar plume          & 126$\times$126$\times$512$\times$1   & 1,001      & 5,305         & 1,475         \\
            steady flow         & square cylinder      & 192$\times$64$\times$48$\times$1     & 1,007      & 13,080        & 2,669         \\
            steady flow         & tornado              & 64$\times$64$\times$64$\times$1      & 1,159      & 6,754         & 1,419         \\
            steady flow         & two swirls           & 64$\times$64$\times$64$\times$1      & 1,031      & 9,201         & 1,951         \\ \hline
            unsteady flow       & solar plume          & 126$\times$126$\times$512$\times$27  & 1,028      & 5,316         & 1,531         \\
            unsteady flow       & tornado              & 64$\times$64$\times$64$\times$49     & 1,005      & 6,329         & 1,163         \\ \hline
            time-varying scalar & earthquake           & 256$\times$256$\times$96$\times$599  & 300        & 175,446       & 9,372         \\
            time-varying scalar & ionization           & 600$\times$248$\times$248$\times$200 & 300        & 148,819       & 7,384         \\
        \end{tabular}
    }
    \label{tab:datasets}
\end{table}

\vspace{-0.05in}
\subsection{Datasets and Preprocessing}

Table~\ref{tab:datasets} lists the experimented flow field datasets.
We primarily investigate stream surfaces extracted from steady flow.
These datasets encompass a variety of flow phenomena, including
a B{\'e}nard flow representing liquid movement between two parallel planes,
a synthesized flow showing five critical points,
a down-flow solar plume,
a square cylinder with fluid flowing around it,
a procedurally derived tornado flow, and
two swirls formed by wake vortices.
The surface instances were generated using random seeding curves that follow the binormal directions~\cite{Tao-TVCG2018}.
These surfaces were constructed based on the easy integral surface solution~\cite{McLoughlin-CGI2009}.
For solar plume and tornado datasets, we also explore stream surfaces extracted from individual timesteps of their unsteady flow.
Note that our SurfPatch is not limited to stream surfaces and can be applied to isosurfaces as well.
Therefore, we additionally explore isosurfaces extracted from different timesteps and selected isovalues of two time-varying datasets: earthquake and ionization.

We experimented with SurfPatch on a workstation with an Intel Core i9-13900HX CPU, 128 GB RAM, and an NVIDIA GeForce RTX 4090 GPU with 24 GB video memory.
During preprocessing, each dataset underwent mesh simplification~\cite{Garland-QEM}, where we set the simplification threshold $\epsilon=0.5$. We applied several methods to extract vertex features, which were then reduced to 2D using different DR techniques. Note that all DR methods discussed in this paper are randomly initialized. Random initialization prevents bias and helps preserve key features of high-dimensional data. To ensure fairness, we used the same random initialization as SurfNet, maintaining consistency and avoiding discrepancies in comparison. Given the larger size of the finest vertex-level features compared to the other two coarser levels, we implemented a GPU-based solution to expedite the DR (Isomap, MDS, t-SNE, UMAP) process, significantly reducing the time to process a single surface mesh.

\begin{table}[htb]

    \caption{Average computation time (in seconds) per surface for each step of the SurfPatch method and total preprocessing (PP) time (in hours) for all surface instances.}
    \vspace{-0.125in}
    \centering
    \huge
    \resizebox{\columnwidth}{!}{
        \begin{tabular}{ccrr|rr|r|r}
            data                &                      & mesh           & HKS     & DR     & DR    &            & PP \\
            type                & dataset              & simplification & feature & (CPU)  & (GPU) & clustering & time        \\\hline
            steady flow         & B{\'e}nard flow      & 0.251          & 3.062   & 8.024  & 0.008 & 0.003      & 3.082               \\
            steady flow         & five critical points & 0.063          & 1.243   & 6.367  & 0.006 & 0.001      & 2.246               \\
            steady flow         & solar plume          & 0.039          & 1.146   & 6.117  & 0.006 & 0.001      & 2.021               \\
            steady flow         & square cylinder      & 0.085          & 1.329   & 6.275  & 0.006 & 0.001      & 2.278               \\
            steady flow         & tornado              & 0.104          & 2.847   & 7.454  & 0.007 & 0.002      & 3.319               \\
            steady flow         & two swirls           & 0.218          & 2.979   & 7.875  & 0.008 & 0.003      & 3.017               \\ \hline
            unsteady flow       & solar plume          & 0.043          & 1.003   & 5.995  & 0.006 & 0.001      & 2.000               \\
            unsteady flow       & tornado              & 0.097          & 2.750   & 7.255  & 0.007 & 0.001      & 2.795               \\ \hline
            time-varying scalar & earthquake           & 0.292          & 4.395   & 91.005 & 0.120 & 0.004      & 7.960               \\
            time-varying scalar & ionization           & 0.264          & 4.000   & 82.403 & 0.081 & 0.003      & 7.207               \\
        \end{tabular}
    }
    \label{tab:time}

\end{table}

The computation time for each step of the SurfPatch algorithm is summarized in Table~\ref{tab:time}. This table shows the average time required for processing a single surface. The experimental data includes stream surfaces extracted from steady or unsteady flow, as well as isosurfaces with different isovalues and timesteps. Among the steps mentioned, mesh simplification and HKS feature extraction~\cite{mesh-signatures} are the most time-consuming, so they are handled as preprocessing steps. In contrast, DR with GPU acceleration and clustering are performed on the fly, reducing the processing time per surface to under 0.125 seconds.

\begin{figure}[htb]
    \centering
    $\begin{array}{c@{\hspace{0.1in}}c}
            \includegraphics[width=0.45\linewidth, height=0.5in]{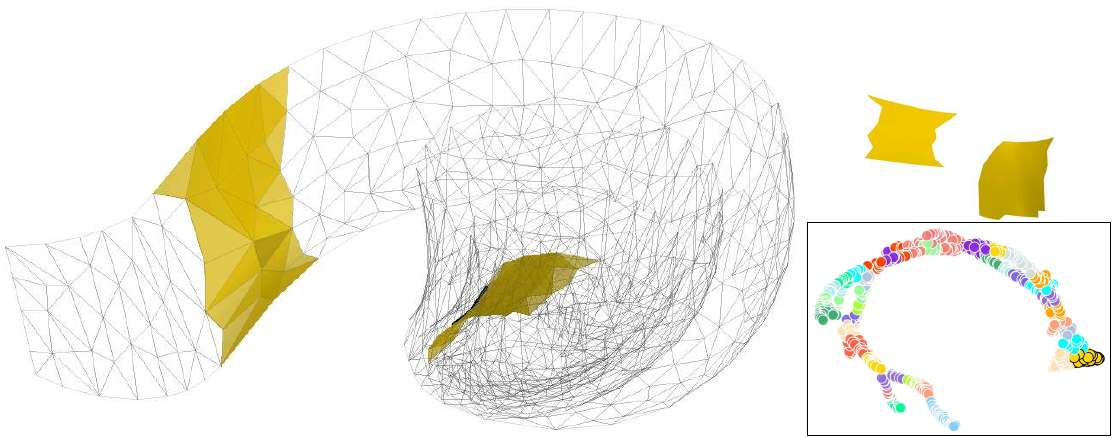}    &
            \includegraphics[width=0.45\linewidth, height=0.5in]{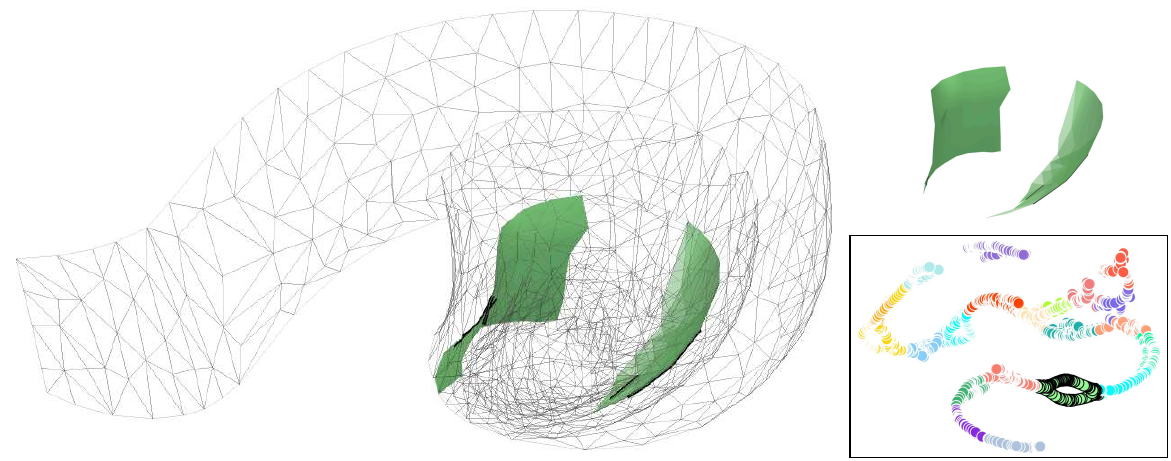}                                                  \\
            \mbox{\footnotesize (a) k-means}                              & \mbox{\footnotesize (b) DBSCAN}             \\
            \includegraphics[width=0.45\linewidth, height=0.5in]{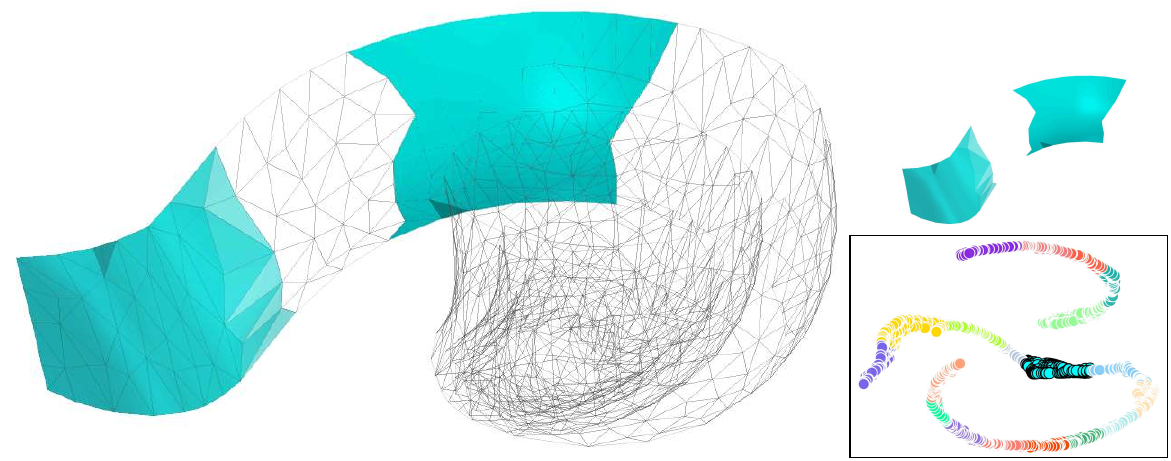} &
            \includegraphics[width=0.45\linewidth, height=0.5in]{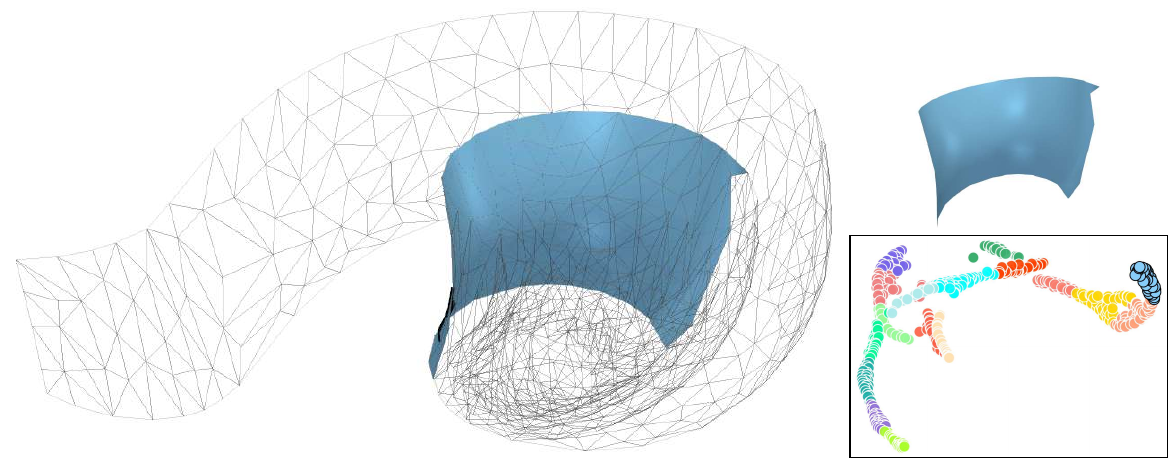}                                                     \\
            \mbox{\footnotesize (c) mean shift}                           & \mbox{\footnotesize (d) AHC with CC (ours)} \\
        \end{array}$
    \vspace{-0.1in}
    \caption{Comparing vertex feature clustering methods in patch generation using the tornado dataset. t-SNE generates all projection views where each point represents a vertex.}
    \vspace{-0.1in}
    \label{patch_cluster}
\end{figure}

\vspace{-0.05in}
\subsection{Vertex-Level Classification}
\label{subsec:vlc}

{\bf Vertex feature clustering.}
Utilizing HKS features as the vertex-level input, we continue to assess the effectiveness of various clustering methods in generating patches based on the t-SNE-reduced HKS features. The assessment criteria include continuity and granularity of the generated patches.
We compare our hierarchy-based AHC with CC against partition-based and density-based clustering methods:
\begin{myitemize}
    \vspace{-0.05in}
    \item k-means partitions points into $k$ clusters where each point belongs to the cluster with the nearest mean (i.e., cluster centroid). We determine the number of clusters using the method proposed by Schubert~\cite{KMeans}.
    \item DBSCAN~\cite{Ester-KDD96} is a density-based clustering algorithm that groups closely packed points and marks points as outliers that lie alone in low-density regions.
    \item Mean shift~\cite{Fukunaga-TIT75} locates a density function's maxima (i.e., modes) given points sampled from that function, and clustering is performed by assigning points to the nearest data distribution mode.
    \vspace{-0.05in}
\end{myitemize}

We compare patch generation results using different vertex feature clustering methods.
A surface generates numerous patches, so we select a single patch from each clustering method, as shown in Figure~\ref{patch_cluster}.
The results reveal that AHC performs better in generating continuous and fine-grained patches.
In contrast, partition-based (k-means) and density-based (DBSCAN and mean shift) clustering methods cannot produce meaningful and acceptable patches, primarily because they group discontinuous surface parts.
AHC with CC considers spatial neighbors and connectivity information.
All other methods lack CC and merely cluster vertices based on their spatial relationships, explaining the discontinuity observed in their patches.
In AHC, CC guides the step-by-step selection of the two most similar clusters for merging.
However, k-means, DBSCAN, and mean shift algorithms partition or identify dense regions based on intrinsic data properties.
They do not form clusters hierarchically and, therefore, cannot integrate CC.

\begin{figure}[htb]
    \centering
    $\begin{array}{c@{\hspace{0.1in}}c}
            \includegraphics[width=0.45\linewidth]{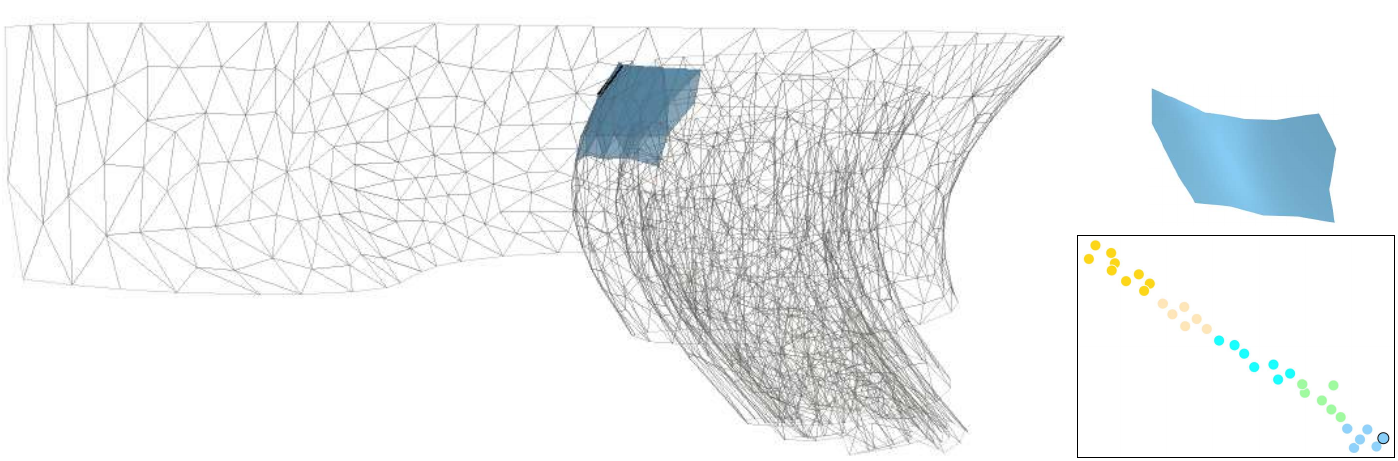} &
            \includegraphics[width=0.45\linewidth]{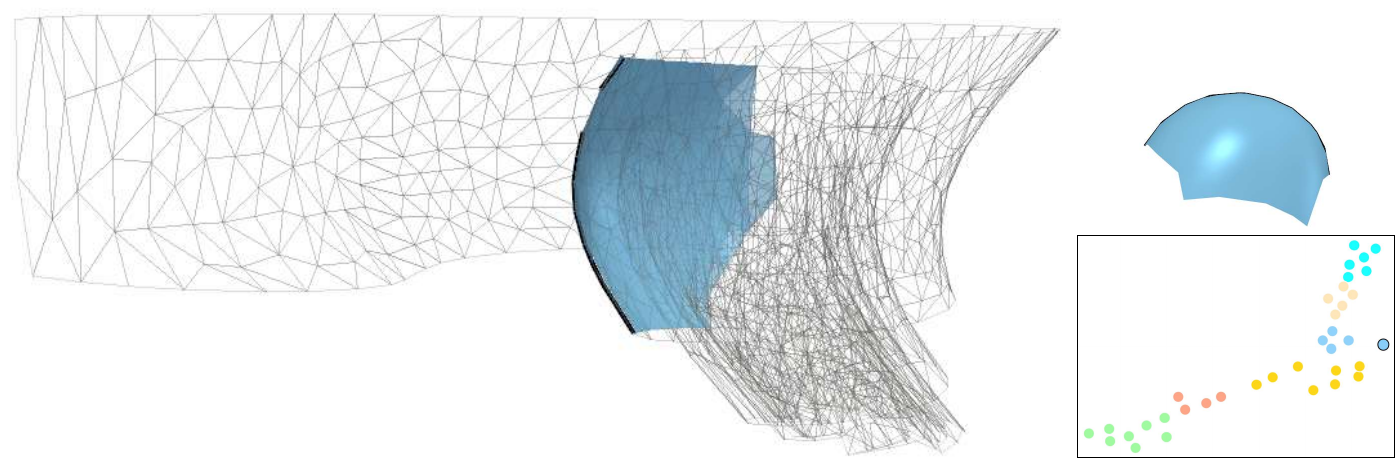}                                          \\
            \mbox{\footnotesize (a) $\delta_1=10$}                & \mbox{\footnotesize (b) $\delta_1=30$} \\
            \includegraphics[width=0.45\linewidth]{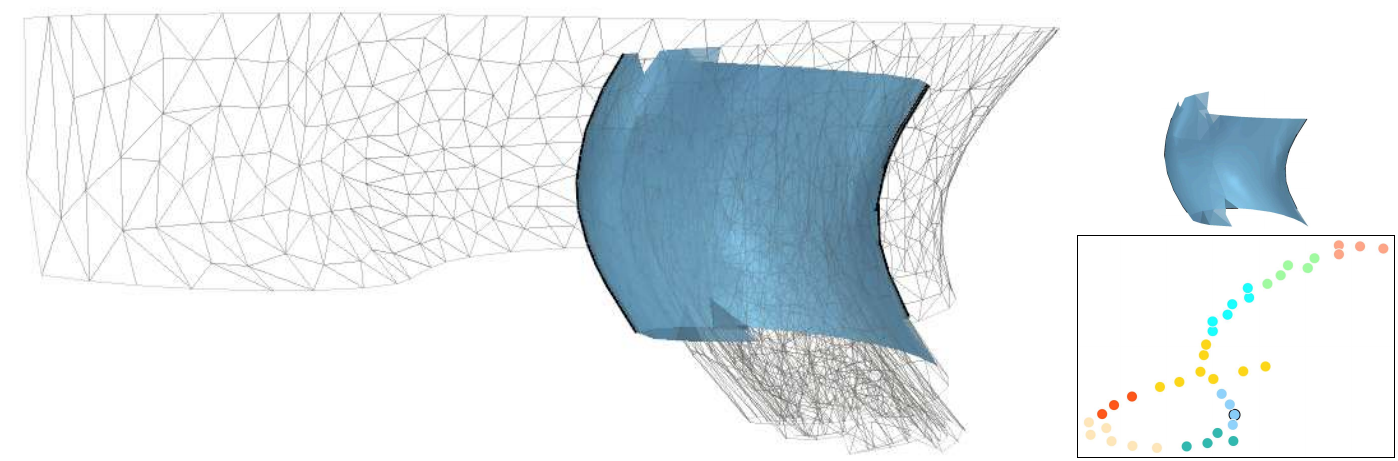} &
            \includegraphics[width=0.45\linewidth]{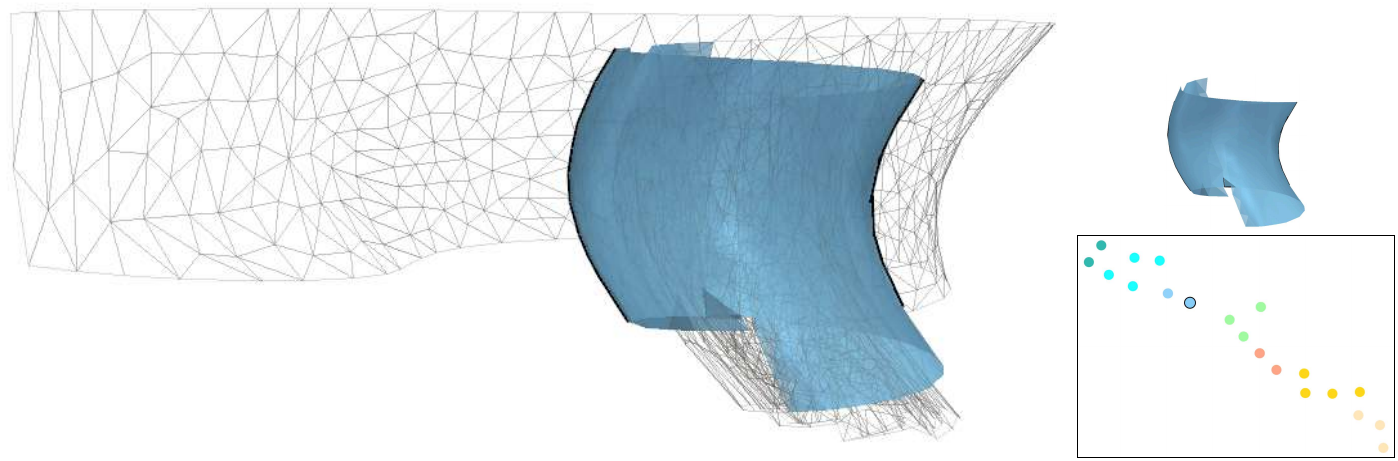}                                          \\
            \mbox{\footnotesize (c) $\delta_1=50$}                & \mbox{\footnotesize (d) $\delta_1=70$} \\
        \end{array}$
    \vspace{-0.1in}
    \caption{Multi-scale patches generated under different AHC distance thresholds ($\delta_1$) 
        using the tornado dataset.}
    \vspace{-0.1in}
    \label{patch_multiscale}
\end{figure}

{\bf Multiscale patch generation.}
To enhance the flexibility of patch querying, we generate patches in a multiscale manner by adjusting the AHC's distance threshold ($\delta_1$), which controls the granularity of clustering when grouping vertices to form patches. A lower $\delta_1$ results in smaller, finer patches, while a higher $\delta_1$ produces larger, coarser patches. AHC allows for clustering in two ways: by specifying a fixed number of clusters or by setting $\delta_1$ to control the separation between clusters. This parameter plays a critical role in determining the patch size and the level of detail in the resulting analysis.
Since a fixed number of clusters is deemed undesirable, we opt to vary $\delta_1$ to obtain multiscale patches.
We choose the tornado dataset for demonstration, as its consistent variation from the inner to outer parts of the vortex core showcases gradual multiscale changes.
The results are depicted in Figure~\ref{patch_multiscale}.
We can see that adjusting $\delta_1$ generates patches of varying scales.
A smaller $\delta_1$ produces a finer patch, while a larger one yields a coarser patch.
This flexibility enables runtime multiscale patch querying.

\begin{figure}[htb]
    \centering
    $\begin{array}{c@{\hspace{0.1in}}c}
            \includegraphics[width=0.42\linewidth, height=0.6in]{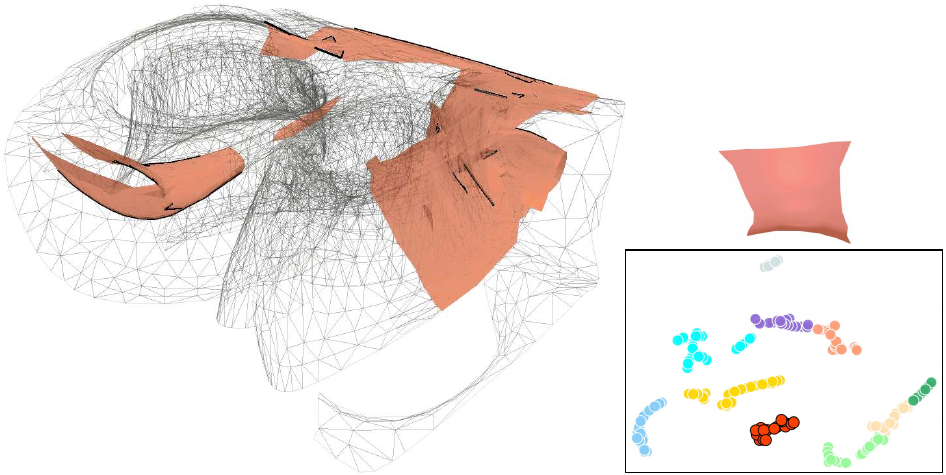}    &
            \includegraphics[width=0.42\linewidth, height=0.6in]{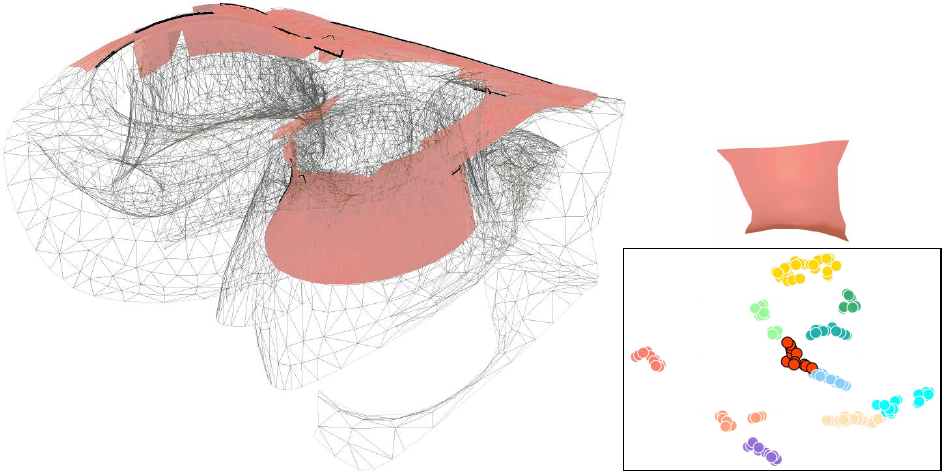}                                                     \\
            \mbox{\footnotesize (a) k-means}                        & \mbox{\footnotesize (b) DBSCAN}                \\
            \includegraphics[width=0.42\linewidth, height=0.6in]{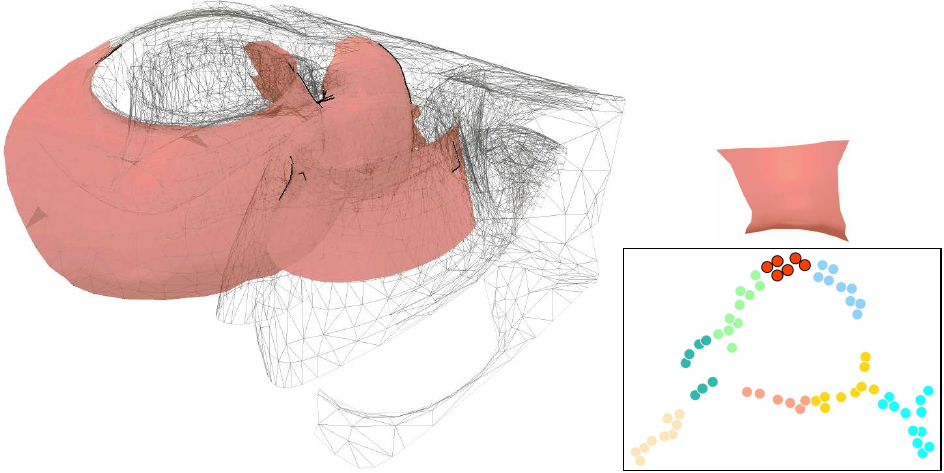} &
            \includegraphics[width=0.42\linewidth, height=0.6in]{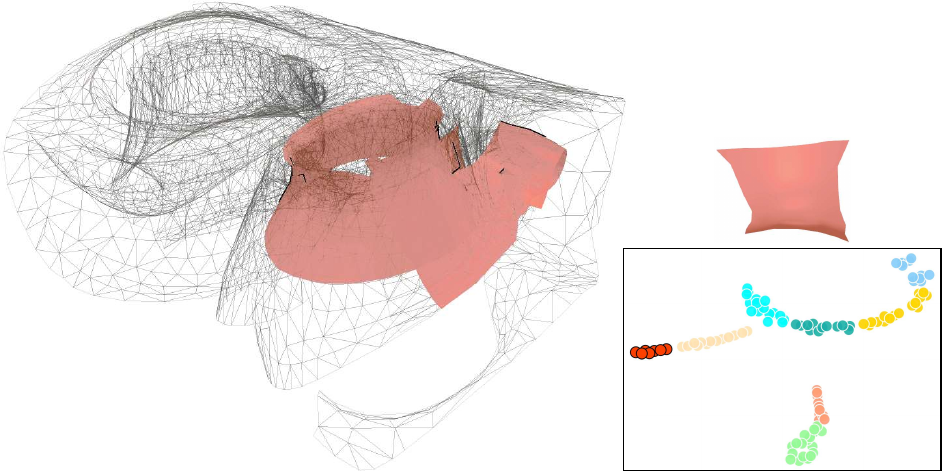}                                                        \\
            \mbox{\footnotesize (c) mean shift}                           & \mbox{\footnotesize (d) AHC without CC (ours)} \\
        \end{array}$
    \vspace{-0.1in}
    \caption{Comparing patch classification methods in patch matching using the two swirls dataset.}
    \vspace{-0.1in}
    \label{pl-clustering}
\end{figure}

\vspace{-0.05in}
\subsection{Patch-Level Matching}
\label{subsec:plm}

As discussed in Section~\ref{sec:patch_matching}, we consolidate vertex-level features to form patch-level embeddings using UMAP aggregation.
Subsequently, we perform DR before clustering, as previous work~\cite{Han-SurfNet} shows that swapping the order does not produce meaningful results.
AHC without CC is used for patch classification.
Following this, we explore interesting patch-matching cases across multiple surfaces and validate patch-matching under various matching tolerances.
Here, we show patch classification and patch matching exploration results. The remaining results on vertex feature aggregation, patch-level DR, and patch size and matching tolerance are in the appendix.

\begin{figure}[htb]
    \centering
    \includegraphics[width=0.85\linewidth, height=0.5in]{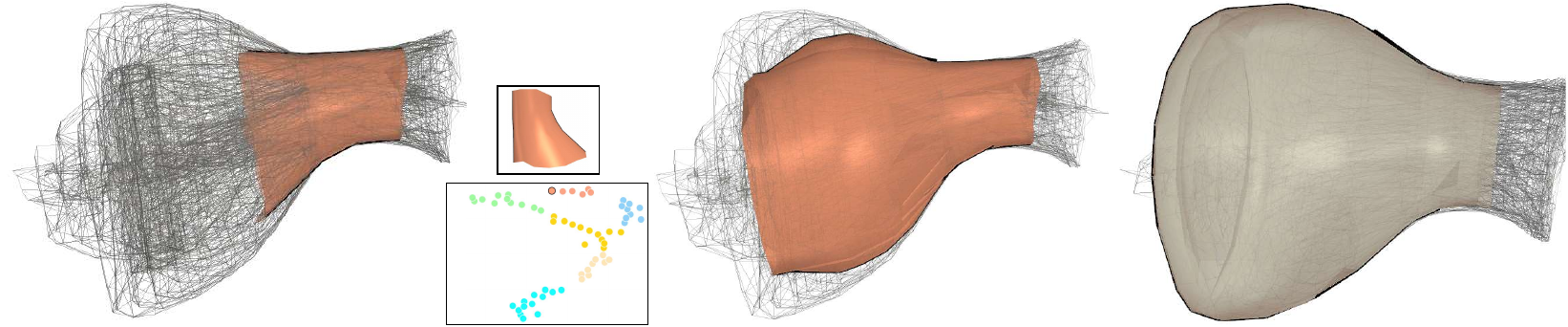}\\
    \mbox{\footnotesize (a) bottle shape}\\
    \includegraphics[width=0.85\linewidth, height=0.5in]{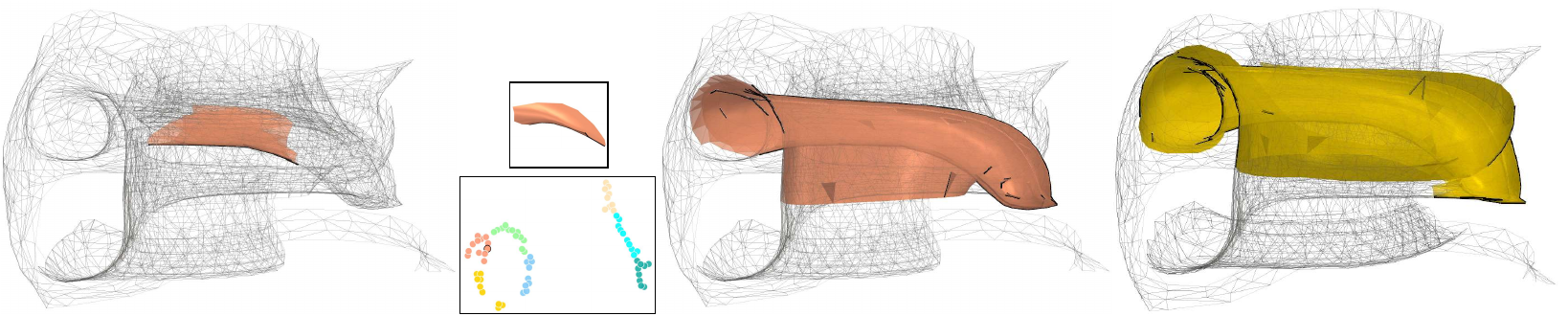}\\
    \mbox{\footnotesize (b) tube shape}\\
    \includegraphics[width=0.85\linewidth, height=0.5in]{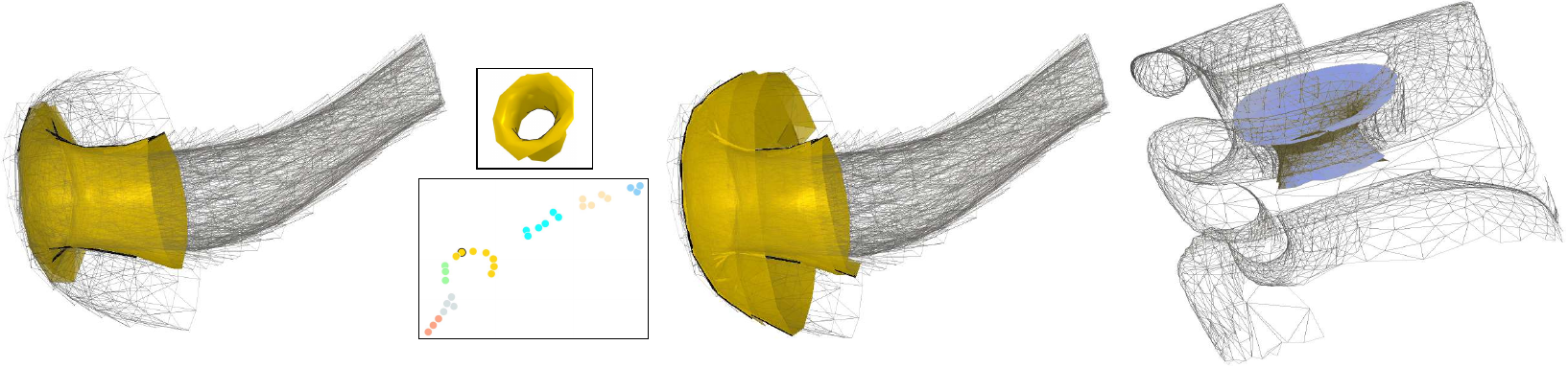} \\
    \mbox{\footnotesize (c) mushroom shape}\\
    \vspace{-0.1in}
    \caption{SurfPatch's patch-matching results using the two swirls dataset.
        Each subfigure displays the selected patch and its location on the surface, followed by the matching results on the same and different surfaces.}
    \vspace{-0.1in}
    \label{patchMatching-1}
\end{figure}

\begin{figure}[htb]
    \centering
    \includegraphics[width=0.85\linewidth, height=0.5in]{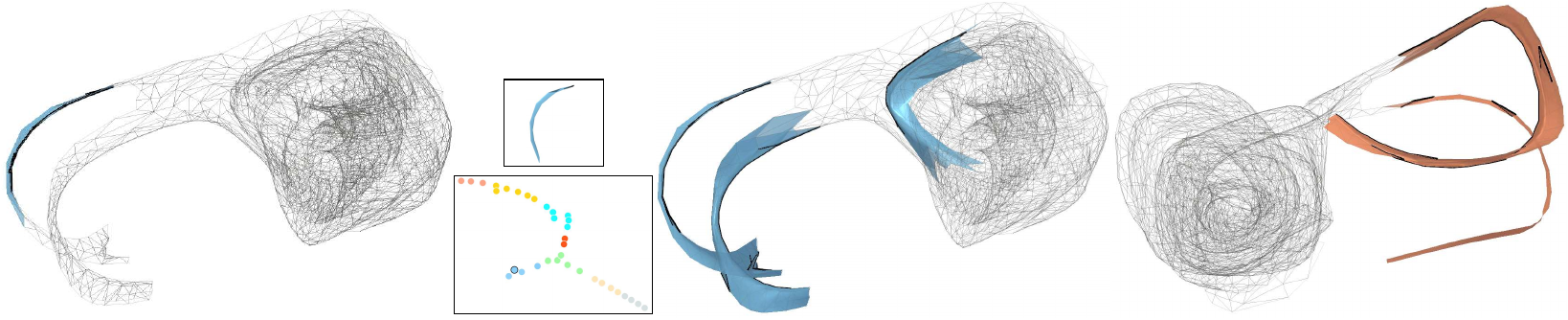}\\
    \mbox{\footnotesize (a) B{\'e}nard flow --- stripe shape}\\
    \includegraphics[width=0.85\linewidth, height=0.5in]{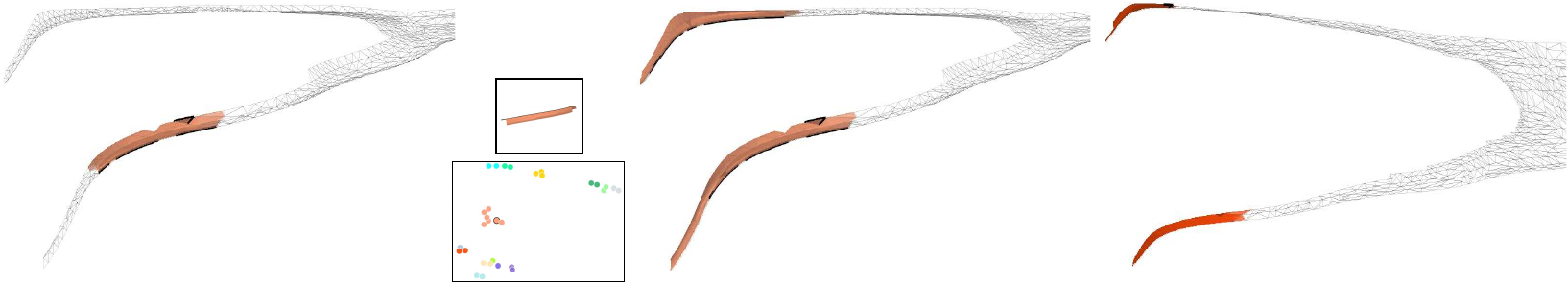}\\
    \mbox{\footnotesize (b) square cylinder --- tentacle shape}\\
    \includegraphics[width=0.85\linewidth, height=0.5in]{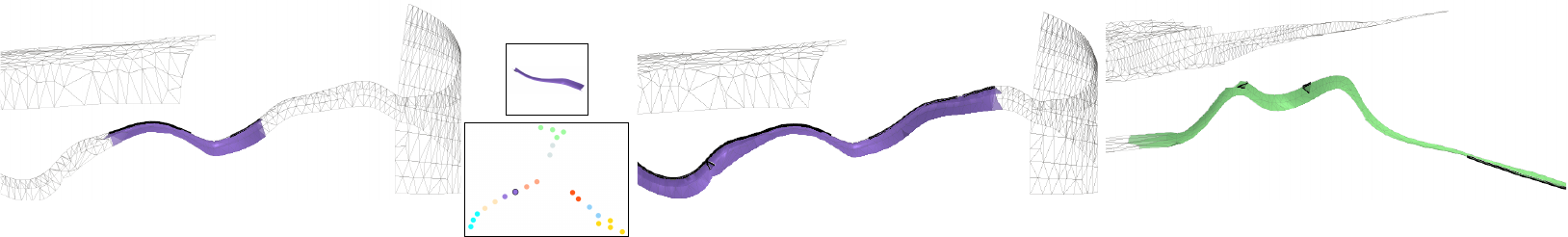} \\
    \mbox{\footnotesize (c) solar plume --- ribbon shape}\\
    \includegraphics[width=0.85\linewidth, height=0.5in]{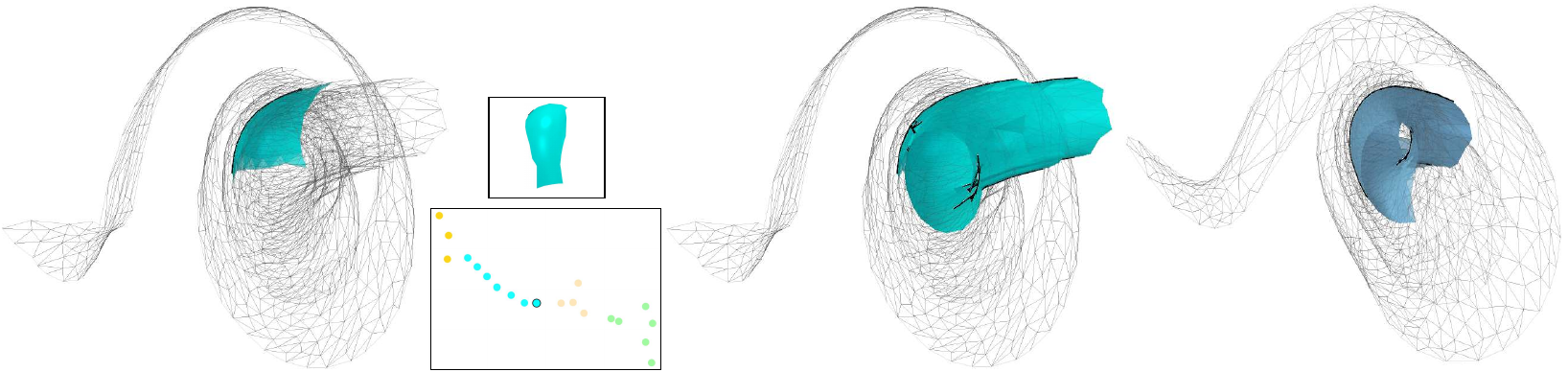} \\
    \mbox{\footnotesize (d) tornado --- vortex shape}\\
    \includegraphics[width=0.85\linewidth, height=0.5in]{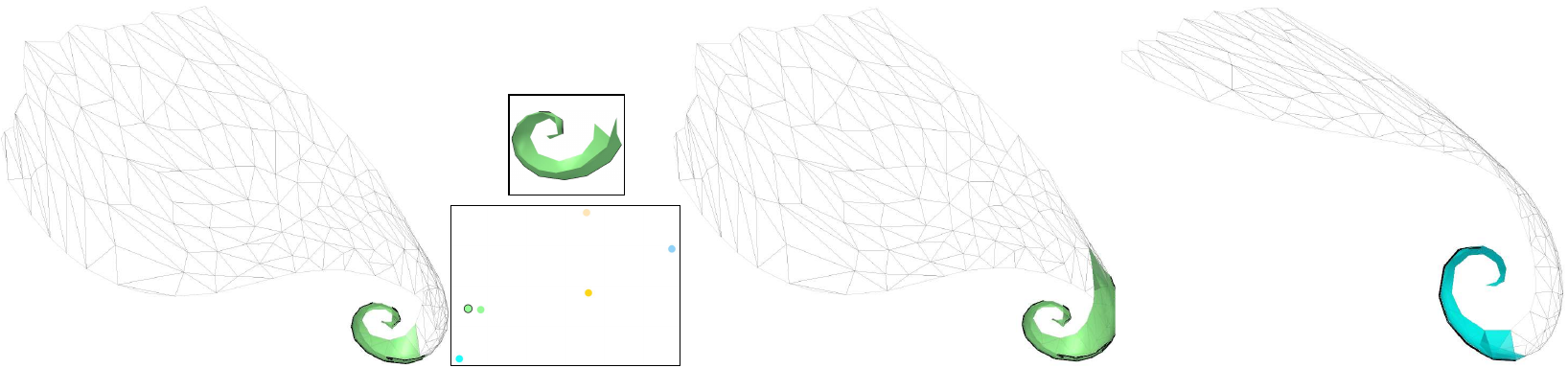} \\
    \mbox{\footnotesize (e) five critical points --- hook shape}\\
    \vspace{-0.1in}
    \caption{SurfPatch's patch-matching results using different datasets.
        Each subfigure displays the selected patch and its location on the surface, followed by the matching results on the same and different surfaces. We show close-up views for clear examination in (b) and (c).}
    \vspace{-0.1in}
    \label{patchMatching-2}
\end{figure}

\begin{figure*}[htbp]
    \centering
    $\begin{array}{c@{\hspace{0.1in}}c}
            \includegraphics[height=0.6in, width=3.225in]{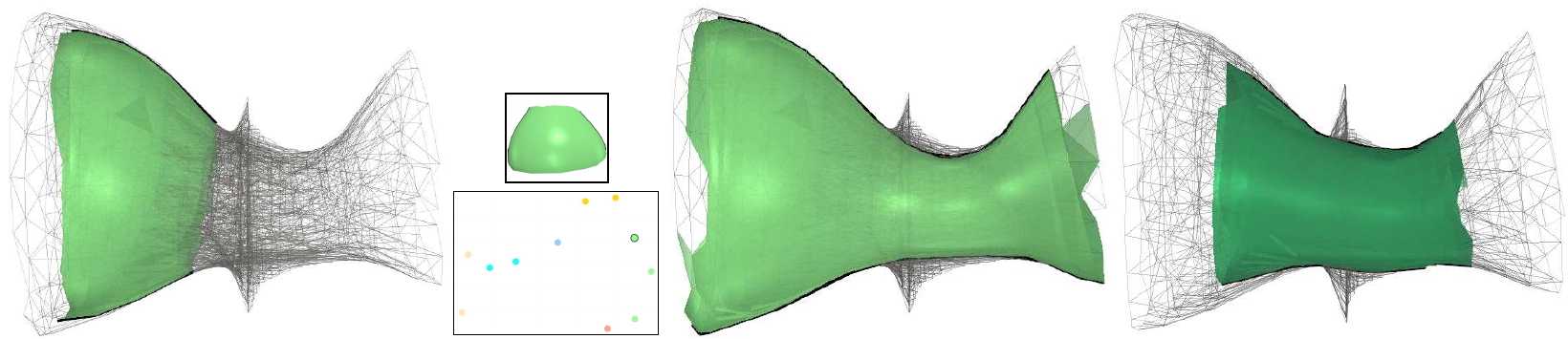} &
            \includegraphics[height=0.6in, width=3.225in]{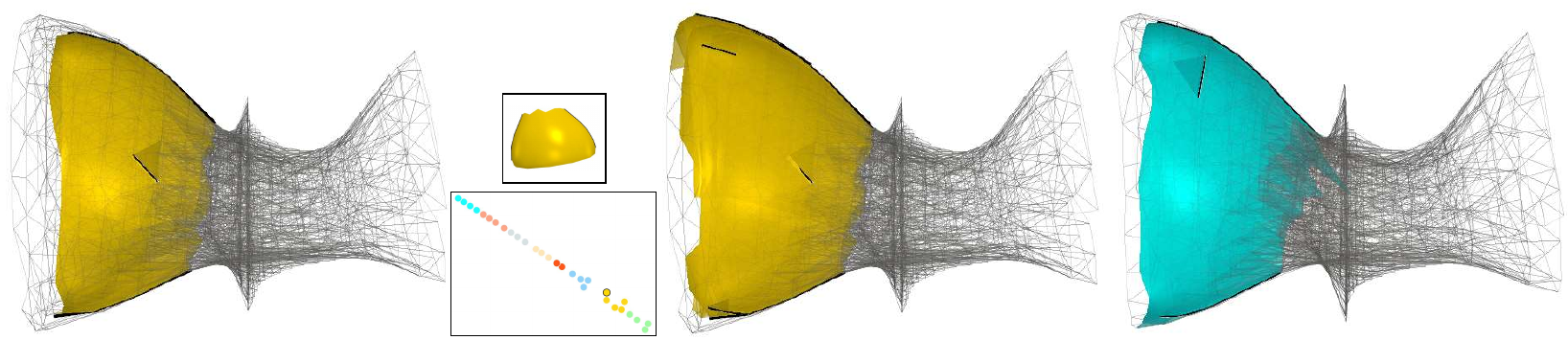}                                                      \\
            \mbox{\footnotesize (a) two swirls (SurfNet)}                 & \mbox{\footnotesize (b) two swirls (SurfPatch)}
        \end{array}$
    $\begin{array}{c@{\hspace{0.1in}}c}
            \includegraphics[height=0.6in, width=3.225in]{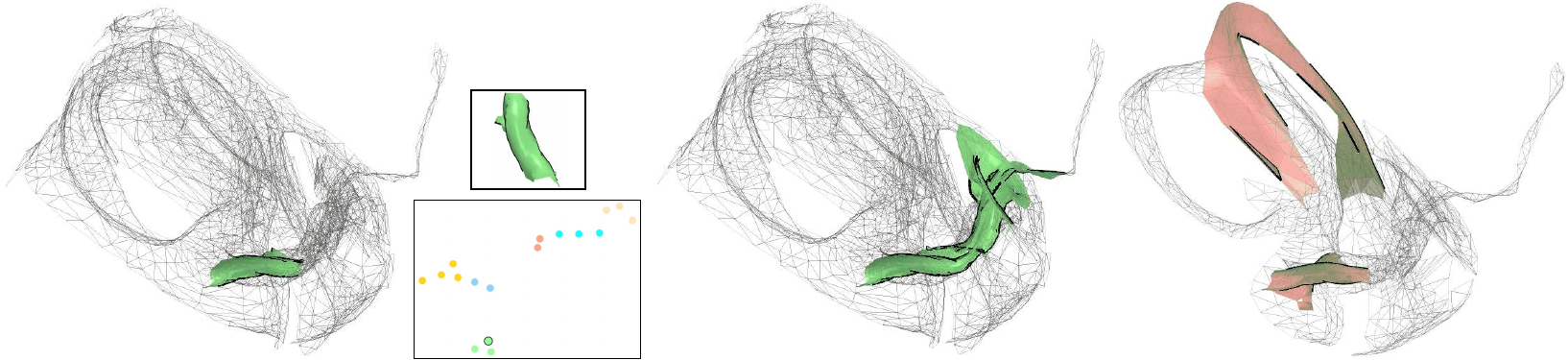} &
            \includegraphics[height=0.6in, width=3.225in]{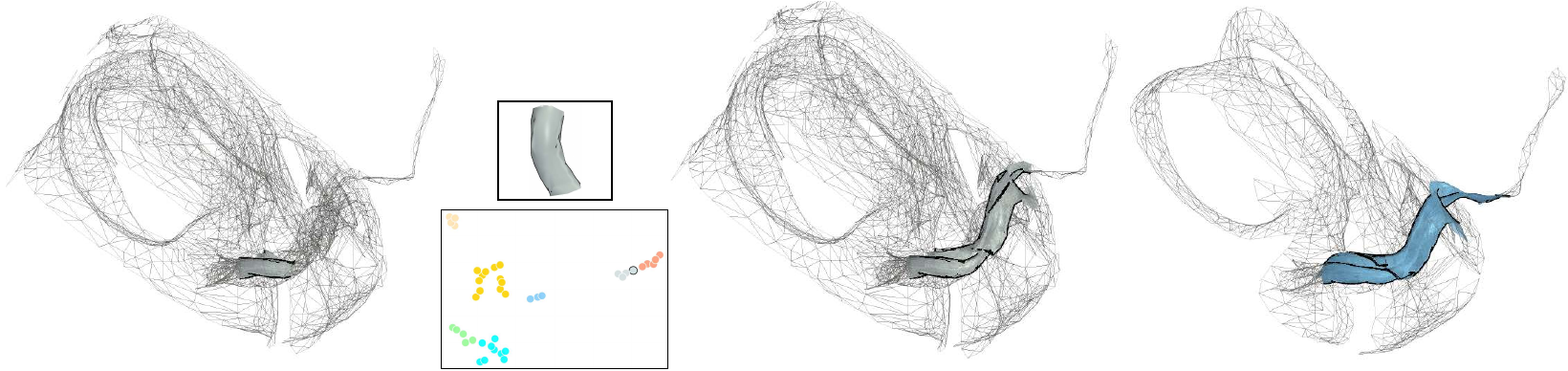}                                                      \\
            \mbox{\footnotesize (c) B{\'e}nard (SurfNet)}                & \mbox{\footnotesize (d) B{\'e}nard (SurfPatch)}
        \end{array}$
    $\begin{array}{c@{\hspace{0.1in}}c}
            \includegraphics[height=0.6in, width=3.225in]{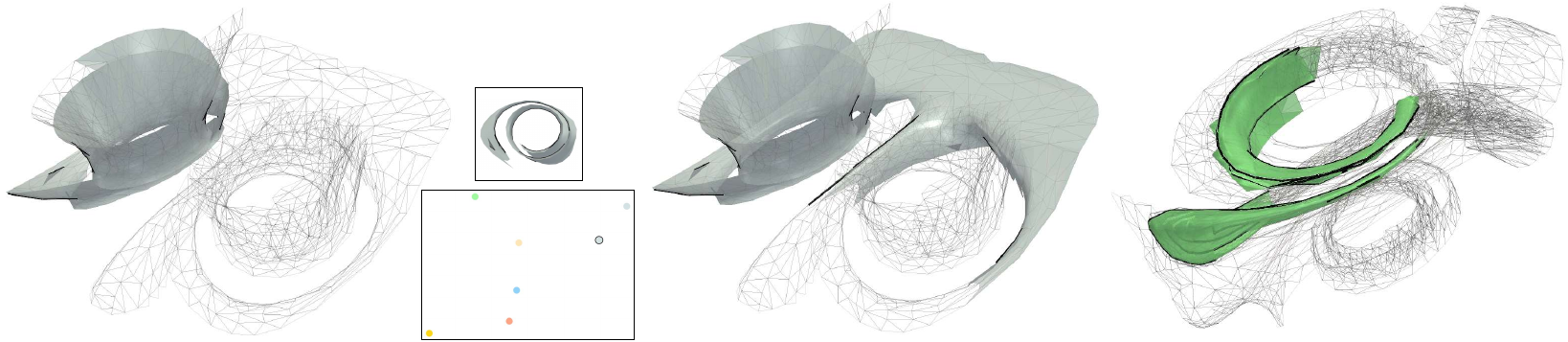} &
            \includegraphics[height=0.6in, width=3.225in]{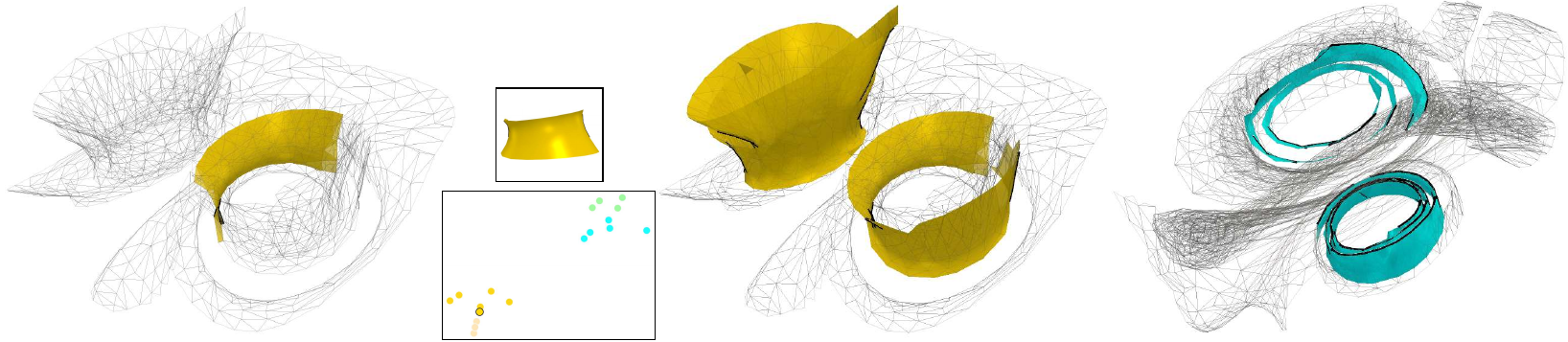}                                                      \\
            \mbox{\footnotesize (e) two swirls (SurfNet)}                     & \mbox{\footnotesize (f) two swirls (SurfPatch)}
        \end{array}$
    $\begin{array}{c@{\hspace{0.1in}}c}
            \includegraphics[height=0.6in, width=3.225in]{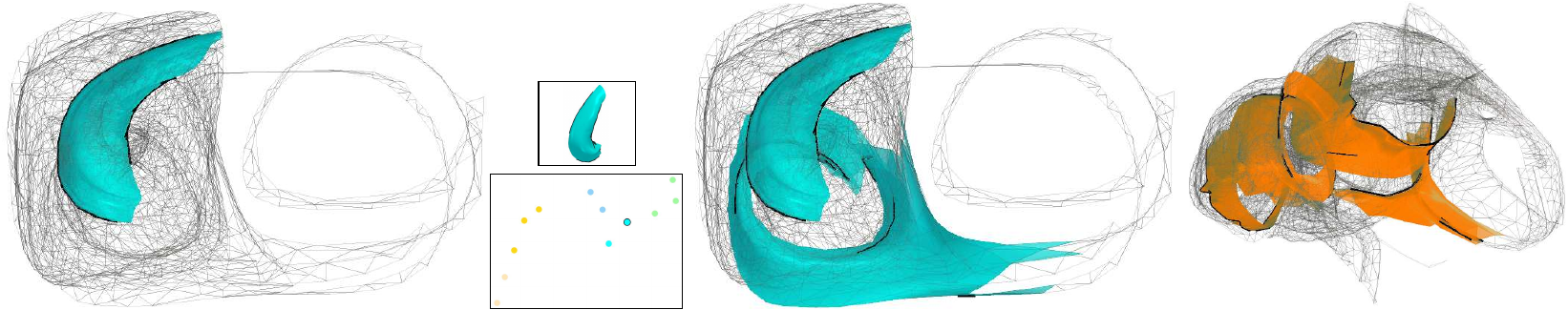} &
            \includegraphics[height=0.6in, width=3.225in]{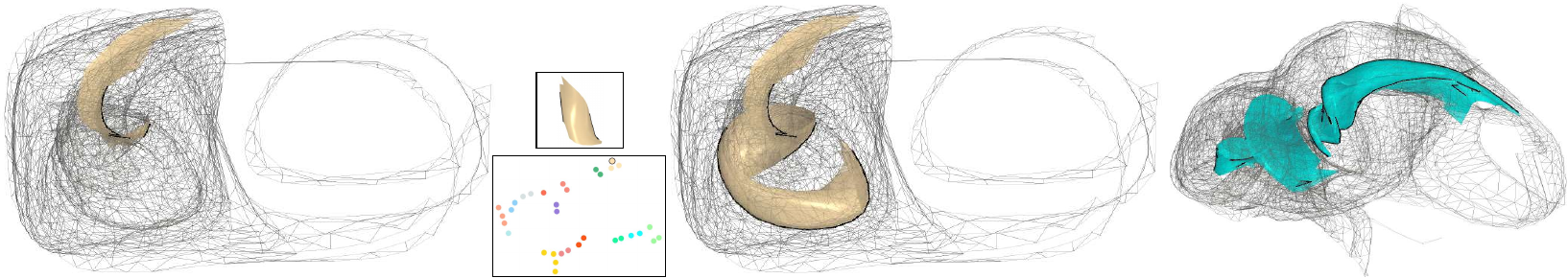}                                                      \\
            \mbox{\footnotesize (g) B{\'e}nard (SurfNet)}                   & \mbox{\footnotesize (h) B{\'e}nard (SurfPatch)}
        \end{array}$
    \vspace{-0.1in}
    \caption{Patch matching results. Top two rows: Matching results on similar surfaces. Bottom two rows: Matching results on dissimilar surfaces. Each subfigure displays the selected patch and its location on the surface, followed by the matching results on the same and different surfaces.}
    \vspace{-0.1in}
    \label{SP&SN}
\end{figure*}

\begin{figure}[htbp]
    \centering
    $\begin{array}{c@{\hspace{0.1in}}c}
            \includegraphics[width=0.45\linewidth]{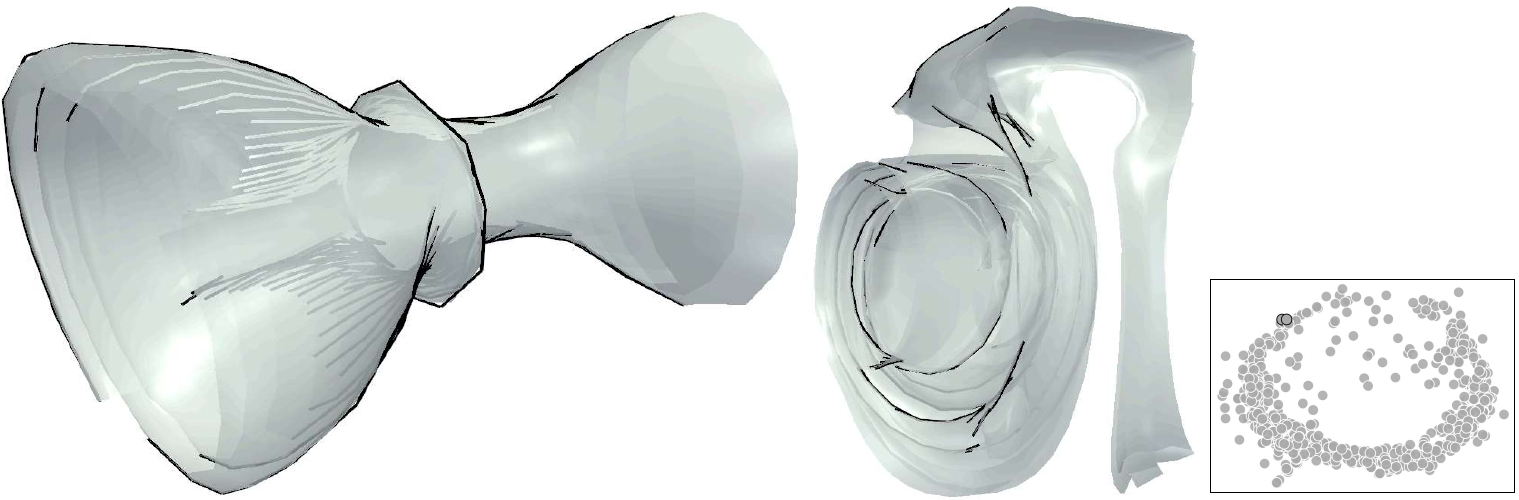} &
            \includegraphics[width=0.45\linewidth]{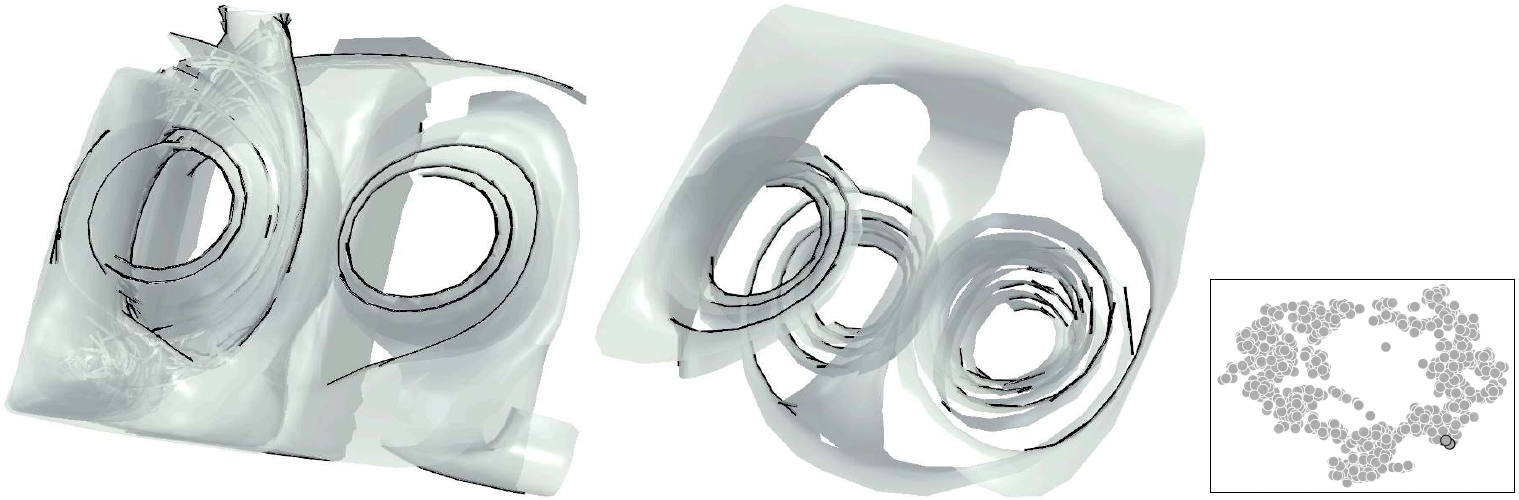}                                              \\
            \mbox{\footnotesize (a) Isomap DR}                         & \mbox{\footnotesize (b) MDS DR}         \\
            \includegraphics[width=0.45\linewidth]{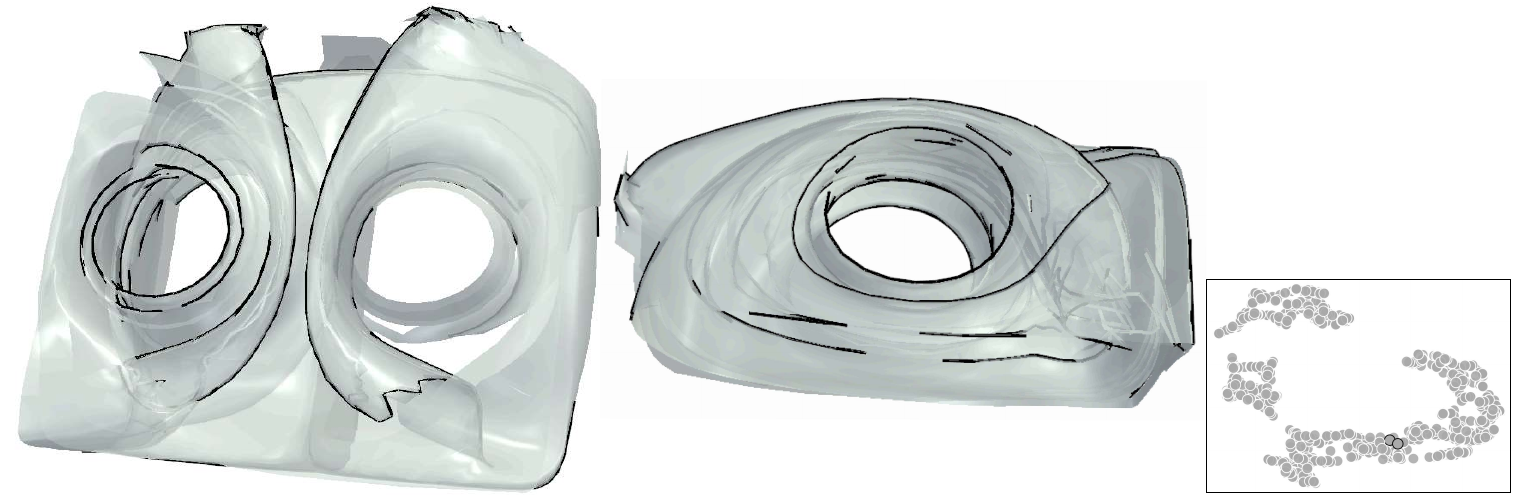}   &
            \includegraphics[width=0.45\linewidth]{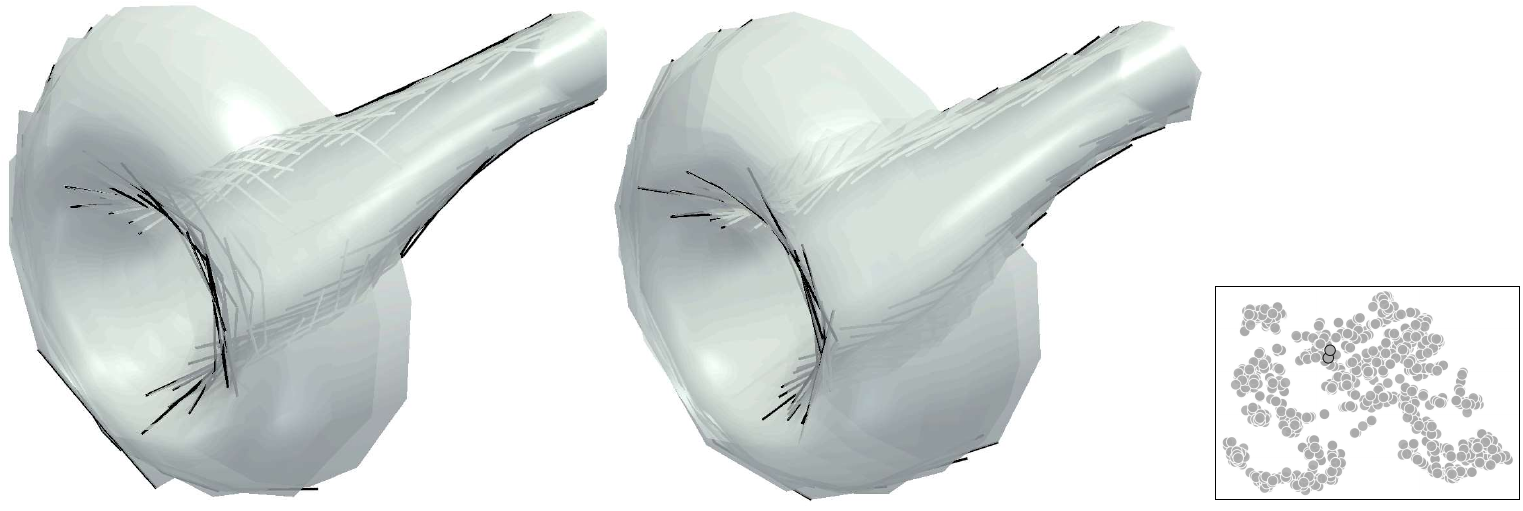}                                             \\
            \mbox{\footnotesize (c) t-SNE DR}                          & \mbox{\footnotesize (d) UMAP DR (ours)} \\
        \end{array}$
    \vspace{-0.1in}
    \caption{Comparing surface-level DR methods using the two swirls dataset. UMAP generates all projection views where each point represents a surface.}
    \vspace{-0.1in}
    \label{sl-DR}
\end{figure}

\begin{figure}[htbp]
    \centering
    $\begin{array}{c@{\hspace{0.1in}}c}
            \includegraphics[width=0.425\linewidth, height=0.95in]{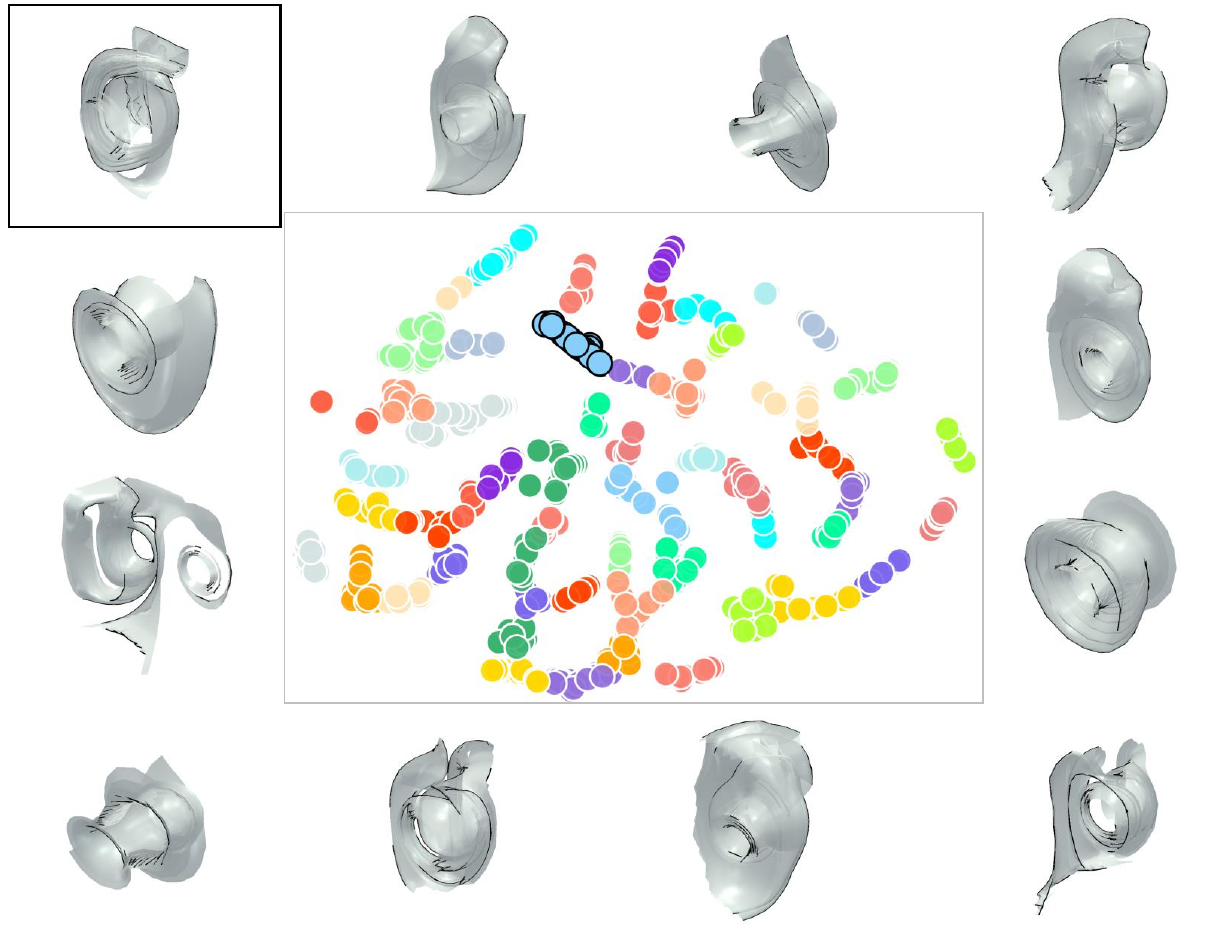}    &
            \includegraphics[width=0.425\linewidth, height=0.95in]{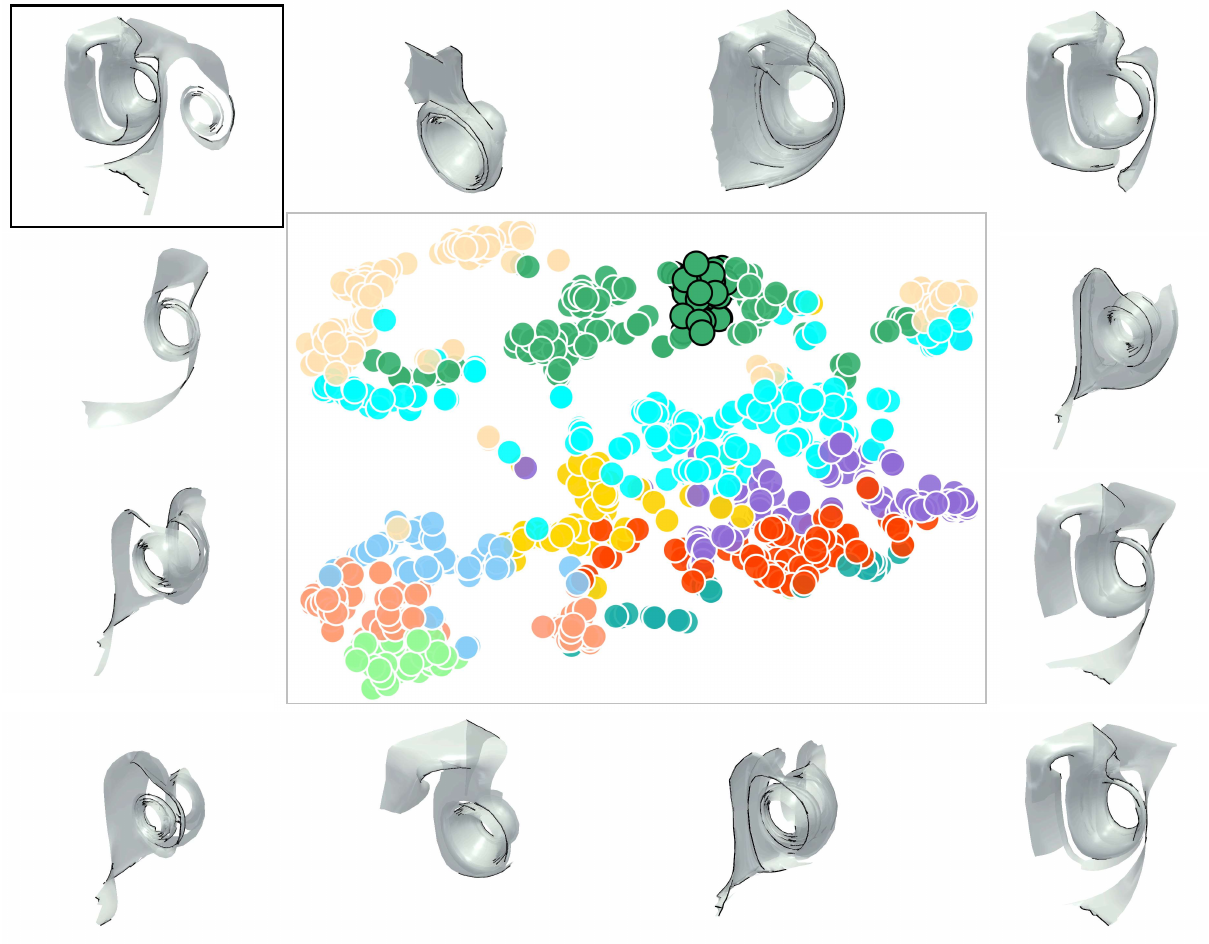}                                                     \\
            \mbox{\footnotesize (a) k-means}                               & \mbox{\footnotesize (b) DBSCAN}                \\
            \includegraphics[width=0.425\linewidth, height=0.95in]{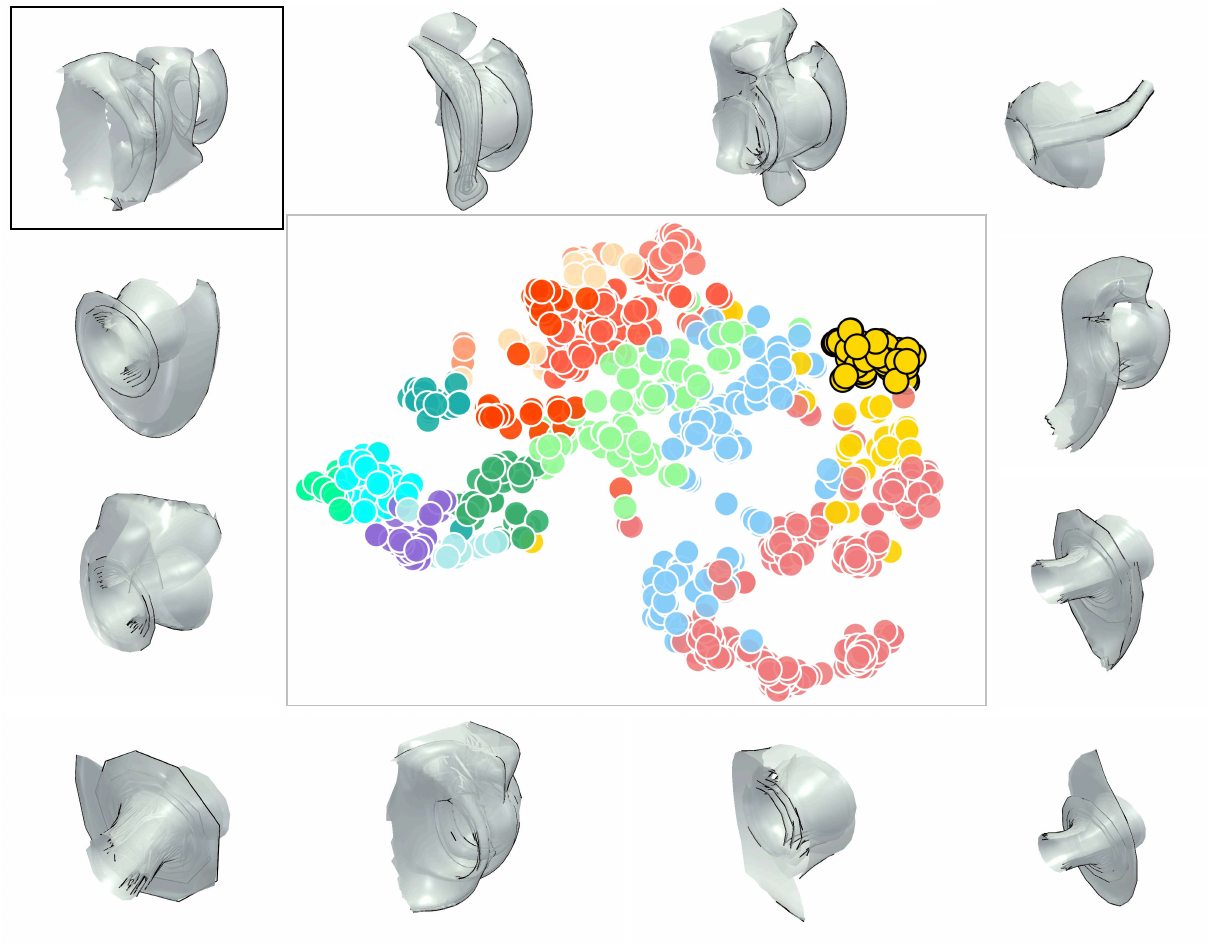} &
            \includegraphics[width=0.425\linewidth, height=0.95in]{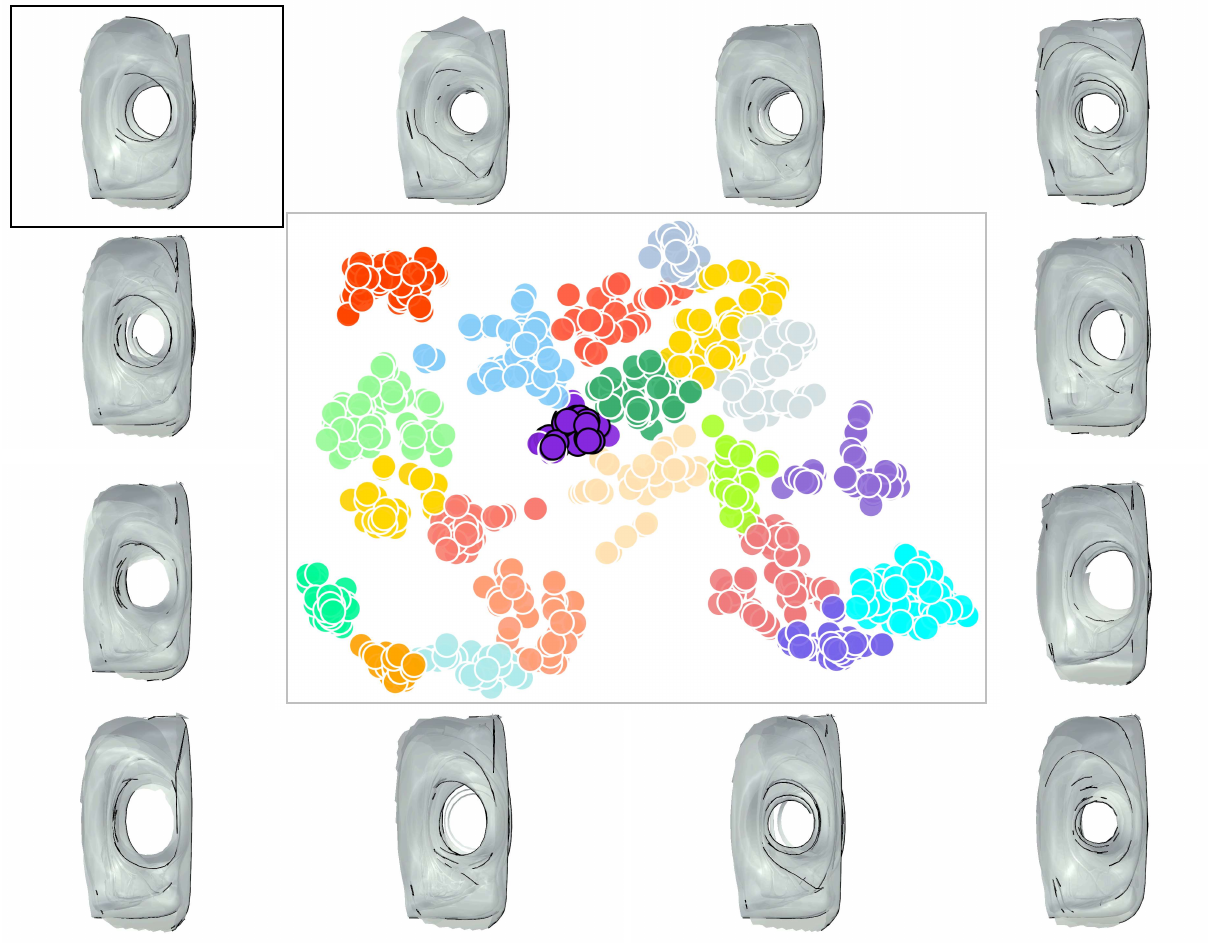}                                                        \\
            \mbox{\footnotesize (c) mean shift}                            & \mbox{\footnotesize (d) AHC without CC (ours)} \\
        \end{array}$
    \vspace{-0.1in}
    \caption{Comparing surface classification methods using the two swirls dataset. UMAP generates all projection views where each point represents a surface.}
    \vspace{-0.1in}
    \label{sl-clustering}
\end{figure}

{\bf Patch classification.}
We compare AHC without CC with k-means, DBSCAN, and mean shift on the UMAP-reduced patch-level embedding.
%
As shown in Figure~\ref{pl-clustering}, we select a patch in the inner swirl to query similar patches within the same surface.
Even though the matched points highlighted in black boundaries are well-packed for each clustering method, their patch-matching results differ.
For k-means and DBSCAN, their matching results are scattered, showing no continuous pattern.
Mean shift performs better than k-means and DBSCAN, but its matching result still does not identify the other patches in the same swirl; instead, it matches the larger patches in the adjacent swirl.
AHC generates its matching result that forms a continuous pattern, including almost all patches in the same swirl, which is most desirable.
We attribute this to the versatility of AHC, which is unconstrained by cluster shapes, enabling the discovery of arbitrary-shape clusters.
Unlike DBSCAN and mean shift, which may struggle with clusters of varying densities, AHC progressively merges adjacent data points based on the distance threshold, thus accommodating clusters with different densities.
Moreover, when constructing clusters, AHC demonstrates greater resilience to outliers by merging adjacent data points, rendering it less sensitive to outliers than k-means and mean shift.
In contrast, DBSCAN could misclassify outliers as noise.

{\bf Patch matching exploration.}
We explore patch matching on surfaces reported in Table~\ref{tab:datasets}.
Our experiments suggest setting $\delta_1=50$ (``patch size'') and $\delta_2=50$ (``matching tolerance''), as they lead to a reasonable number of patches and meaningful matching results.
Figures~\ref{patchMatching-1}~and~\ref{patchMatching-2} show representative examples, highlighting interesting results on similar and dissimilar surfaces.
Figure~\ref{patchMatching-1} (a) and (b) and Figure~\ref{patchMatching-2} (a), (b), (d), and (e) depict patch matching on the same and another similar surface where similarly shaped patches are consistently identified, confirming the effectiveness of SurfPatch.
In Figure~\ref{patchMatching-1} (c) and Figure~\ref{patchMatching-2} (c), we perform patch matching on both the same and another dissimilar surface drawn from a different surface cluster, and the results still remain meaningful.
For example, in Figure~\ref{patchMatching-1} (c), a mushroom-shaped patch identified on the first surface finds its counterpart on the second.
These results underscore our approach's ability to identify similar flow patterns across disparate surfaces.

Figure~\ref{SP&SN} compares SurfPatch with SurfNet on patch matching.
For SurfNet, we acquire node embeddings via the GCN and treat them as vertex features.
Since SurfNet does not provide patch-level representations, subsequent feature aggregation, clustering, and projection at the patch level follow the solutions presented for SurfPatch.
We select similar and dissimilar surfaces for two datasets in our comparison.
We set $\delta_1=70$ for SurfPatch's patch size, matching SurfNet's large patches to ensure fairness.
Results show that SurfPatch consistently produces more meaningful matched patches.
Specifically, SurfPatch matches patches in analogous positions on similar surfaces, while SurfNet struggles with similar shapes because its graph convolution aggregates surface information without ensuring that the patches are continuous. Additionally, their surface features contain many zero values and have significant values only in some dimensions, leading to larger errors during patch-level matching. 
Even on dissimilar surfaces, SurfPatch identifies more similar patches than SurfNet.

\begin{table*}[htbp]
    \caption{Comparison of Hausdorff Distance, Chamfer Distance, and RMSE across EdgeConv, SurfNet, and SurfPatch methods.}
    \vspace{-0.125in}
    \centering
    \resizebox{2\columnwidth}{!}{
        \begin{tabular}{cc|ccc|ccc|ccc}
           data   &      & \multicolumn{3}{c|}{EdgeConv} & \multicolumn{3}{c|}{SurfNet} & \multicolumn{3}{c}{SurfPatch}                                                                                                                              \\
                      type        &       dataset               & Hausdorff                    & Chamfer                     & RMSE                        & Hausdorff & Chamfer & RMSE & Hausdorff   & Chamfer     & RMSE        \\ \hline
            steady flow         & B{\'e}nard flow          & 0.735±0.057                            & 0.446±0.205                           & 0.754±0.090                            & 0.485±0.092        & 0.394±0.105      & 0.628±0.088   & 0.196±0.022 & 0.025±0.005 & 0.240±0.008 \\
            steady flow         & five critical points & 0.625±0.064                            & 0.386±0.135                           & 0.705±0.065                            & 0.433±0.074        & 0.312±0.086      & 0.582±0.072   & 0.158±0.013 & 0.009±0.004 & 0.200±0.003 \\
            steady flow         & solar plume          & 0.634±0.058                            & 0.388±0.140                           & 0.709±0.082                            & 0.448±0.090        & 0.320±0.105      & 0.586±0.080   & 0.161±0.006 & 0.011±0.003 & 0.212±0.004 \\
            steady flow         & square cylinder      & 0.612±0.055                            & 0.375±0.096                           & 0.684±0.071                            & 0.421±0.064        & 0.310±0.092      & 0.525±0.066   & 0.147±0.005 & 0.008±0.001 & 0.197±0.002 \\
            steady flow         & tornado              & 0.655±0.054                            & 0.402±0.154                           & 0.730±0.095                            & 0.450±0.075        & 0.325±0.085      & 0.585±0.075   & 0.160±0.008 & 0.009±0.001 & 0.210±0.003 \\
            steady flow         & two swirls           & 0.684±0.065                            & 0.415±0.150                           & 0.732±0.120                            & 0.462±0.080        & 0.375±0.122      & 0.614±0.090   & 0.185±0.019 & 0.012±0.001 & 0.225±0.005 \\ \hline
            unsteady flow       & solar plume          & 0.630±0.062                            & 0.384±0.125                           & 0.712±0.081                            & 0.452±0.092        & 0.323±0.102      & 0.584±0.083   & 0.160±0.005 & 0.012±0.003 & 0.215±0.005 \\
            unsteady flow       & tornado              & 0.658±0.055                            & 0.398±0.132                           & 0.722±0.092                            & 0.447±0.074        & 0.328±0.080      & 0.582±0.078   & 0.159±0.007 & 0.009±0.002 & 0.208±0.004 \\ \hline
            time-varying scalar & earthquake           & 0.742±0.068                            & 0.504±0.225                           & 0.766±0.092                            & 0.503±0.095        & 0.405±0.124      & 0.655±0.085   & 0.205±0.018 & 0.038±0.012 & 0.285±0.055 \\
            time-varying scalar & ionization           & 0.739±0.065                            & 0.488±0.203                           & 0.763±0.087                            & 0.498±0.096        & 0.399±0.118      & 0.657±0.086   & 0.208±0.032 & 0.036±0.010 & 0.288±0.037 \\ 
        \end{tabular}
    }
    \label{tab:quantitative}
    \vspace{-0.15in}
\end{table*}

\vspace{-0.05in}
\subsection{Surface-Level Clustering}
We perform surface-level clustering to guide users in selecting surfaces of interest for patch querying. Based on the findings in Section~\ref{sec:surface_clustering}, we evaluate various DR techniques and clustering methods using surface-level embeddings derived from patch-level ones. The two swirls dataset comprises the most diverse shapes, making it suitable for comparisons.

    {\bf Surface-level DR.}
We compare four DR techniques: Isomap, MDS, t-SNE, and UMAP. We expect that surfaces adjacent to each other in the 2D projection should be similar. We select two neighboring surfaces from the projection view for validation, representing the general DR outcome.
To assist users in exploring the dataset, our approach clusters surfaces and identifies ``representative'' surfaces, which are cluster centroids. These representatives provide different-shaped surfaces and reduce the search scope. Users can first browse these representative surfaces to identify interesting ones. Once an interesting surface is found, they can explore other surfaces within the cluster, allowing for targeted exploration. This strategy balances diversity and relevance, enabling users to quickly locate distinct and representative surfaces for further analysis.
In Figure~\ref{sl-DR}, t-SNE projection reveals three distinct clusters, while Isomap and MDS give similar circular patterns, and UMAP yields a relatively compact projection. For the two neighboring surfaces selected from the projection, UMAP has the most similar surfaces, followed by MDS. Isomap and t-SNE give the least similar ones. We attribute UMAP's superior performance to its optimization process, which prioritizes the preservation of local neighborhoods and is advantageous in capturing intricate surface similarities.

    {\bf Surface classification.}
We compare AHC without CC with k-means, DBSCAN, and mean shift on the UMAP-reduced surface-level embedding. We adjust each method's parameter to yield similar numbers of surface clusters across different methods.
Figure~\ref{sl-clustering} presents a subset of sample surfaces, with representative surfaces (i.e., cluster centroids) highlighted with black bounding boxes.
The result with k-means tends to group more points to form larger clusters interspersed with smaller ones. DBSCAN and mean shift produce clusters of varying sizes but merge some distinct clusters, resulting in less accurate separation. Conversely, AHC yields more evenly distributed clusters.
Regarding surface similarity, with k-means, some sample surfaces resemble the representative, while others do not. DBSCAN and mean shift give the worst results, with sample surfaces significantly deviating from representatives. With AHC, sample surfaces show the best similarity to the representative one.
Although dimensionality-reduced surface-level embedding may make neighboring points appear similar, inappropriate clustering along the borders could group dissimilar surfaces. AHC emerges as the most suitable method for surface-level clustering as it discovers clusters of arbitrary shapes and accommodates clusters with different densities, unconstrained by their shapes.

\begin{figure*}[htbp]
    \centering
    $\begin{array}{c@{\hspace{0.025in}}c@{\hspace{0.025in}}c@{\hspace{0.025in}}c}
            \includegraphics[height=0.825in]{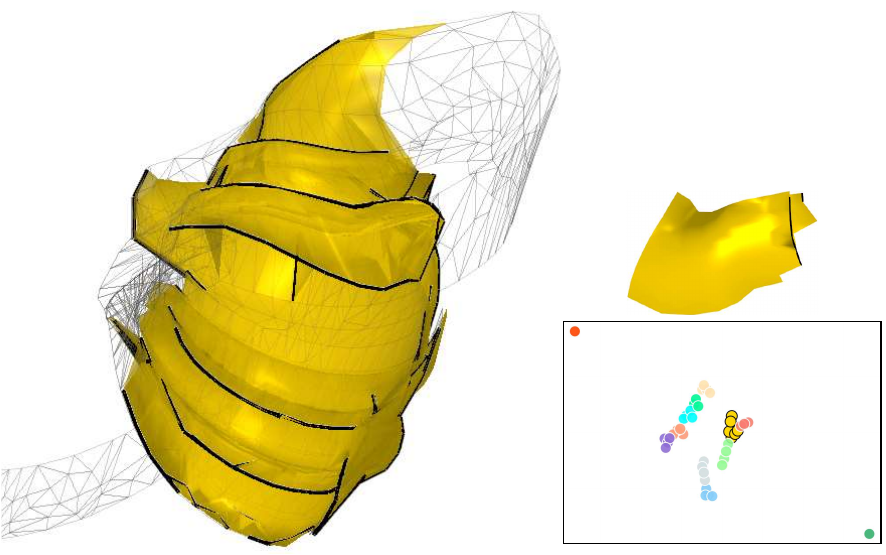} &
            \includegraphics[height=0.825in]{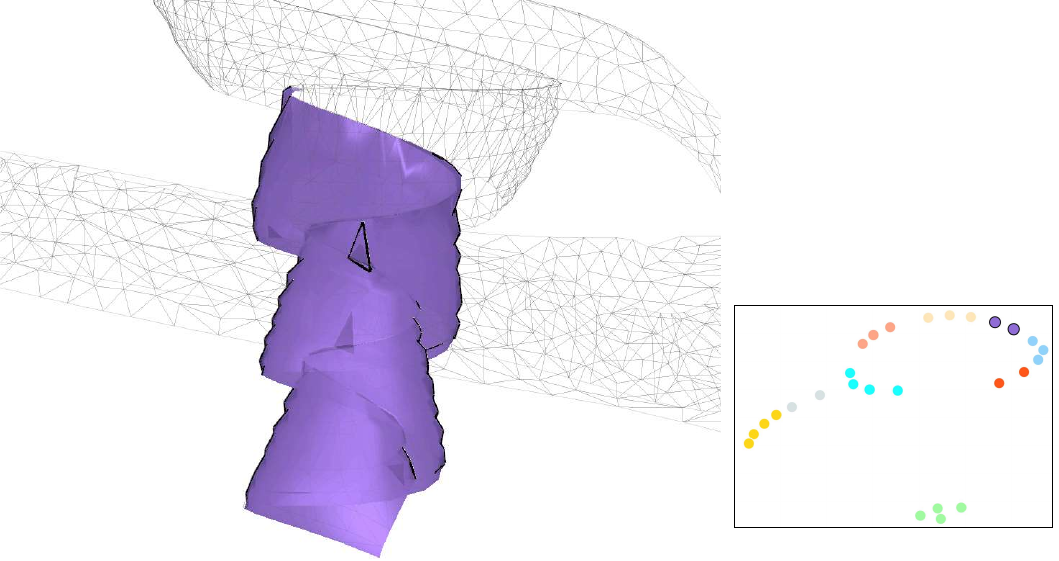}  &
            \includegraphics[height=0.825in]{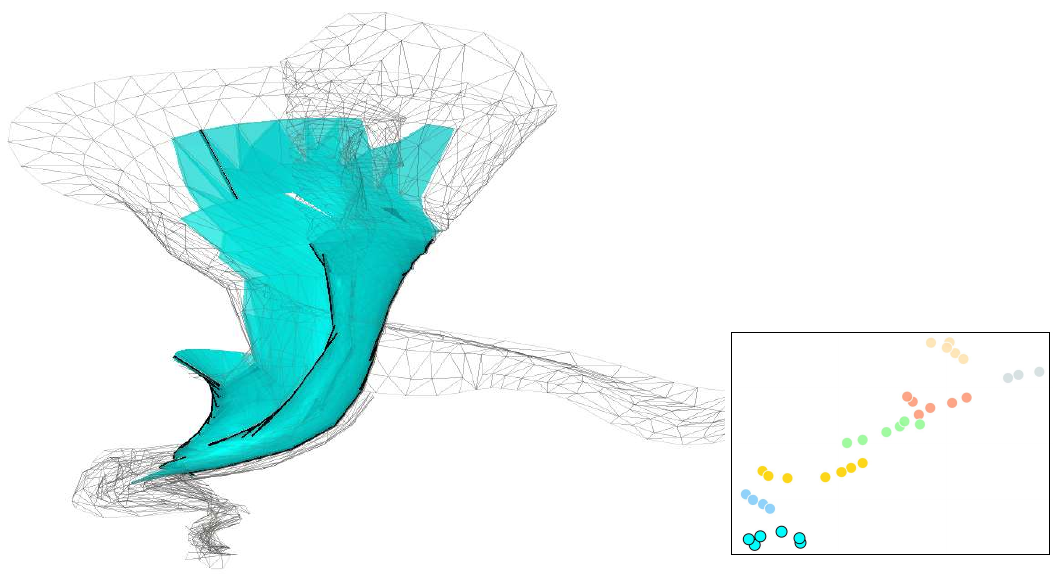}  &
            \includegraphics[height=0.825in]{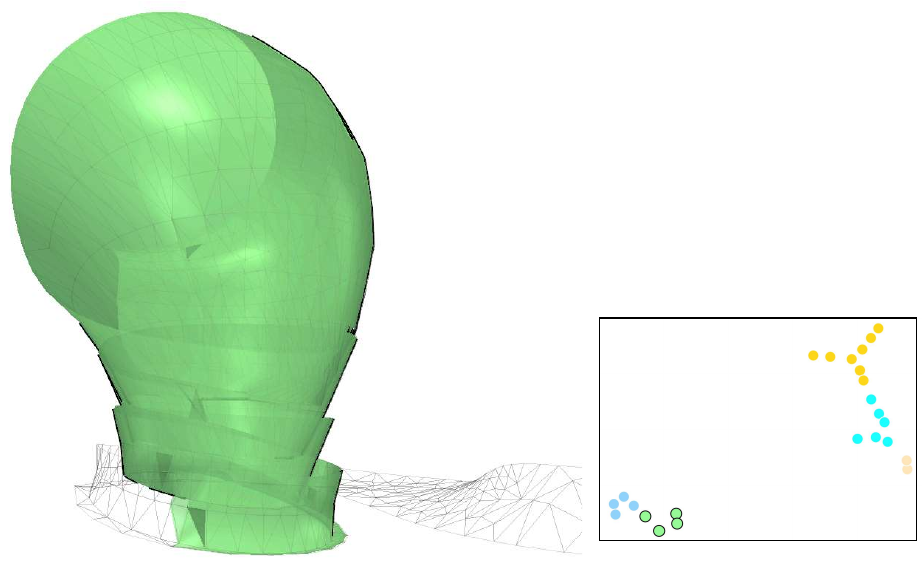}                                                                                                     \\
            \mbox{\footnotesize (a) $t=12$}                          & \mbox{\footnotesize (b) $t=1$} & \mbox{\footnotesize (c) $t=6$} & \mbox{\footnotesize (d) $t=9$}
        \end{array}$
    $\begin{array}{c@{\hspace{0.025in}}c@{\hspace{0.025in}}c@{\hspace{0.025in}}c}
            \includegraphics[height=0.825in]{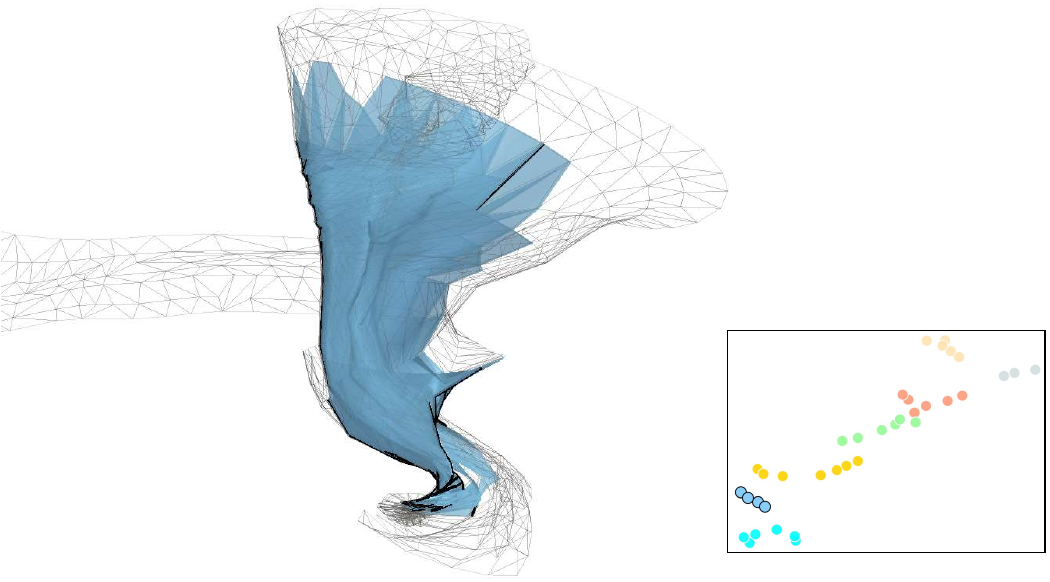} &
            \includegraphics[height=0.825in]{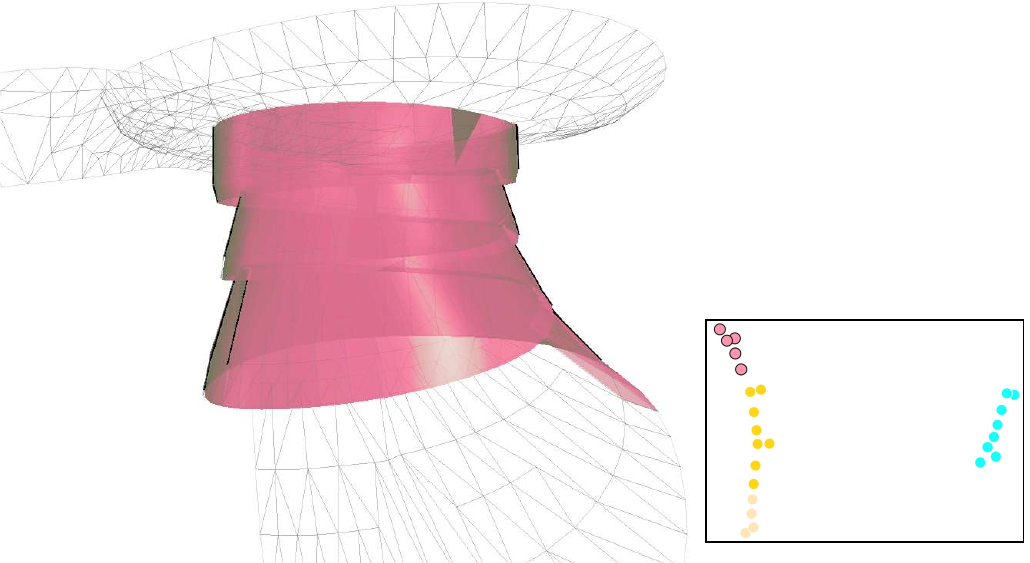} &
            \includegraphics[height=0.825in]{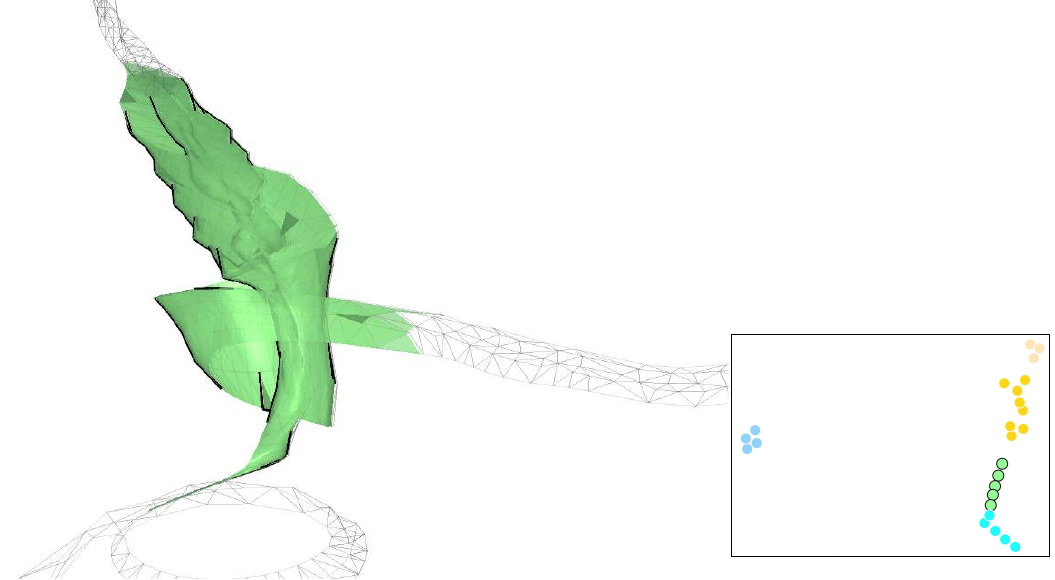} &
            \includegraphics[height=0.825in]{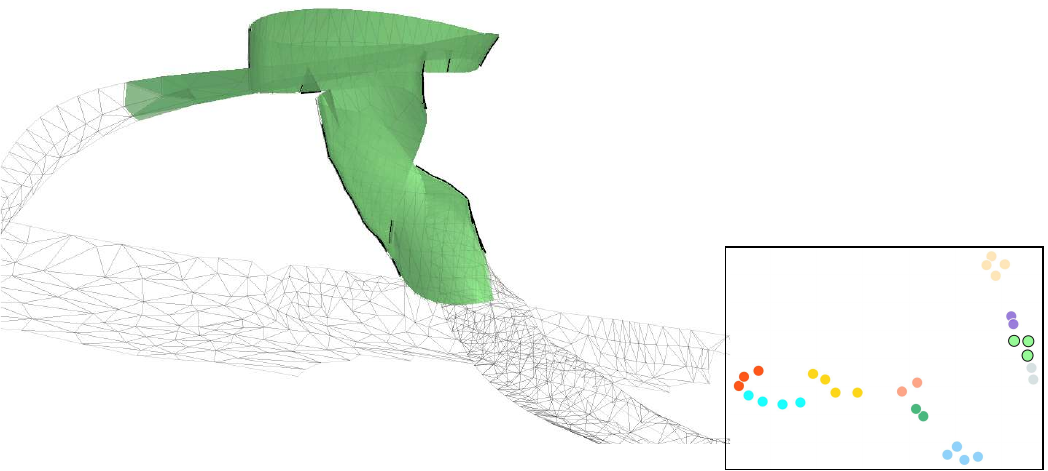}                                                                                                       \\
            \mbox{\footnotesize (e) $t=18$}                          & \mbox{\footnotesize (f) $t=20$} & \mbox{\footnotesize (g) $t=24$} & \mbox{\footnotesize (h) $t=27$}
        \end{array}$
    $\begin{array}{c@{\hspace{0.025in}}c@{\hspace{0.025in}}c@{\hspace{0.025in}}c}
            \includegraphics[height=0.825in]{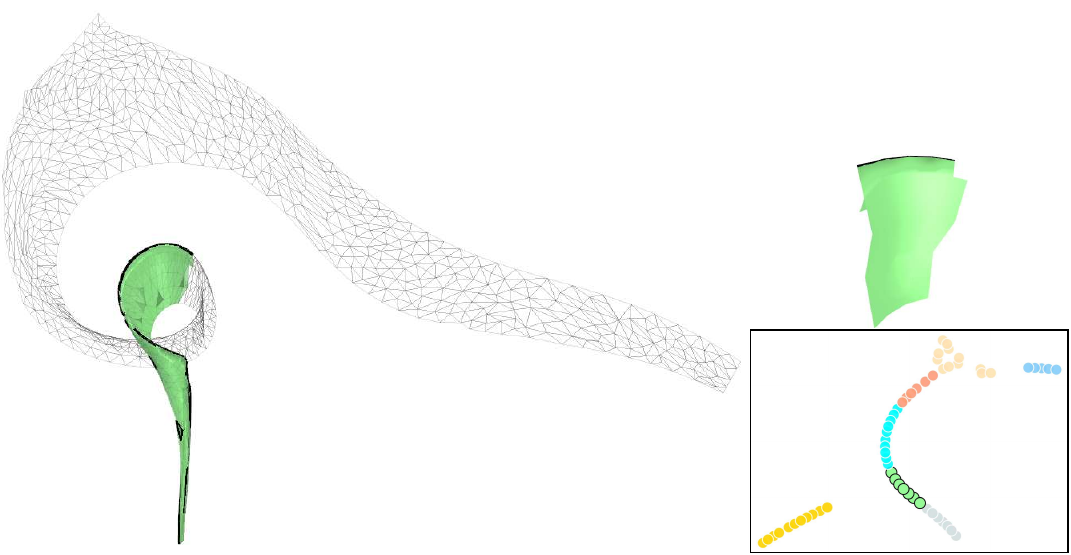} &
            \includegraphics[height=0.825in]{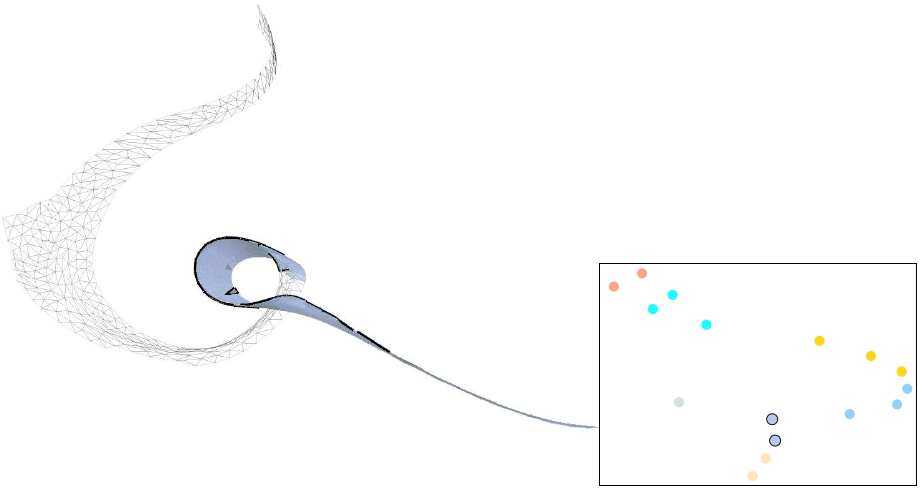}  &
            \includegraphics[height=0.825in]{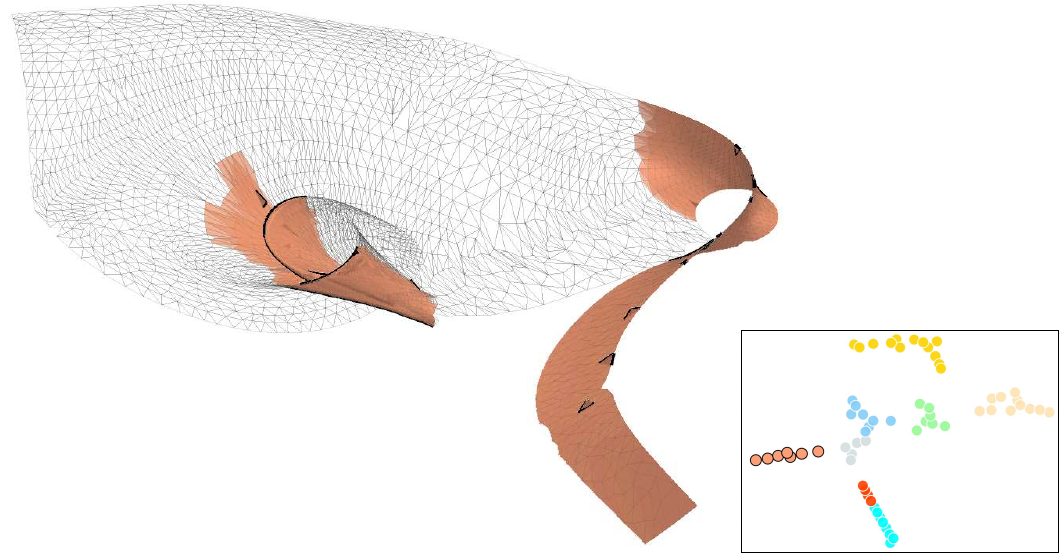}  &
            \includegraphics[height=0.825in]{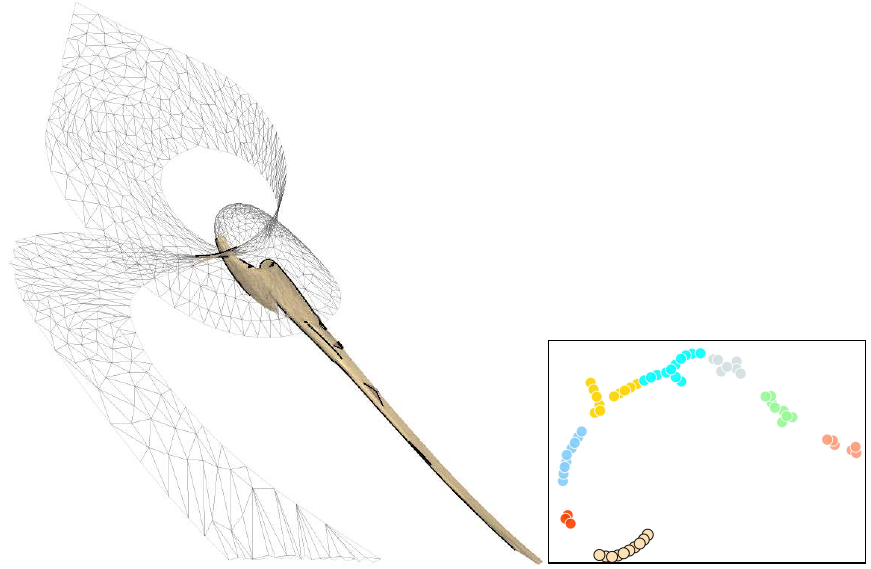}                                                                                                     \\
            \mbox{\footnotesize (a) $t=24$}                            & \mbox{\footnotesize (b) $t=1$} & \mbox{\footnotesize (c) $t=6$} & \mbox{\footnotesize (d) $t=12$}
        \end{array}$
    $\begin{array}{c@{\hspace{0.025in}}c@{\hspace{0.025in}}c@{\hspace{0.025in}}c}
            \includegraphics[height=0.825in]{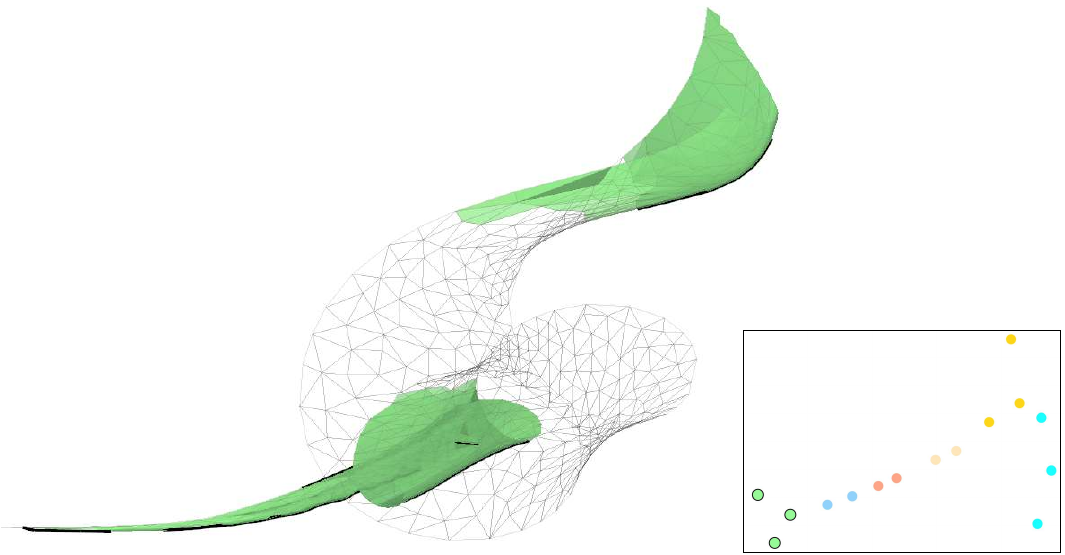} &
            \includegraphics[height=0.825in]{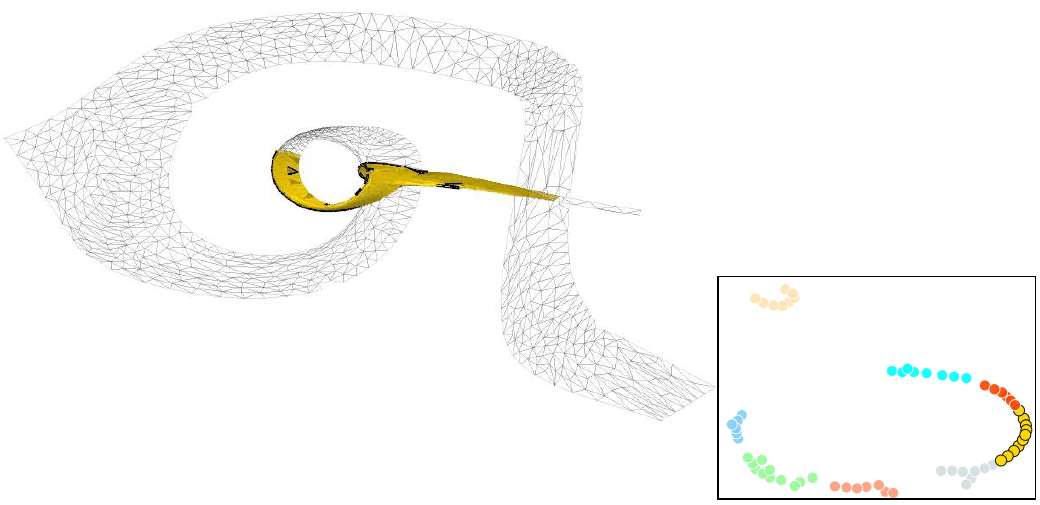} &
            \includegraphics[height=0.825in]{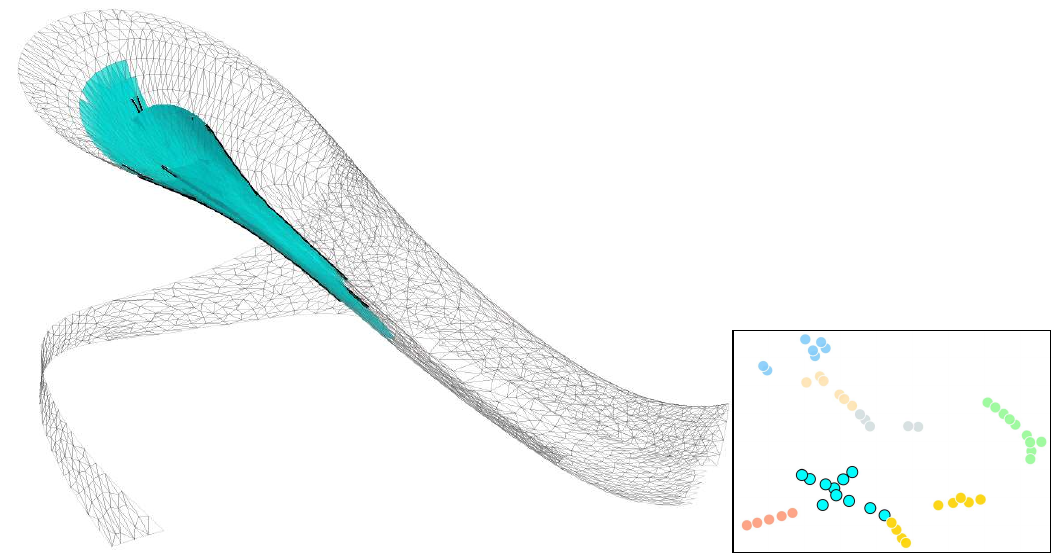} &
            \includegraphics[height=0.825in]{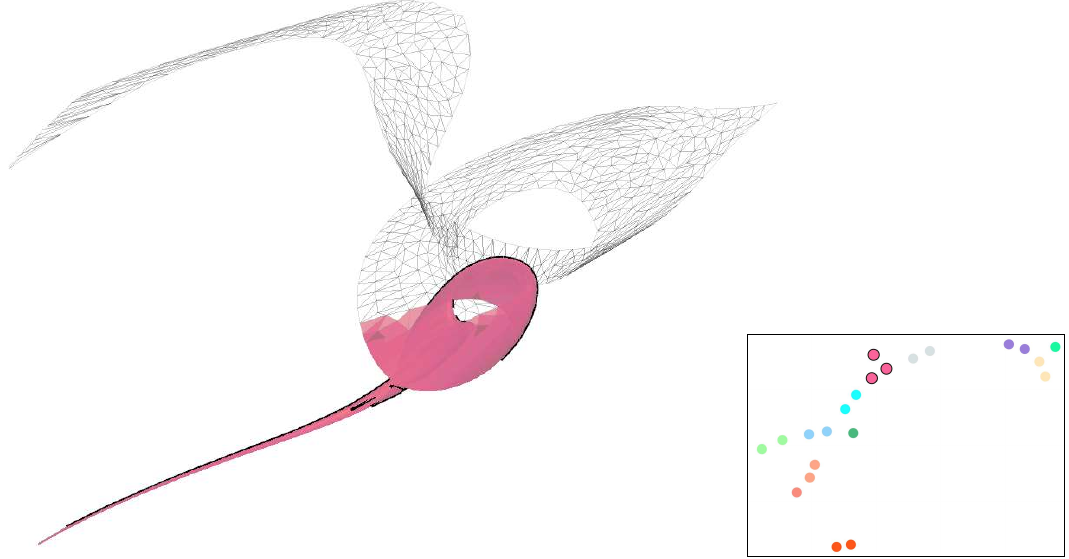}                                                                                                       \\
            \mbox{\footnotesize (e) $t=20$}                            & \mbox{\footnotesize (f) $t=36$} & \mbox{\footnotesize (g) $t=40$} & \mbox{\footnotesize (h) $t=48$}
        \end{array}$
    \vspace{-0.1in}
    \caption{Patch matching results on unsteady flow solar plume (top two rows) and tornado (bottom two rows). The selected stream surface patch is shown in (a), and the matching results on the same and different timesteps ($t$) are shown in (a) to (h).}
    \vspace{-0.1in}
    \label{time-streamsurface}
\end{figure*}

\begin{figure*}[htbp]
    \centering
    $\begin{array}{c@{\hspace{0.025in}}c@{\hspace{0.025in}}c@{\hspace{0.025in}}c}
            \includegraphics[height=0.825in]{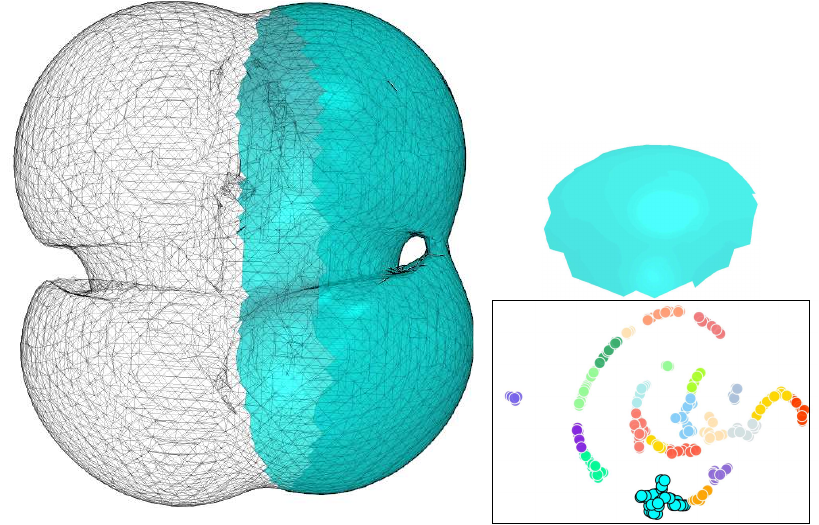}  &
            \includegraphics[height=0.825in]{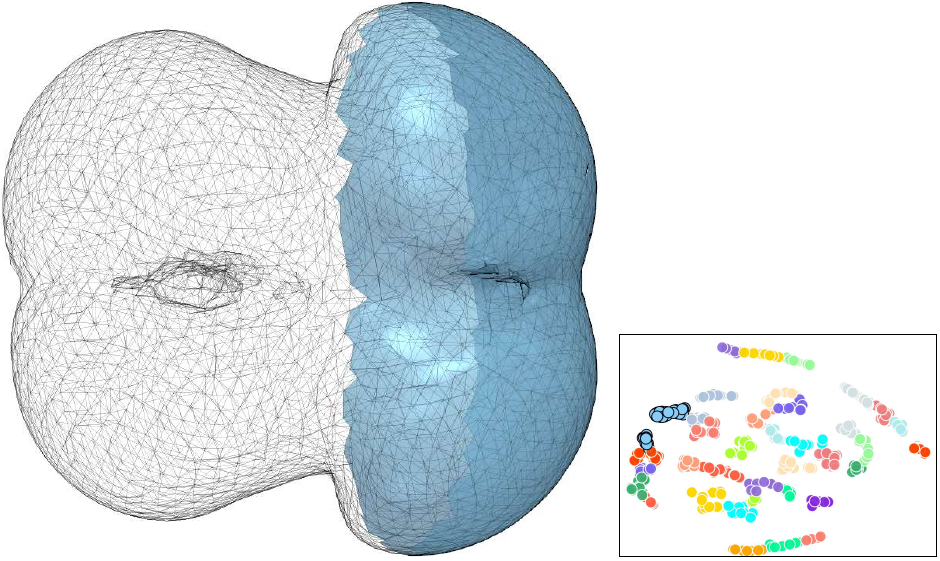}  &
            \includegraphics[height=0.825in]{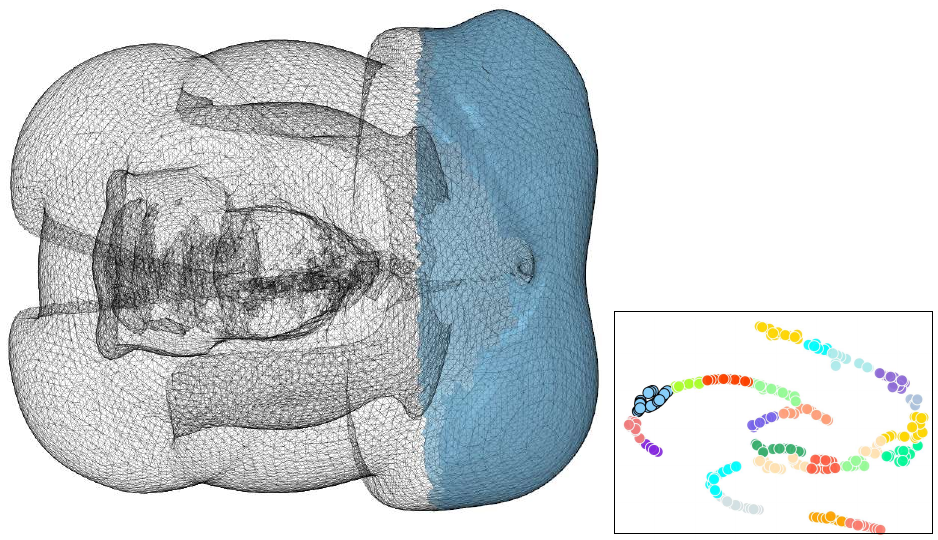} &
            \includegraphics[height=0.825in]{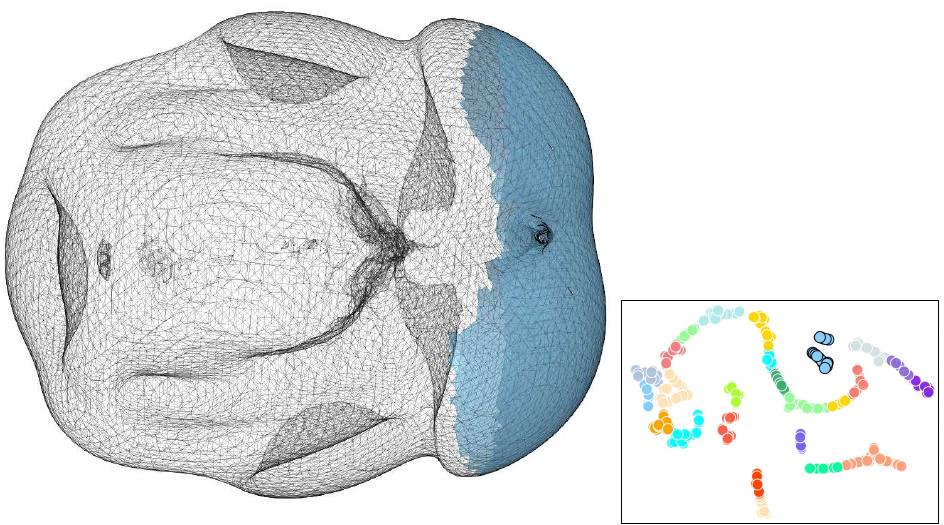}                                                                                                                                    \\
            \mbox{\footnotesize (a) $v=0.04$, $t=1$}            & \mbox{\footnotesize (b) $v=0.08$, $t=1$} & \mbox{\footnotesize (c) $v=0.04$, $t=30$} & \mbox{\footnotesize (d) $v=0.08$, $t=30$}
        \end{array}$
    $\begin{array}{c@{\hspace{0.025in}}c@{\hspace{0.025in}}c@{\hspace{0.025in}}c}
            \includegraphics[height=0.825in]{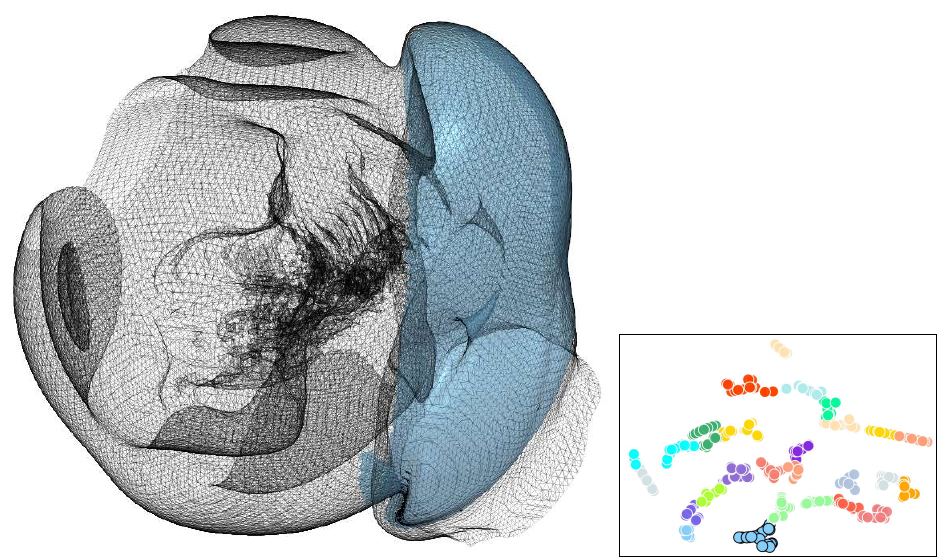} &
            \includegraphics[height=0.825in]{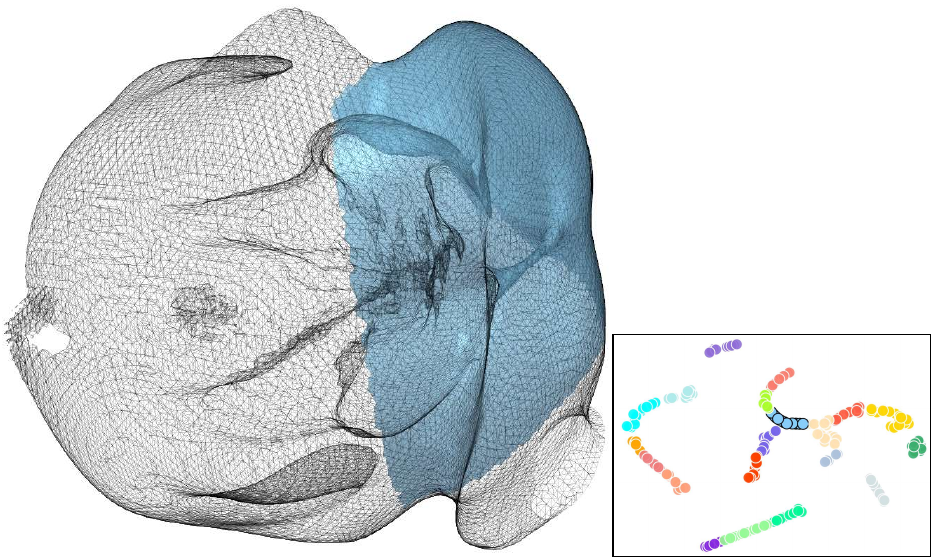} &
            \includegraphics[height=0.825in]{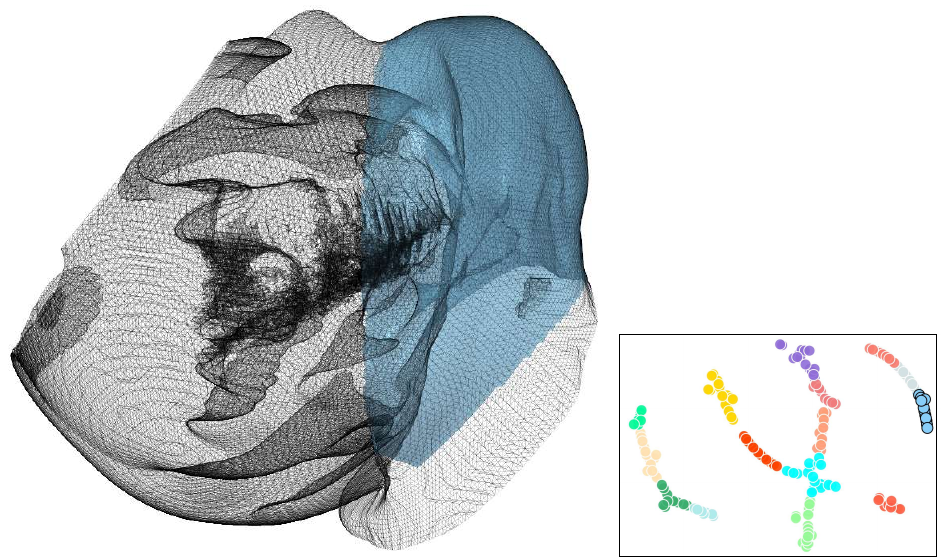} &
            \includegraphics[height=0.825in]{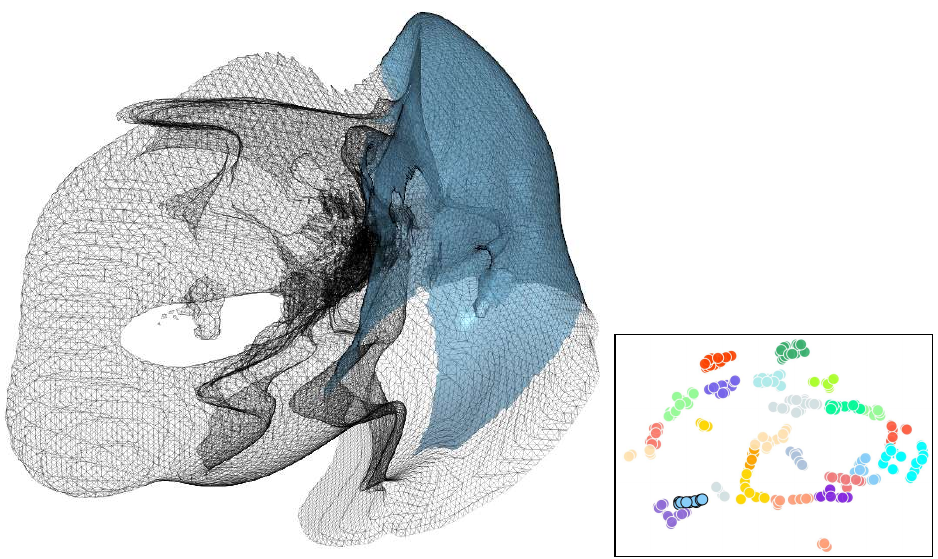}                                                                                                                                     \\
            \mbox{\footnotesize (e) $v=0.04$, $t=45$}           & \mbox{\footnotesize (f) $v=0.08$, $t=45$} & \mbox{\footnotesize (g) $v=0.04$, $t=60$} & \mbox{\footnotesize (h) $v=0.08$, $t=60$}
        \end{array}$
    $\begin{array}{c@{\hspace{0.025in}}c@{\hspace{0.025in}}c@{\hspace{0.025in}}c}
            \includegraphics[height=0.825in]{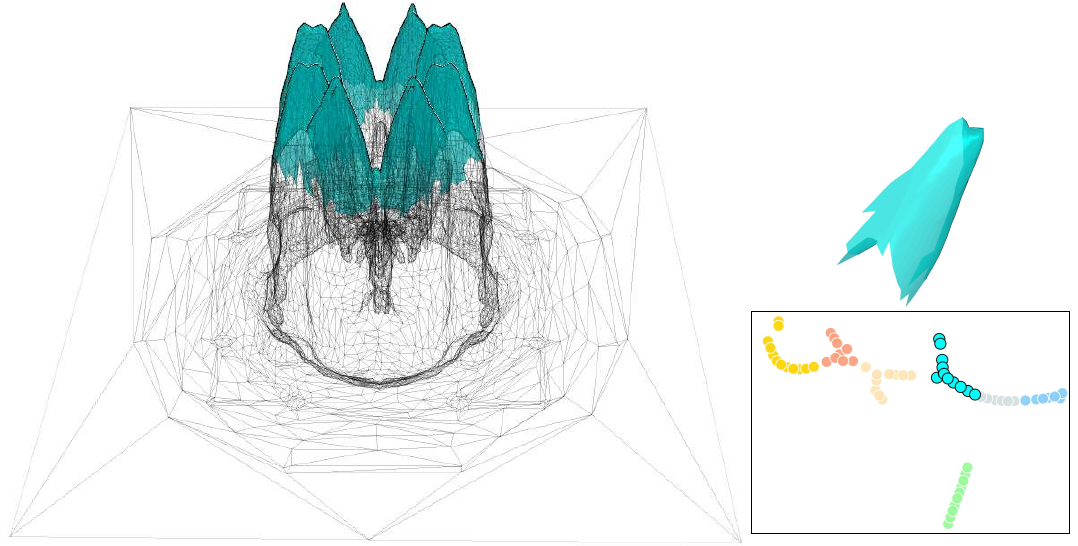}  &
            \includegraphics[height=0.825in]{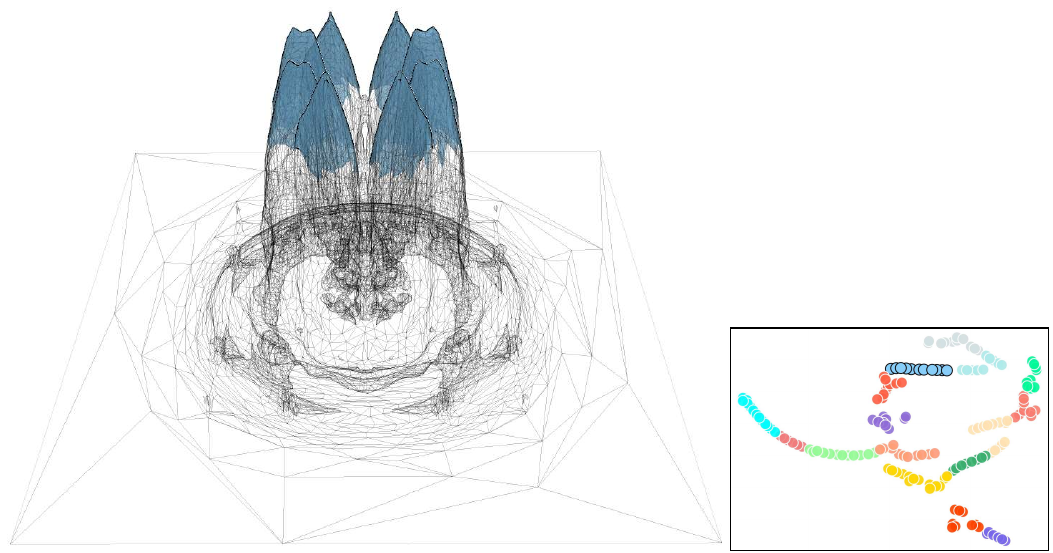}  &
            \includegraphics[height=0.825in]{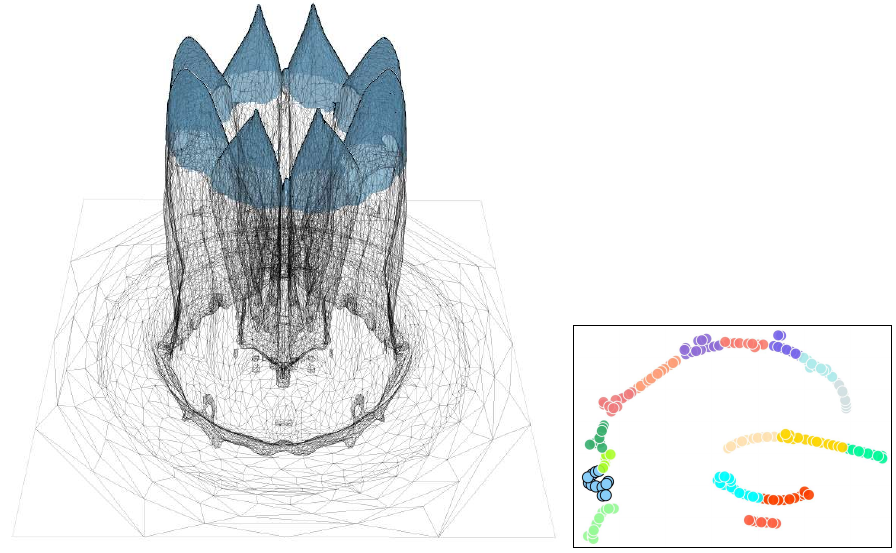} &
            \includegraphics[height=0.825in]{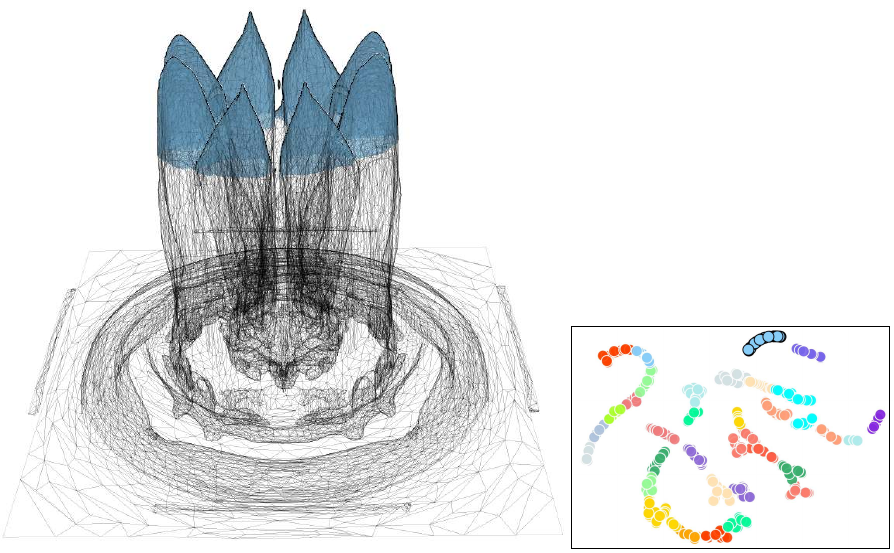}                                                                                                                                 \\
            \mbox{\footnotesize (a) $v=0.5$, $t=1$}            & \mbox{\footnotesize (b) $v=0.7$, $t=1$} & \mbox{\footnotesize (c) $v=0.5$, $t=30$} & \mbox{\footnotesize (d) $v=0.7$, $t=30$}
        \end{array}$
    $\begin{array}{c@{\hspace{0.025in}}c@{\hspace{0.025in}}c@{\hspace{0.025in}}c}
            \includegraphics[height=0.825in]{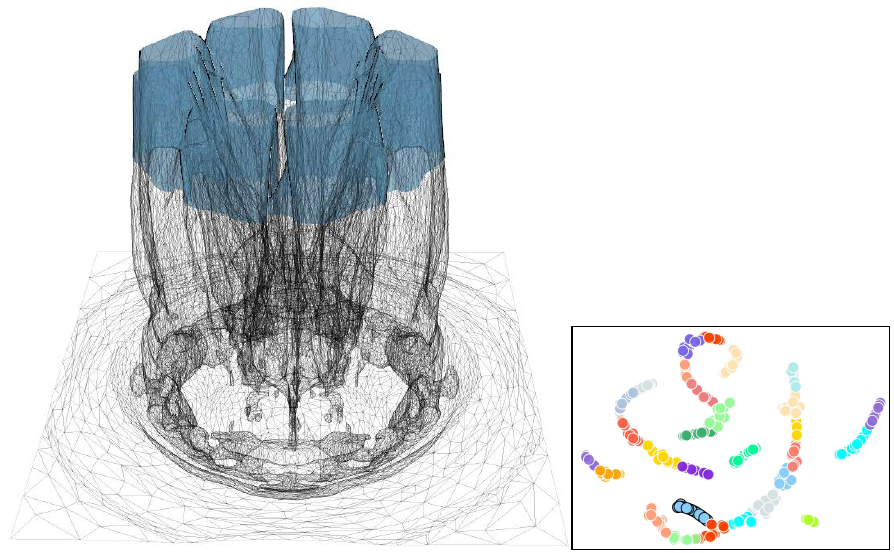}  &
            \includegraphics[height=0.825in]{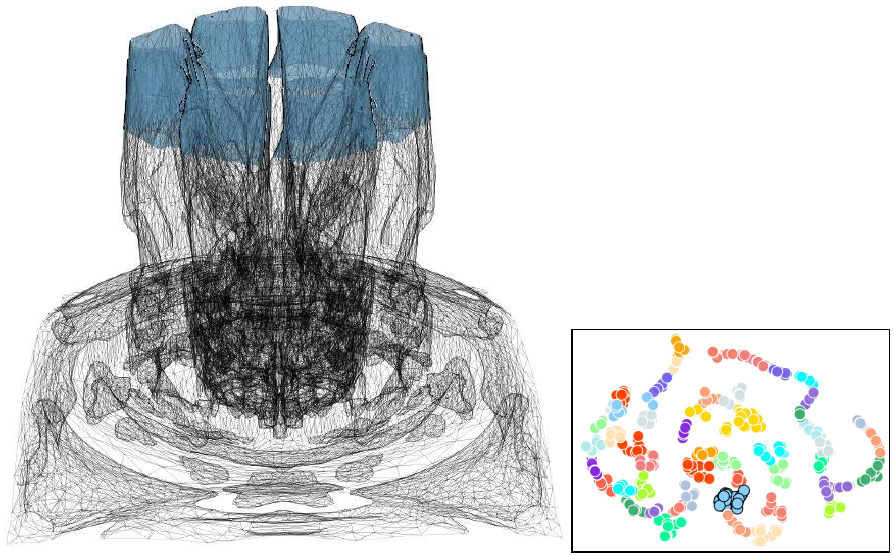}  &
            \includegraphics[height=0.825in]{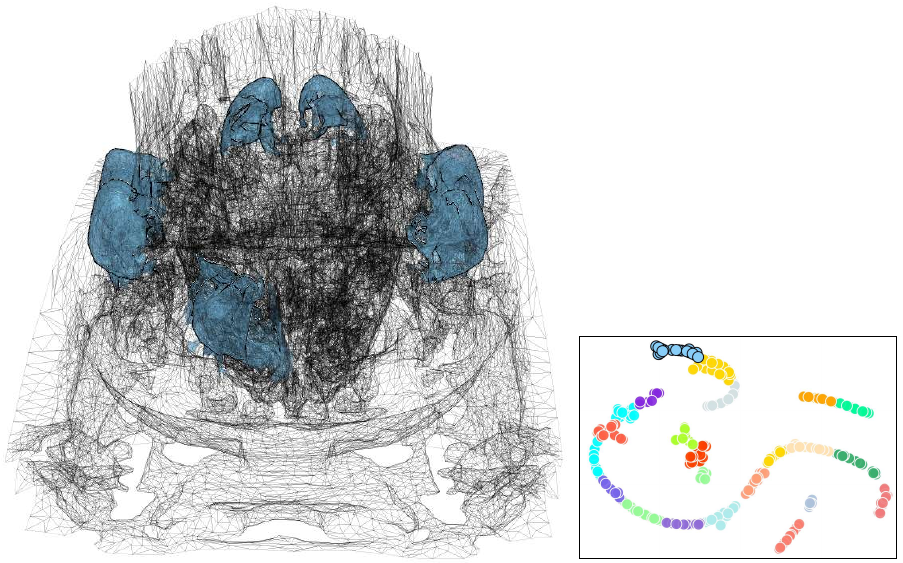} &
            \includegraphics[height=0.825in]{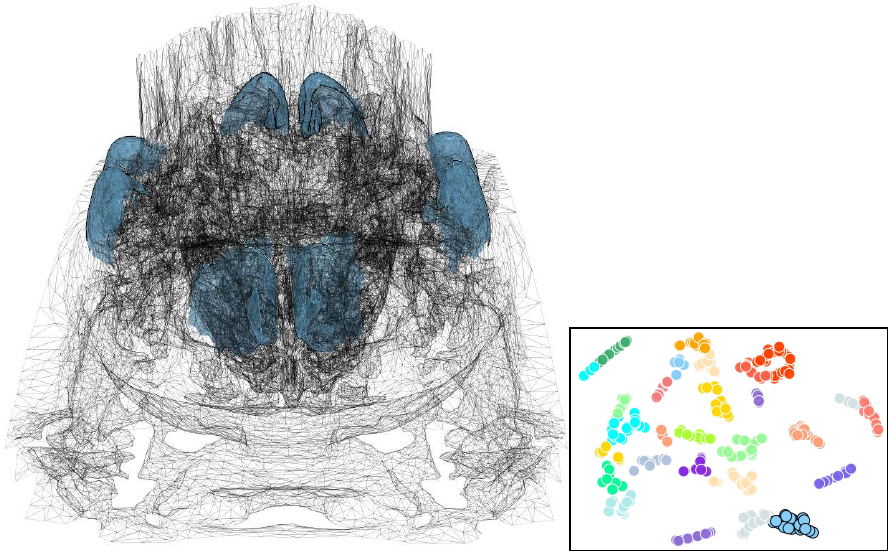}                                                                                                                                    \\
            \mbox{\footnotesize (e) $v=0.5$, $t=60$}            & \mbox{\footnotesize (f) $v=0.7$, $t=60$} & \mbox{\footnotesize (g) $v=0.5$, $t=100$} & \mbox{\footnotesize (h) $v=0.7$, $t=100$}
        \end{array}$
    \vspace{-0.1in}
    \caption{Patch matching results on time-varying scalar earthquake (top two rows) and ionization (bottom two rows). The selected isosurface patch is shown in (a), and the matching results on the same and different isovalues ($v$) and timesteps ($t$) are shown in (a) to (h).}
    \vspace{-0.1in}
    \label{time-isosurface}
\end{figure*}

\vspace{-0.05in}
\subsection{Quantitative Evaluation}
\label{subsec:quantitative}

We evaluate SurfPatch using three metrics: Hausdorff distance for global differences, Chamfer distance for local differences, and {\em root mean squared error} (RMSE) for overall mean deviation. Table~\ref{tab:quantitative} summarizes the results across different datasets. For evaluation, 100 surfaces are randomly selected from each dataset. For each surface, a patch is randomly chosen along with its matching results using the suggested $\delta_1$ and $\delta_2$. The vertex coordinates of these patches are normalized, and three metrics are computed. The results demonstrate that SurfPatch outperforms EdgeConv and SurfNet in identifying similar patches. Consistent findings were observed when varying patch size and matching tolerance, aligning with our previous experiments.

\vspace{-0.05in}
\subsection{Unsteady Flow and Time-Varying Scalar Fields}
\label{subsec:more}

In addition to stream surfaces extracted from steady flow, we evaluate SurfPatch on stream surfaces extracted from unsteady flow (solar plume and tornado). The patch-matching results are illustrated in Figure~\ref{time-streamsurface}. We can see that SurfPatch can find similar patches across different timesteps, which is desirable.
For isosurfaces, we test on the earthquake and ionization datasets, and isosurfaces are extracted from different isovalues and timesteps. The patch-matching results are illustrated in Figure~\ref{time-isosurface}. The results demonstrate that SurfPatch can effectively match patches across isovalues and timesteps, showcasing its broad applicability.

\vspace{-0.05in}
\subsection{Ad-Hoc Expert Feedback}

We conducted an ad-hoc evaluation with a domain expert who is not a co-author of this work.
The expert has over 12 years of research experience in computation fluid dynamics (CFD), cardiovascular flow, and machine learning surrogate models.
We demonstrated the SurfPatch's visual interface to him, explaining essential functions and features.

The expert also used the SurfPatch interface to analyze the tornado and two swirls flow datasets. In the tornado dataset, the expert considered that SurfPatch effectively identified the funnel-shaped surface, clearly revealing the location and size of the vortex core. In the two swirls dataset, he used SurfPatch to identify unique ``mushroom''-shaped surface patches formed by the swirling motion. This allowed the expert to observe the rotation axis and locate vortex positions. Two main vortices, rotating in opposite directions and of similar size, were detected. SurfPatch also discovered a ``tube'' shape along the helical streamline, highlighting the vortex's progression from upstream to downstream, and a ``bottle''-shaped surface patch enveloping the two main vortices, illustrating the downstream expansion of the vortices.

The expert's evaluation indicates that SurfPatch effectively identifies and visualizes complex flow structures, supporting in-depth analysis of vortex features. However, it may require expert interpretation to understand subtle flow features fully, and its applicability to extremely complex or turbulent datasets needs further investigation.

\vspace{-0.05in}
\subsection{Limitations}

We acknowledge the following limitations of our SurfPatch.
    {\bf Flexibility.}
SurfPatch allows users to adjust ``patch size'' and ``matching tolerance.''
However, the current setup lacks automatic parameter tuning, leading to a trial-and-error process, which could entail significant user effort when dealing with numerous surfaces or patches.
Our framework could benefit from automatic tuning to streamline operations and enhance user experience.
    {\bf Performance.}
Even though we accelerate the DR process via GPU, loading a large simplified surface still takes considerable time.
For example, 100 surfaces with over 2,000 vertices each (like one from the B{\'e}nard flow) take around 2 seconds to load, as noticed by the expert.
This delay is primarily due to the computational complexity of DR techniques and clustering algorithms.
Future efforts should optimize the implementation to reduce processing time and enhance performance.
    {\bf Generalization.}
For flow datasets, we only use stream surfaces and do not show the generalization of SurfPatch to other surfaces, such as path surfaces, streak surfaces, and time surfaces. Future research should further demonstrate SurfPatch's generalizability.

\vspace{-0.05in}
\section{Conclusions and Future Work}

We have presented SurfPatch, a novel patch-matching framework for exploratory SBFV. 
Placing a unique focus on the partial surfaces we call {\em patches}, SurfPatch enables fine-grained matching and querying flow features and patterns exhibited by a single and across multiple stream surfaces. 
We achieve this goal by advocating a bottom-up approach to building the multiscale connections between vertices and patches and between patches and surfaces. 
This approach leads to the three pillars: vertex-level classification, patch-level matching, and surface-level clustering. 
We provide design choices in each pillar and conduct careful comparisons to validate our selected methods or techniques. 
While the primary goal of SurfPatch is to support patch-level matching and querying, we enrich our results with vertex- and surface-level comparisons for a convincing presentation. 
Finally, we design an intuitive visual interface and provide a suite of interactions for users to perform exploratory stream surface visualization, a critical need neglected by many existing SBFV works. 
We also demonstrate the usefulness of SurfPatch on isosurfaces extracted from scalar fields.

For future work, we would like to extend SurfPatch from handling stream surfaces only to path surfaces, streak surfaces, and time surfaces. 
These surfaces have self-intersections, making partitioning patches and obtaining matching results difficult. 
We have only performed an ad-hoc evaluation with a CFD expert so far. 
We will conduct a comprehensive task-driven evaluation with CFD experts and students with flow visualization backgrounds for further improvements.

\vspace{-0.05in}
\acknowledgments{This research was supported in part by the U.S.\ National Science Foundation through grants IIS-1955395, IIS-2101696, OAC-2104158, and IIS-2401144, and the U.S.\ National Institutes of Health through grant 1R01HL177814-01. The authors would like to thank the anonymous reviewers for their insightful comments.}


\setcounter{section}{0}
\setcounter{figure}{0}
\setcounter{table}{0}

\vspace{-0.05in}
\section*{Appendix}

\section{Vertex Feature Extraction}

We evaluate different vertex feature extraction methods that impact the continuity and granularity of patch generation. To produce surface patches, we reduce the feature space's dimensionality using t-SNE and then classify vertex-level features through AHC. Identical parameter settings are employed for a fair comparison.
We compare HKS features with geometric features and features derived from deep learning methods:
\begin{myitemize}
    \vspace{-0.05in}
    \item Point position is a commonly used geometric feature representing each vertex's 3D position.
    \item EdgeConv~\cite{Wang-TOG2019} is a spectral-free geometric deep learning method that extracts the $n$-D features on each vertex by aggregating their nearest neighbors.
    \item SurfNet~\cite{Han-SurfNet} is a GCN-based method that learns node embedding from the graph structure through their point positions.
    \vspace{-0.05in}
\end{myitemize}

\begin{figure}[htb]
    \centering
    $\begin{array}{c@{\hspace{0.1in}}c}
            \includegraphics[width=0.45\linewidth]{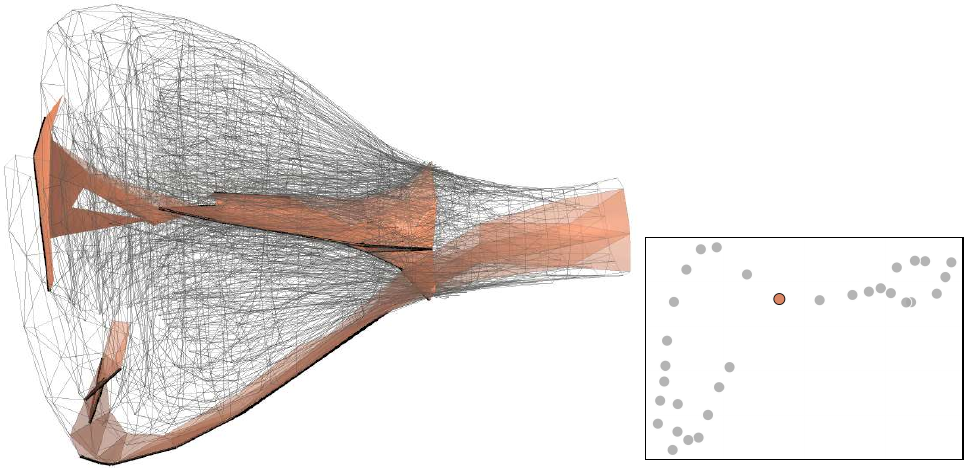}   &
            \includegraphics[width=0.45\linewidth]{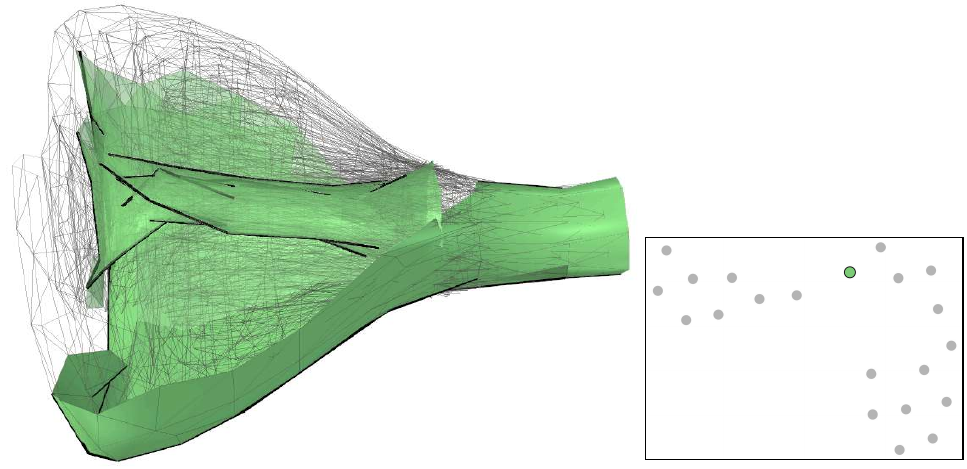}                                      \\
            \mbox{\footnotesize (a) point position}                  & \mbox{\footnotesize (b) EdgeConv}   \\
            \includegraphics[width=0.45\linewidth]{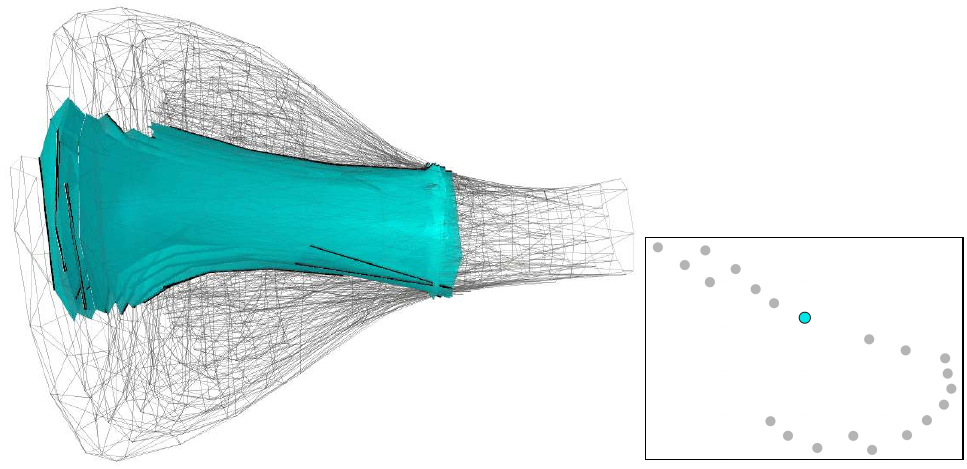} &
            \includegraphics[width=0.45\linewidth]{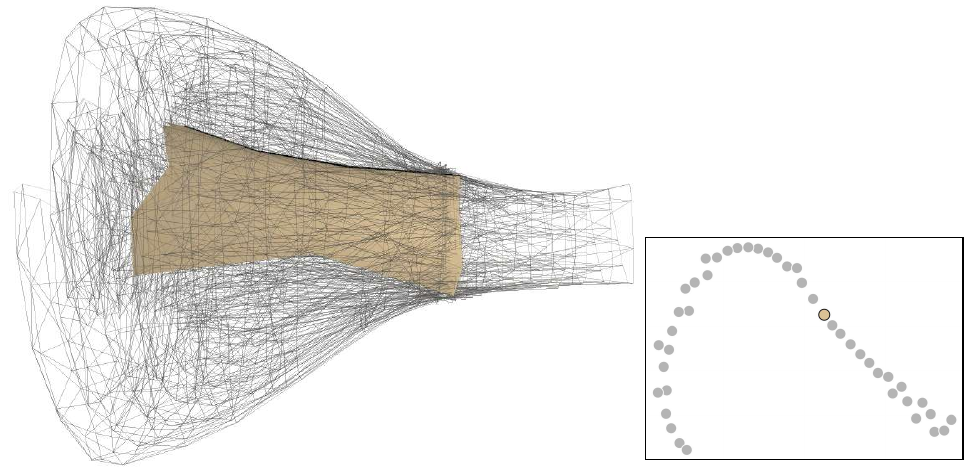}                                           \\
            \mbox{\footnotesize (c) SurfNet}                         & \mbox{\footnotesize (d) HKS (ours)} \\
        \end{array}$
    \vspace{-0.1in}
    \caption{Comparing vertex feature extraction methods in patch-level embedding using the two swirls dataset.}
    \label{feature_compare}
\end{figure}

We utilize our framework to generate patch results based on features extracted using different methods to validate their efficacy.
As shown in Figure~\ref{feature_compare}, the UMAP projection view represents the patch-level embedding aggregated from vertex-level features via UMAP aggregation.
The results indicate that the HKS feature is the best for producing continuous and fine-grained patches.
HKS captures the intrinsic surface geometry as a geometry-aware feature, making it well-suited for patch generation.
In contrast, using the point position alone lacks geometry awareness and fails to generate acceptable patches.
Features based on EdgeConv and SurfNet also prove unsuitable for patch generation, as they tend to generate large patches and do not capture fine-grained surface details.

\vspace{-0.05in}
\section{Vertex Feature Aggregation}

We compare UMAP aggregation with the following aggregations based on other DR techniques:
\begin{myitemize}
    \vspace{-0.05in}
    \item Isomap~\cite{tenenbaum2000global} computes a quasi-isometric, low-dimensional embedding for high-dimensional data by estimating the intrinsic geometry through neighbors on the manifold, offering efficiency and broad applicability.
    \item MDS~\cite{MDS} aims to place high-dimensional data points in a low-dimensional space to preserve their distances as much as possible.
    \item t-SNE~\cite{vanderMaaten-JMLR08} is a statistical method that ensures similar high-dimensional data points are represented by nearby points in the low-dimensional projection and dissimilar data points by distant points with high probability.
    \vspace{-0.05in}
\end{myitemize}

\begin{figure}[htb]
    \centering
    $\begin{array}{c@{\hspace{0.1in}}c}
            \includegraphics[width=0.45\linewidth]{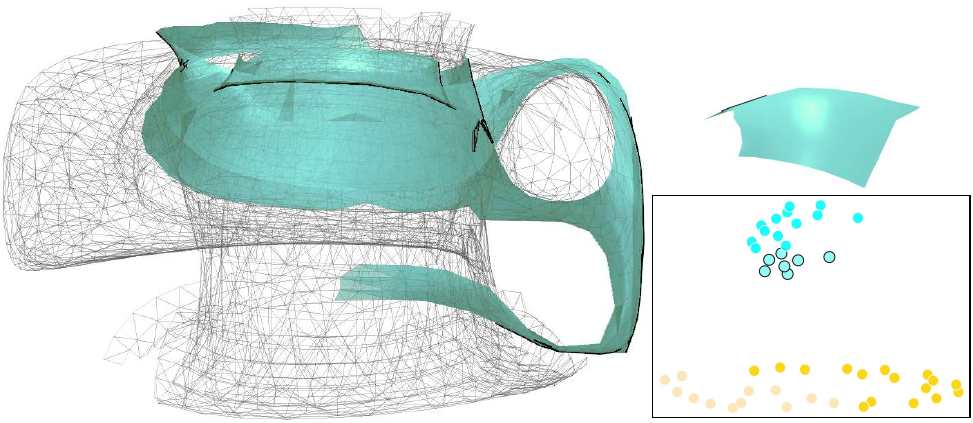} &
            \includegraphics[width=0.45\linewidth]{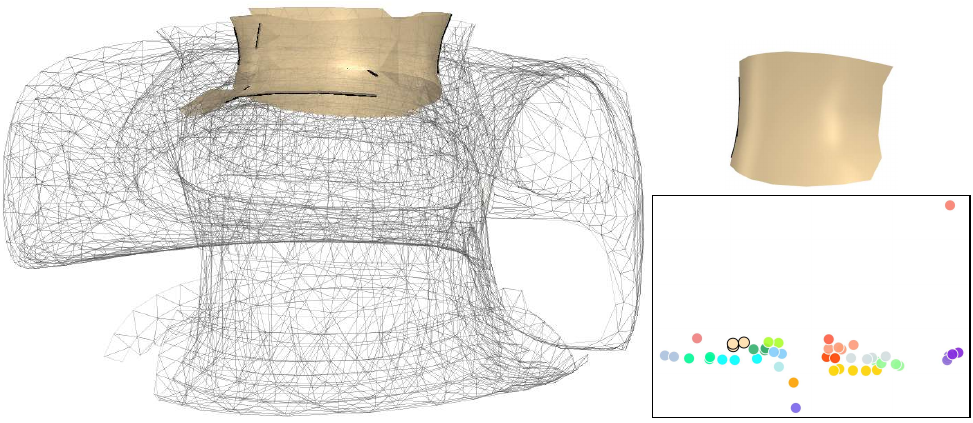}                                                       \\
            \mbox{\footnotesize (a) Isomap aggregation}                       & \mbox{\footnotesize (b) MDS aggregation}         \\
            \includegraphics[width=0.45\linewidth]{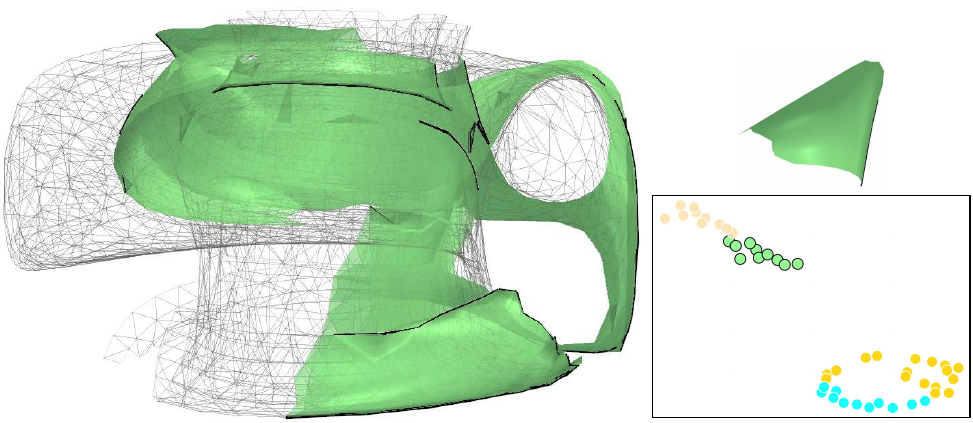}   &
            \includegraphics[width=0.45\linewidth]{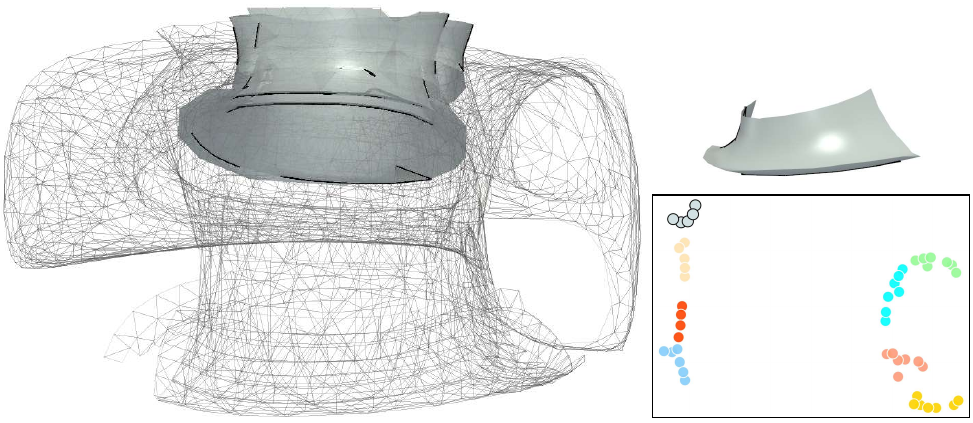}                                                      \\
            \mbox{\footnotesize (c) t-SNE aggregation}                        & \mbox{\footnotesize (d) UMAP aggregation (ours)} \\
        \end{array}$
    \vspace{-0.1in}
    \caption{Comparing vertex feature aggregation via DR in patch classification using the two swirls dataset.}
    \label{app_exp2}
\end{figure}

In Figure~\ref{app_exp2}, we compare the aggregation results using various DR techniques.
We use AHC with CC and set the same threshold 
for a fair comparison.
The results highlight the efficacy of our UMAP aggregation in generating well-separated clusters and successfully matching the inner circular flow pattern.
In contrast, Isomap and t-SNE yield fewer clusters, with patch-matching results encompassing dissimilar patches.
MDS, on the other hand, results in poorly separated clusters, and its patch-matching outcomes only identify partially similar patches.

\begin{figure}[htbp]
    \centering
    $\begin{array}{c@{\hspace{0.1in}}c}
            \includegraphics[width=0.45\linewidth]{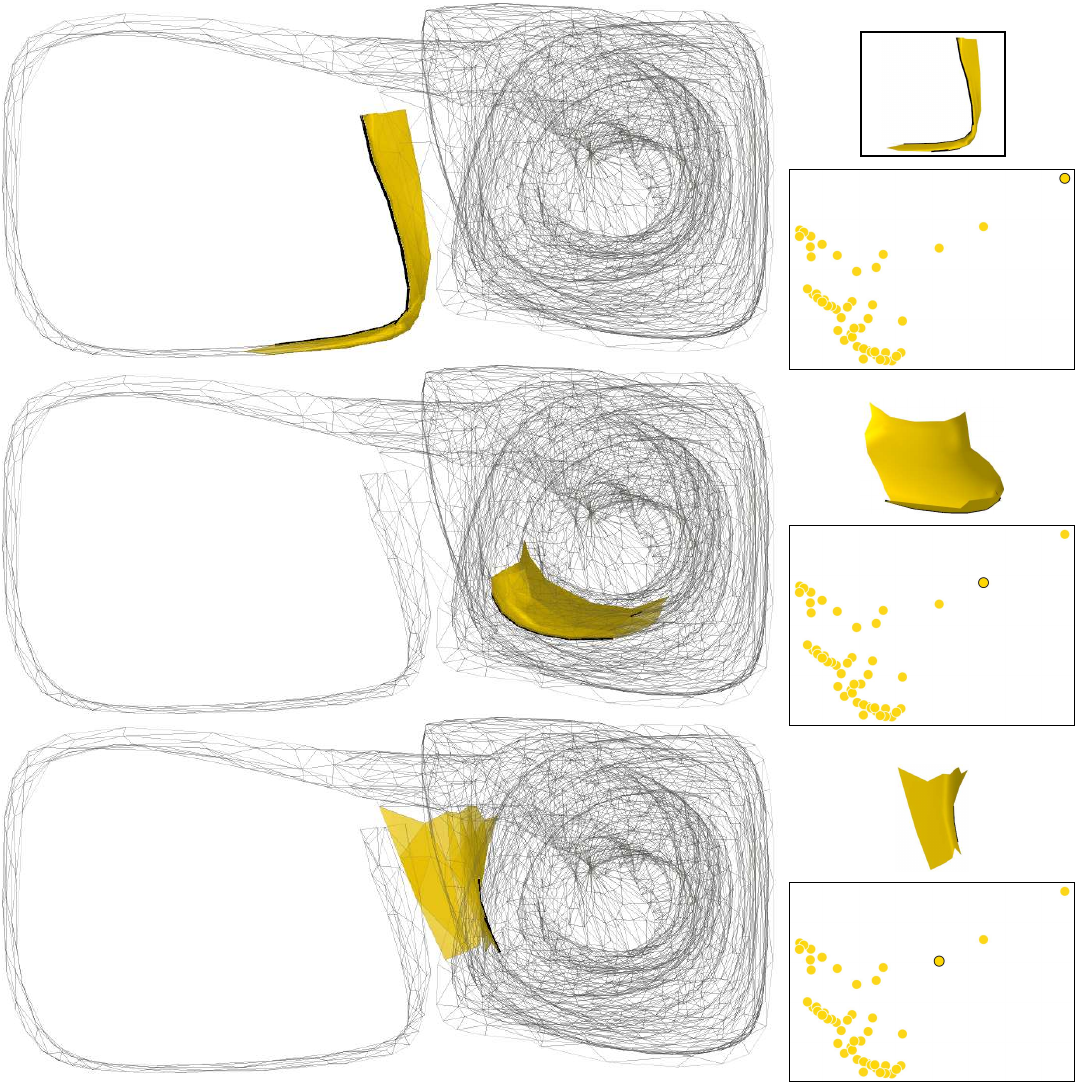} &
            \includegraphics[width=0.45\linewidth]{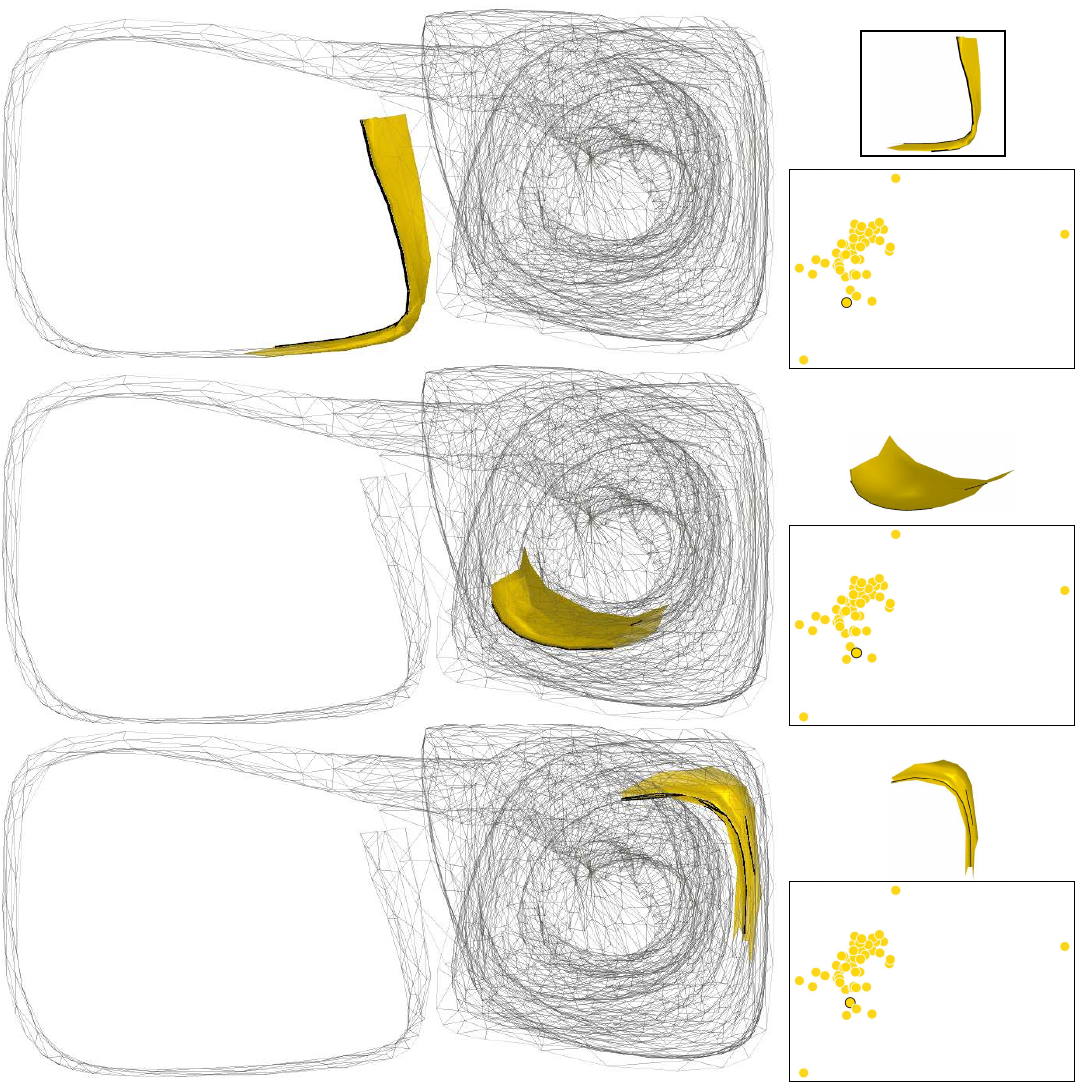}                                              \\
            \mbox{\footnotesize (a) Isomap DR}                         & \mbox{\footnotesize (b) MDS DR}         \\
            \includegraphics[width=0.45\linewidth]{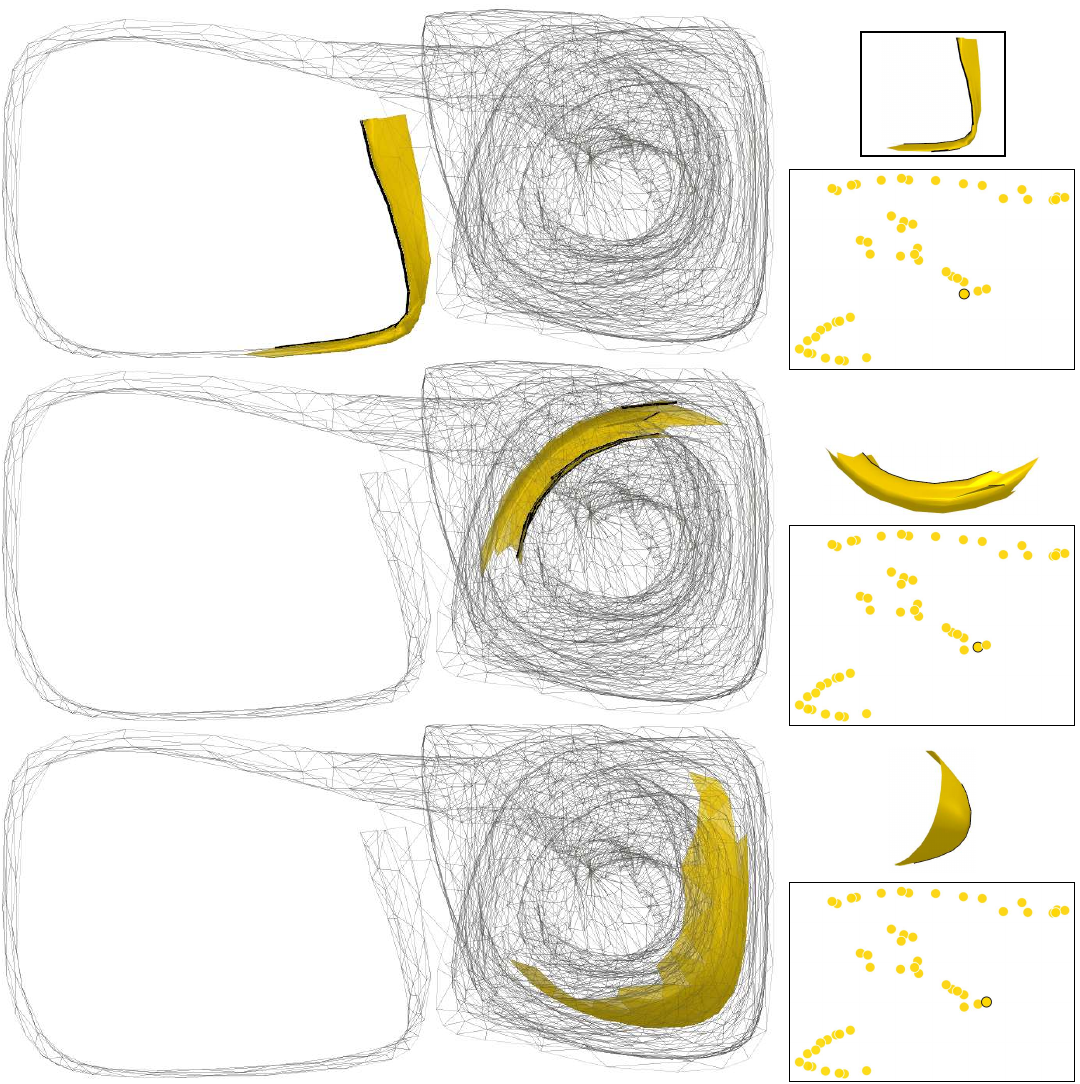}   &
            \includegraphics[width=0.45\linewidth]{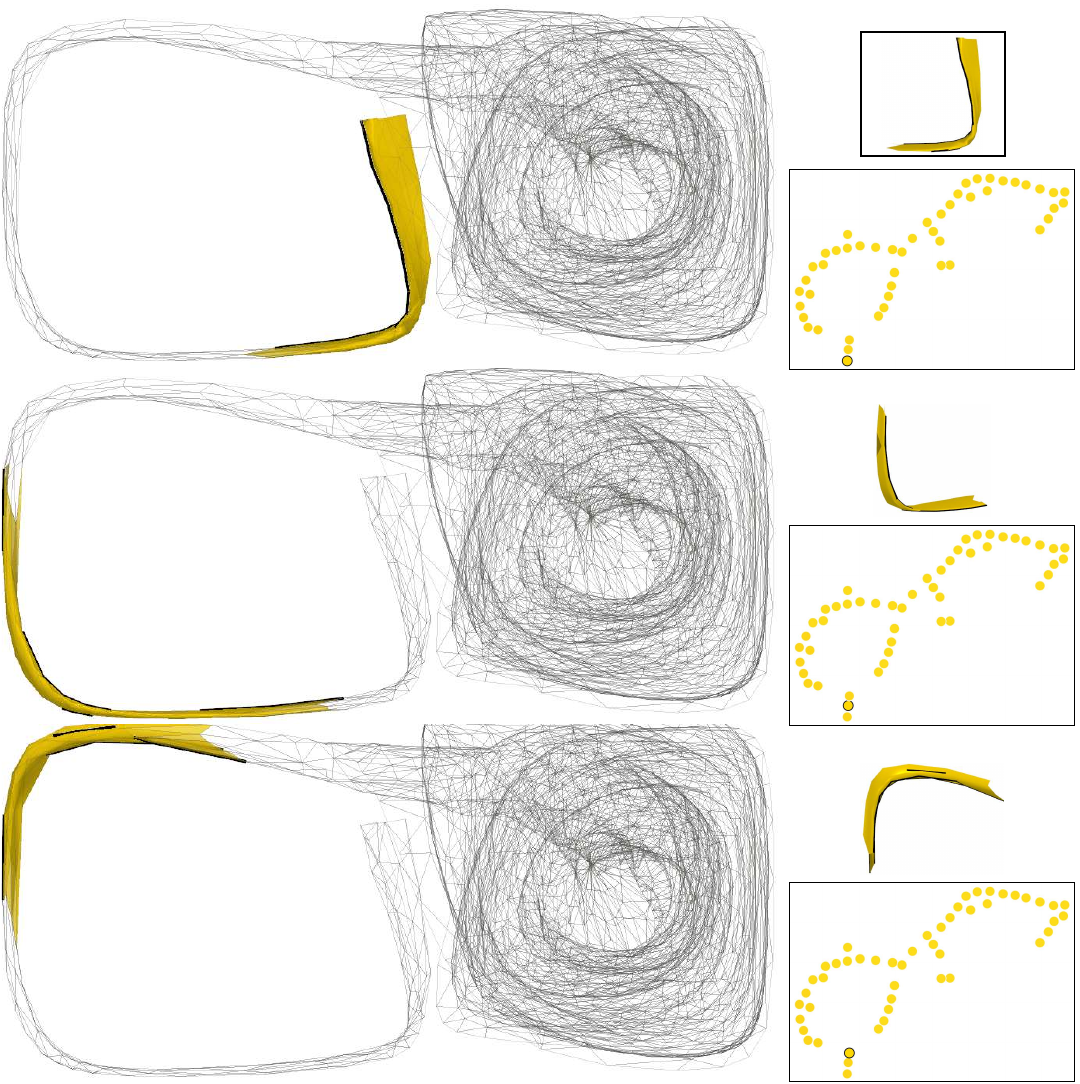}                                             \\
            \mbox{\footnotesize (c) t-SNE DR}                          & \mbox{\footnotesize (d) UMAP DR (ours)} \\
        \end{array}$
    \vspace{-0.1in}
    \caption{Comparing patch-level DR methods in patch matching using the two swirls dataset.
        Each subfigure shows the selected patch at the top, followed by the two closest patches according to their projections.}
    \label{pl-dim-red}
\end{figure}

\vspace{-0.05in}
\section{Vertex Merge Strategy}

\begin{figure}[htb]
    \centering
    $\begin{array}{c@{\hspace{0.1in}}c}
            \includegraphics[width=0.45\linewidth]{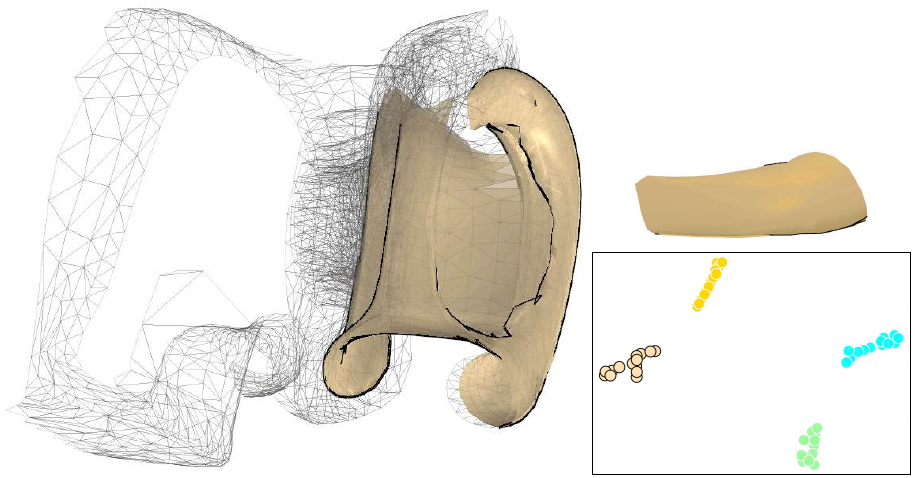} &
            \includegraphics[width=0.45\linewidth]{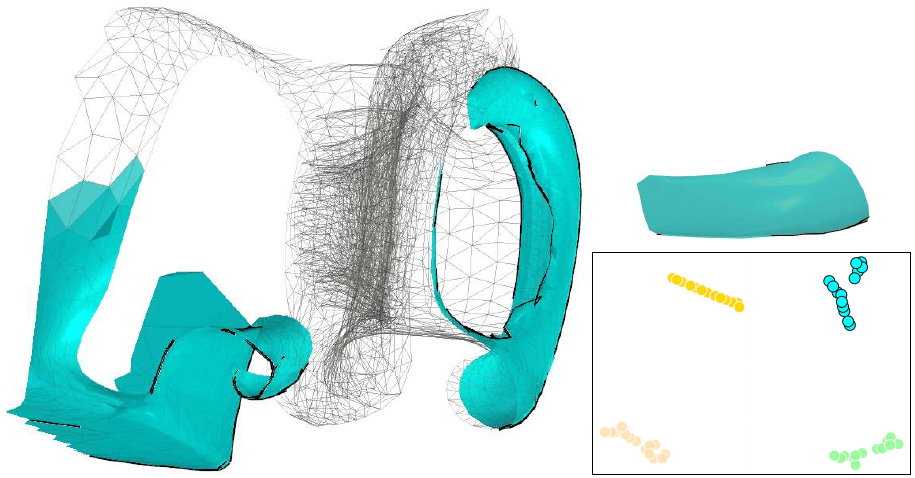}                                            \\
            \mbox{\footnotesize (a) average linkage}                 & \mbox{\footnotesize (b) complete linkage} \\
            \includegraphics[width=0.45\linewidth]{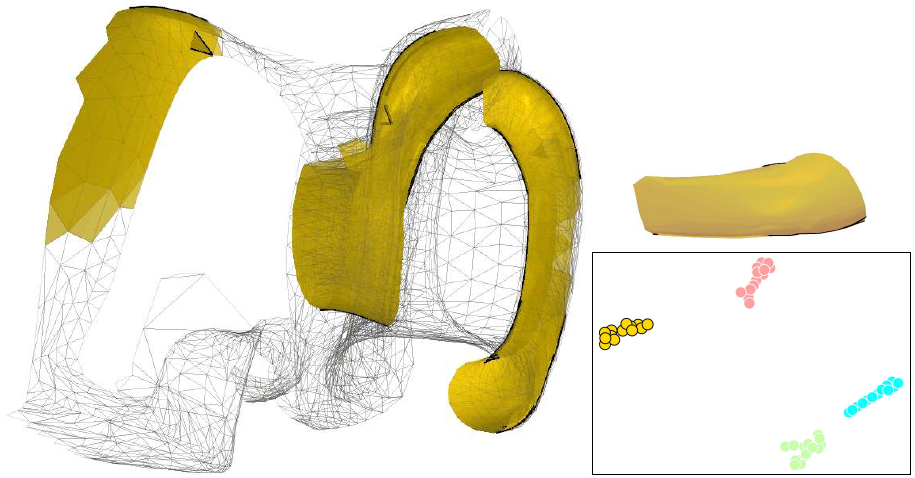}  &
            \includegraphics[width=0.45\linewidth]{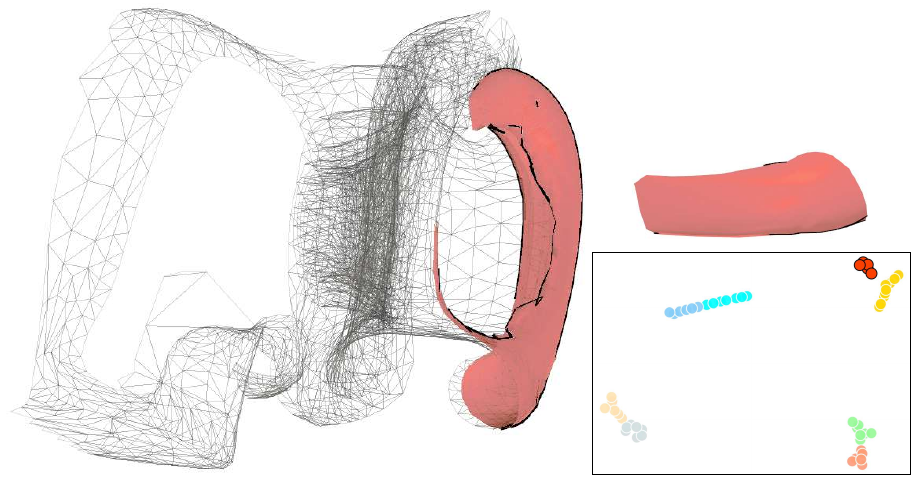}                                                \\
            \mbox{\footnotesize (c) single linkage}                  & \mbox{\footnotesize (d) Ward (ours)}      \\
        \end{array}$
    \vspace{-0.1in}
    \caption{Comparing merge strategies for AHC. Each point in the UMAP projection view represents a patch; the selected patches are highlighted with black boundaries.}
    \label{merge}
\end{figure}

We compare different merge strategies for AHC under the same $\delta$ value in Figure~\ref{merge}.
The results support our selection of the Ward strategy, as it appropriately clusters the semicircular tubular patches. This is attributed to its tendency to create compact clusters while minimizing the variance.
Other strategies, however, yield similar outcomes.
They tend to capture patches with diverse shapes, focus more on measuring distances between clusters, and do not optimize the variance increment within clusters as the Ward strategy does.

\vspace{-0.05in}
\section{Patch-Level DR}

DR is essential for patch matching, enhancing feature separability.
Still, the projection should preserve the proximity of neighboring points from the original high-dimensional space, maintaining similar shapes and structures.
This preservation becomes essential for desirable clustering.
Again, we compare four methods: Isomap, MDS, t-SNE, and UMAP.
As shown in Figure~\ref{pl-dim-red}, we select a bent stripe-like patch for querying similar patches within the same surface.
Intuitively, similar patches are distributed along the surface's stripe part where the selected patch is located. 
The projection views reveal that UMAP generates the most separable projection.
While Isomap and t-SNE follow UMAP closely in achieving separability, MDS yields a clustered 2D projection.
To assess the similarity of adjacent patches, we select patches whose 2D projection points are adjacent to the selected patch.
The Isomap's outcome appears to be the least similar to the selected patch, with adjacent points distant apart.
Although closer, MDS and t-SNE are less desirable as matched patches are located in the inner swirl rather than the stripe-like tail.
UMAP stands out with good performance in separability and similarity, with adjacent patches resembling the selected one and located on the same stripe, making it the most suitable DR technique for patch matching.

\begin{figure}[htbp]
    \centering
    $\begin{array}{c@{\hspace{0.1in}}c}
            \includegraphics[width=0.45\linewidth]{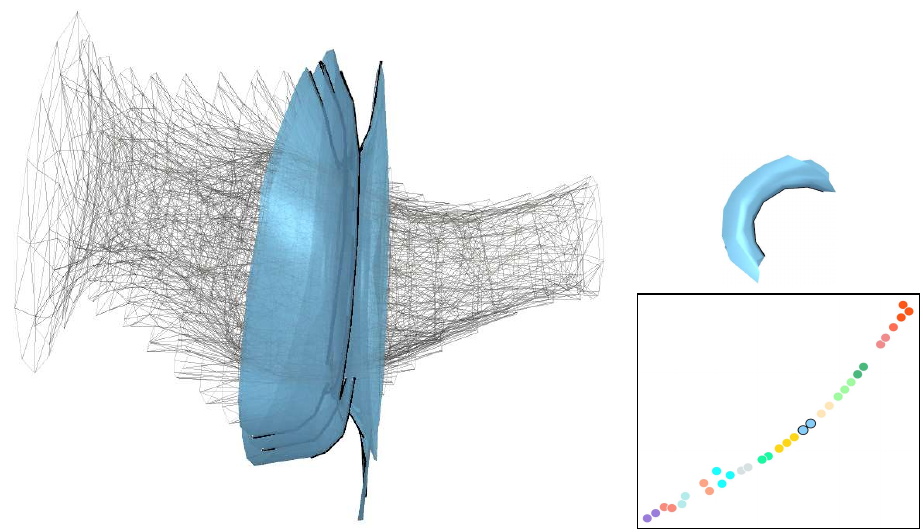} &
            \includegraphics[width=0.45\linewidth]{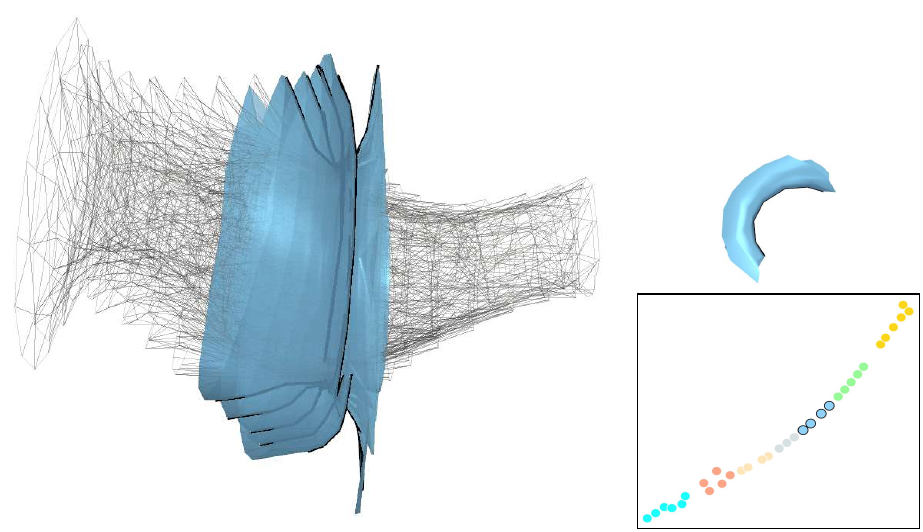}                                          \\
            \mbox{\footnotesize (a) $\delta_2=10$}                    & \mbox{\footnotesize (b) $\delta_2=30$} \\
            \includegraphics[width=0.45\linewidth]{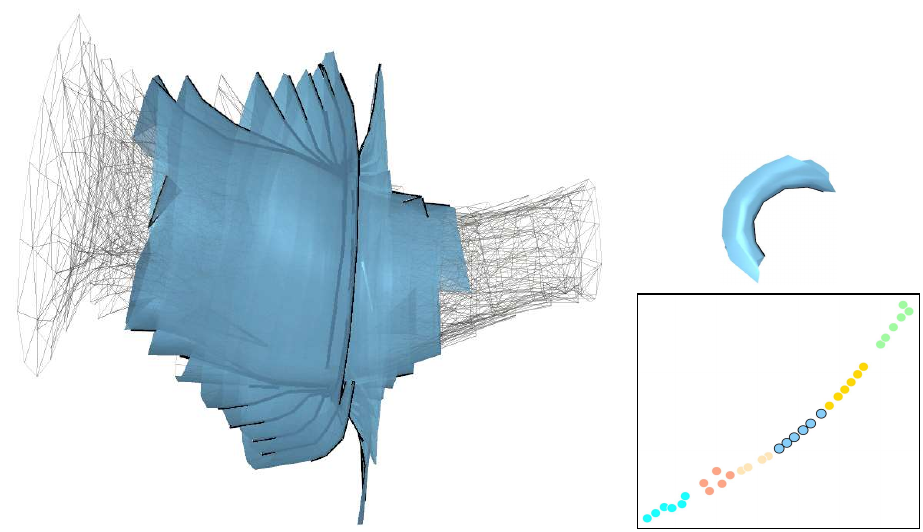} &
            \includegraphics[width=0.45\linewidth]{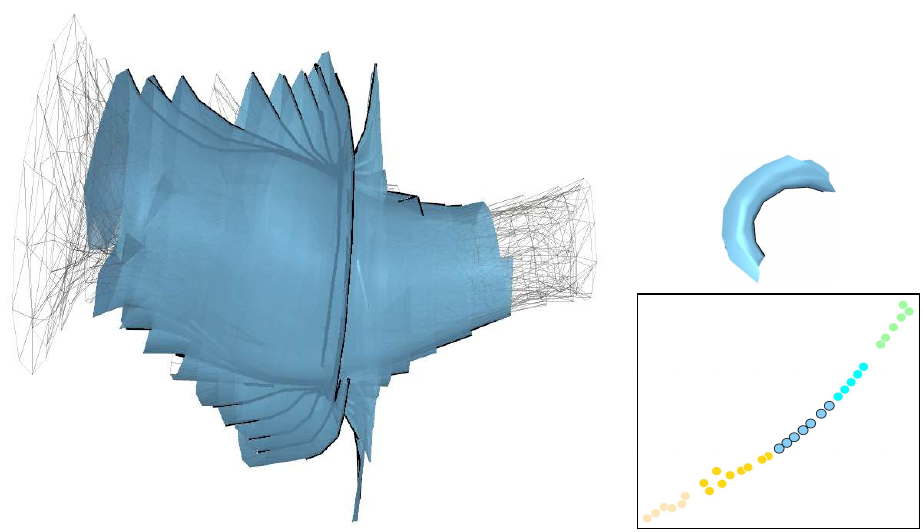}                                          \\
            \mbox{\footnotesize (c) $\delta_2=50$}                    & \mbox{\footnotesize (d) $\delta_2=70$} \\
        \end{array}$
    \vspace{-0.1in}
    \caption{
        Patch matching results under different matching tolerances ($\delta_2$) using the two swirls dataset.
    }
    \label{pl-similarity}
\end{figure}

\vspace{-0.05in}
\section{Patch Size and Matching Tolerance}

\begin{figure}[htbp]
    \centering
    \includegraphics[width=0.9\linewidth]{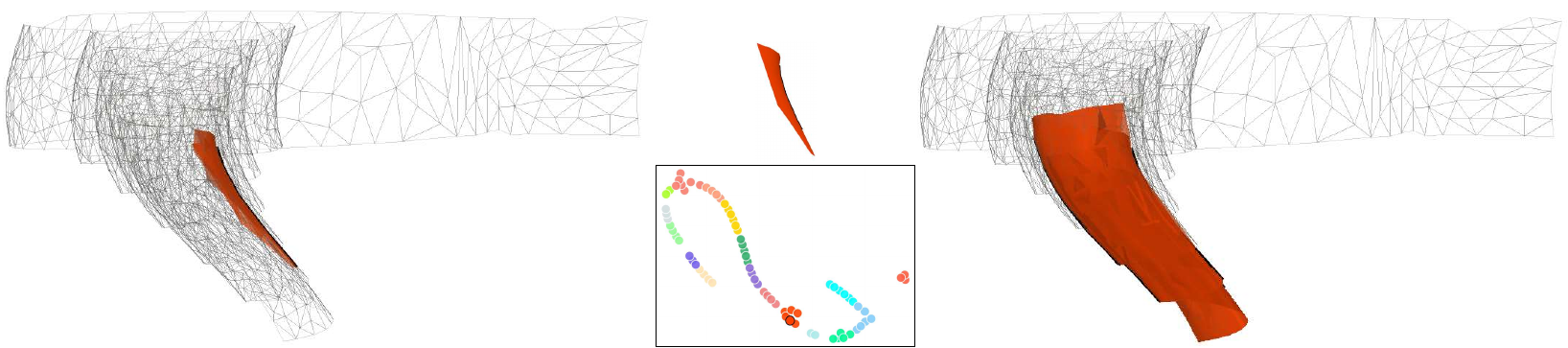} \\
    \mbox{\footnotesize (a) $\delta_1=10$, $\delta_2=10$} \\
    \includegraphics[width=0.9\linewidth]{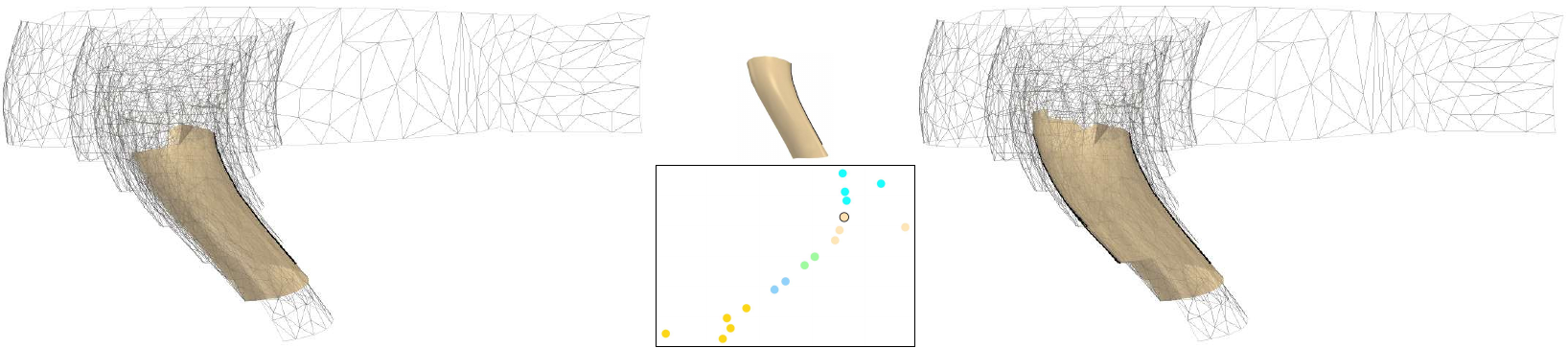} \\
    \mbox{\footnotesize (b) $\delta_1=70$, $\delta_2=30$} \\
    \includegraphics[width=0.9\linewidth]{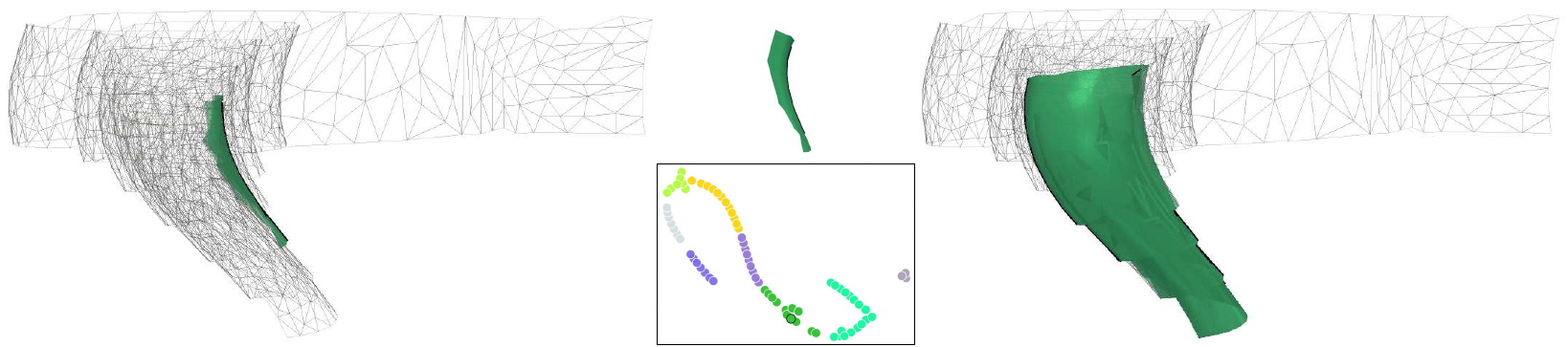}  \\
    \mbox{\footnotesize (c) $\delta_1=10$, $\delta_2=30$} \\
    \includegraphics[width=0.9\linewidth]{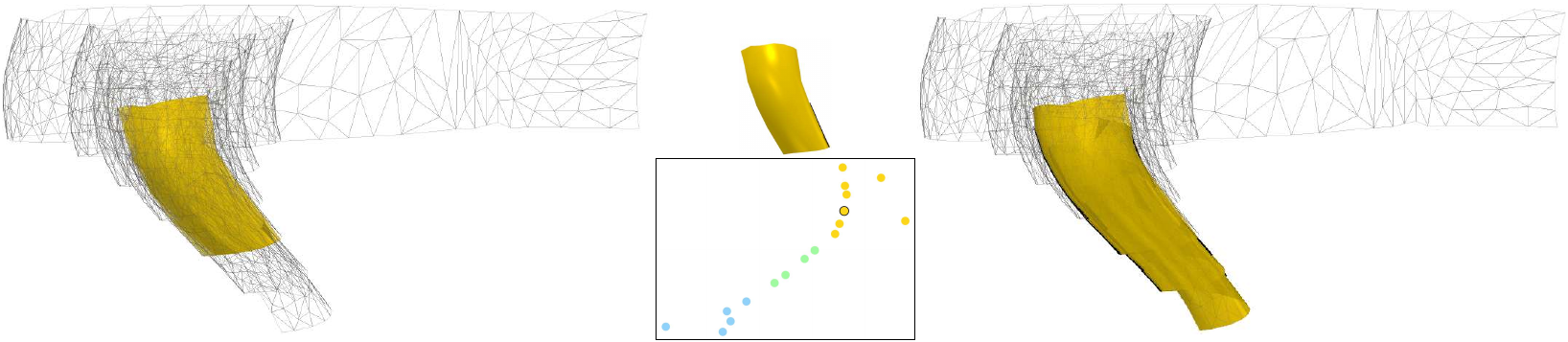} \\
    \mbox{\footnotesize (d) $\delta_1=70$, $\delta_2=50$}
    \vspace{-0.1in}
    \caption{Patch matching results under different combinations of patch size and matching tolerance ($\delta_1$, $\delta_2$) using the tornado dataset.}
    \label{pz_match}
\end{figure}

Figure~\ref{pz_match} illustrates patch-matching results with different patch sizes ($\delta_1$) and matching tolerances ($\delta_2$, which controls how similar patches should be for the query) using the tornado dataset.
We can see that a smaller/larger $\delta_1$ leads to a smaller/larger patch for the query.
Under the same $\delta_1$ setting, a larger $\delta_2$ increases the tolerance, allowing less similar patches to be matched, thereby covering a larger portion of the underlying surface. Nevertheless, different combinations of $\delta_1$ and $\delta_2$ can yield similar query results (Figure~\ref{pz_match} (a), (c), and (d)), capturing the vortex core.

We assess patch matching across various AHC distance thresholds to examine how they affect the outcome.
Using a stream surface with patches of increasing radius from inside out, we vary the value of $\delta_2$ from 10 to 70, thereby altering the range of matching results accordingly, as shown in Figure~\ref{pl-similarity}.
We expect patches in the outer swirl to first match those nearby.
As $\delta_2$ increases, the matching will gradually encompass patches from the inner swirl.
As expected, the matching result expands as $\delta_2$ increases.

\vspace{-0.05in}
\section{Quantitative and Qualitative Evaluation}

\begin{figure*}[tp]
    \centering
    \includegraphics[width=1.0\linewidth]{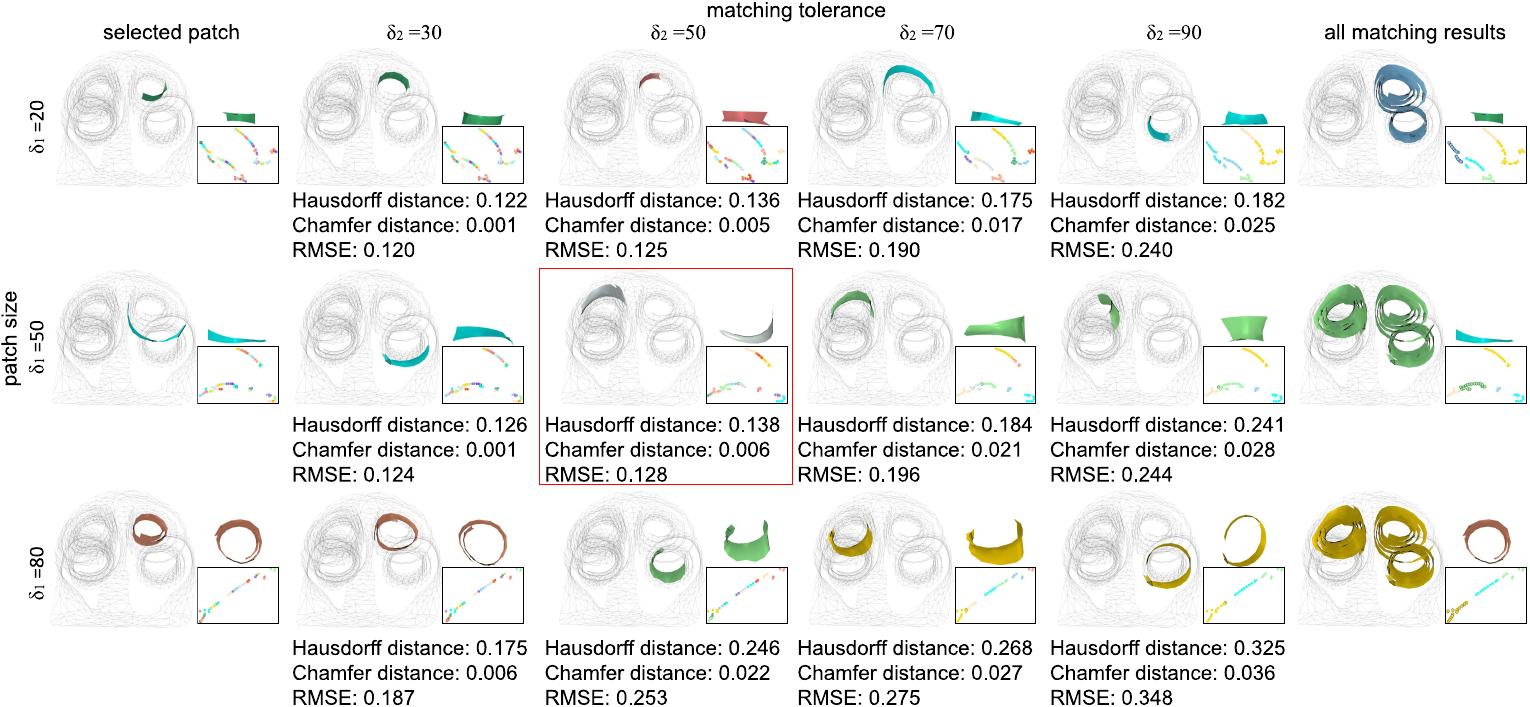} 
    \vspace{-0.2in}
    \caption{Quantitative and qualitative evaluation of patch matching under different $\delta_1$ and $\delta_2$ values using the two swirls dataset. For a selected patch in a row, the number of its matching patches varies across columns. In each column, we identify the most distant patch in the cluster to present the worst-case scenario. In the last column, all matching results are obtained with $\delta_2=90$.}
    \label{fig:metrics}
\end{figure*}

In Figure~\ref{fig:metrics}, we compare patches under different $\delta_1$ and $\delta_2$ values, presenting both quantitative and qualitative results. 
For each row, $\delta_1$ is fixed, meaning the number of partitioned patches remains the same. As $\delta_2$ increases, the number of matching patches within each cluster grows. Both quantitative and qualitative results show that the similarity of matched patches reduces as $\delta_2$ increases. 
For each column, $\delta_2$ is fixed, so the tolerance of matching remains the same. When $\delta_1$ increases, the patches get larger, and the number of partitioned patches shrinks. 
Based on these observations, we set both $\delta_1$ and $\delta_2$ to 50, balancing the number of partitioned patches and the similarity of matched patches while maintaining acceptable quantitative results. This aligns with our suggested configuration reported in Section 4.3 of the paper.

\vspace{-0.05in}
\bibliographystyle{abbrv-doi-hyperref}

\bibliography{template-arxiv}

\end{document}